\documentclass[aps,prd,a4paper,reprint,superscriptaddress,showpacs,floatfix,nofootinbib,amsmath,amssymb]{revtex4-1}
\usepackage{graphicx}
\usepackage{longtable}
\usepackage{hyperref}
\newcommand{\MS}{\overline{\mathrm{MS}}}
\newcommand{\lat}{\mathrm{lat}}
\begin{document}
\title{Nucleon isovector couplings from $N_{\mathrm{f}}=2$ lattice QCD}
\author{Gunnar S.~Bali}
\email{gunnar.bali@ur.de}
\affiliation{Institut f\"ur Theoretische Physik, Universit\"at Regensburg,
              93040 Regensburg, Germany}   
\affiliation{Tata Institute of Fundamental Research, Homi Bhabha Road, Mumbai 400005, India}
\author{Sara~Collins}
\email{sara.collins@ur.de}
\affiliation{Institut f\"ur Theoretische Physik, Universit\"at Regensburg,
              93040 Regensburg, Germany}   
\author{Benjamin~Gl\"a\ss{}le}        
\affiliation{Institut f\"ur Theoretische Physik, Universit\"at Regensburg,
              93040 Regensburg, Germany}   
\author{Meinulf~G\"{o}ckeler}        
\affiliation{Institut f\"ur Theoretische Physik, Universit\"at Regensburg,
              93040 Regensburg, Germany}   
\author{Johannes Najjar}        
\affiliation{Institut f\"ur Theoretische Physik, Universit\"at Regensburg,
              93040 Regensburg, Germany}   
\author{Rudolf~H.~R\"odl}  
\affiliation{Institut f\"ur Theoretische Physik, Universit\"at Regensburg,
              93040 Regensburg, Germany}   
\author{Andreas~Sch\"{a}fer}  
\affiliation{Institut f\"ur Theoretische Physik, Universit\"at Regensburg,
              93040 Regensburg, Germany}   
\author{Rainer~W.~Schiel}  
\affiliation{Institut f\"ur Theoretische Physik, Universit\"at Regensburg,
              93040 Regensburg, Germany}   
\author{Wolfgang~S\"oldner}  
\affiliation{Institut f\"ur Theoretische Physik, Universit\"at Regensburg,
              93040 Regensburg, Germany}   
\author{Andr\'e~Sternbeck}  
\affiliation{Theoretisch-Physikalisches Institut,
Friedrich-Schiller-Universit\"at Jena, Max-Wien-Platz 1, 07743 Jena, Germany}
\affiliation{Institut f\"ur Theoretische Physik, Universit\"at Regensburg,
              93040 Regensburg, Germany}   
\collaboration{RQCD Collaboration}
\date{\today}
\begin{abstract}
We compute the axial, scalar, tensor and pseudoscalar
isovector couplings of the nucleon as well as the induced tensor
and pseudoscalar charges in lattice simulations with $N_{\mathrm{f}}=2$
mass-degenerate non-perturbatively improved Wilson-Sheikholeslami-Wohlert
fermions. The simulations are carried out down to a pion mass
of 150~MeV and linear spatial lattice extents of up to 4.6~fm
at three different lattice spacings ranging from approximately
0.08~fm to
0.06~fm. Possible excited state contamination is carefully
investigated and finite volume effects are studied.
The couplings, determined at these lattice spacings,
are extrapolated to the physical pion mass.
In this limit we find agreement with
experimental results, where these exist, 
with the exception of the magnetic moment.
A proper continuum limit
could not be performed, due to our limited range of
lattice constants, but no significant lattice spacing dependence
is detected. Upper limits on discretization effects are estimated
and these dominate the error budget.
\end{abstract}
\pacs{12.38.Gc, 14.20.Dh, 13.60.Fz, 13.75.-n, 13.85.Tp}
\maketitle

\section{Introduction}
The electron spectrum measured in nuclear $\beta$-decays
led to Pauli's postulate
of an electrically neutral, almost massless particle in his famous
letter presented to a meeting of nuclear physicists in 1930
(reprinted and translated in Ref.~\cite{Pauli:1930pc}).
The existence of this particle was confirmed
with the discovery of the electron-antineutrino some 25 years
later~\cite{Cowan:1992xc}. The axial coupling (or charge)
of the nucleon $g_A=1.2723(23)g_V$~\cite{Agashe:2014kda}
associated with the $\beta$-decay of the neutron into a proton
is experimentally well determined
(see, e.g., Ref.~\cite{Kuhn:2008sy}) and
a parameter of fundamental importance for the structure of baryons.
Also the induced tensor charge
$\tilde{g}_T\approx\mu_p-\mu_n-1=3.7058901(5)$~\cite{Agashe:2014kda}
is well known as it quantifies
the difference between the anomalous magnetic moments of the proton and
the neutron while the vector charge $g_V=1$ is fixed due to
baryon number conservation.\footnote{In the isovector channel that we
consider here there will be corrections to $g_V=1$ to second order in the
isospin breaking parameter~\protect\cite{Ademollo:1964sr}, however, we assume
isospin symmetry.}
Computing these quantities
provides a non-trivial cross-check of
lattice predictions of similar observables.

Little is known about charges related
to flavour changing processes in any other channels
since these do not feature in tree-level standard model interactions.
However, new physics processes analogous to
the standard model nucleon $\beta$-decay or neutrino capture may depend on such
parameters, see, e.g.,
Refs.~\cite{Bhattacharya:2011qm,Ivanov:2012qe,Cirigliano:2013xha}.
This is, in particular, also relevant with respect to dark matter searches.
Only the pseudoscalar charges are, to
some extent, constrained through the effective
field theory description of low energy scattering processes
$n+\pi^+\rightarrow p +\pi^0$, see Ref.~\cite{Gasser:1987rb} and references
therein, as well as by the current algebra
relations discussed below. The
charges $g_T$ and $g_S$ can
at present only be determined
through lattice simulation.

In this article we compute the isovector nucleon
couplings $g_A$, $g_V$, $g_S$, $g_T$,
$g_P$ and the induced charges  $\tilde{g}_T$ and $g_P^*$,
simulating $N_{\mathrm{f}}=2$ QCD down to a nearly physical quark mass.
For calculations of isovector charges one can
rely on standard methods. In particular, quark-line
disconnected contributions to correlation functions
cancel in the isospin symmetric case which we realize
here, i.e.\ we neglect
the mass difference between and the electric charges of
up and down quarks.

We extract the couplings from the following form factors
at $q^2=0$, where --- in contrast to the remainder of this
article --- we employ Minkowski spacetime conventions:
\begin{align}\label{eq:gs}
\langle p|\bar{u}d|n\rangle
&=g_S(q^2)\bar{u}_p({\mathbf p}_{\mathrm{f}})u_n(\mathbf{p}_{\mathrm{i}})\,,\\\label{eq:gp}
\langle p|\bar{u}\gamma_5d|n\rangle
&=g_P(q^2)\bar{u}_p({\mathbf p}_{\mathrm{f}})\gamma_5u_n(\mathbf{p}_{\mathrm{i}})\,,\\\label{eq:gv}
\langle p|\bar{u}\gamma_\mu d|n\rangle
&=\bar{u}_p({\mathbf p}_{\mathrm{f}})\!\!\left[g_V(q^2)\gamma_{\mu}+\frac{\tilde{g}_T(q^2)}{2m_N}i\sigma_{\mu\nu}q^{\nu}\!\right]\!\!u_n(\mathbf{p}_{\mathrm{i}})\,,\\\label{eq:ga}
\langle p|\bar{u}\gamma_\mu\gamma_5 d|n\rangle
&=\bar{u}_p({\mathbf p}_{\mathrm{f}})\!\left[g_A(q^2)\gamma_{\mu}
+\frac{\tilde{g}_P(q^2)}{2m_N}q_\mu\right]\!\gamma_5u_n(\mathbf{p}_{\mathrm{i}})\,,\\
\langle p|\bar{u}\sigma_{\mu\nu} d|n\rangle
&=g_T(q^2)\bar{u}_p({\mathbf p}_{\mathrm{f}})\sigma_{\mu\nu}u_n(\mathbf{p}_{\mathrm{i}})\,,\label{eq:gt}
\end{align}
where $\sigma_{\mu\nu}=\frac{i}{2}[\gamma_{\mu},\gamma_{\nu}]$.
Above, we have assumed isospin
symmetry~\cite{Weinberg:1958ut,Bhattacharya:2011qm}. The proton and
neutron states
$|p\rangle$ and $|n\rangle$ carry
four-momenta $p_{\mathrm{f}}$ and $p_{\mathrm{i}}$, respectively.
$u_{p}$ and $u_n$ denote the proton and neutron spinors, $m_N$ the nucleon mass
and the momentum transfer is $q_0=\sqrt{m_N^2+\mathbf{p}_{\mathrm{f}}^2}-
\sqrt{m_N^2+\mathbf{p}_{\mathrm{i}}^2}$, 
$\mathbf{q}=\mathbf{p}_{\mathrm{f}}-\mathbf{p}_{\mathrm{i}}$.
The virtuality is given as $Q^2=-q^2\ge 0$.
In the isospin symmetric limit the identity
$g_V\equiv g_V(0)=1$ holds for the isovector vector charge~\cite{Ademollo:1964sr} (and
therefore $\lambda\equiv g_A/g_V=g_A$) since
\begin{align}
\langle p|\bar{u}\Gamma d|n\rangle
&=\left\langle p\left|\left(\bar{u}\Gamma u-\bar{d}\Gamma d\right)\right|p\right\rangle
=\left\langle n\left|\left(\bar{d}\Gamma d-\bar{u}\Gamma u\right)\right|n\right\rangle\nonumber\\
&=\left\langle p\left|\left(\frac23\bar{u}\Gamma u-\frac13\bar{d}\Gamma d\right)\right|p\right\rangle\\
&\qquad-\left\langle n\left|\left(\frac23\bar{u}\Gamma u-\frac13\bar{d}\Gamma d\right)\right|n\right\rangle\,.\nonumber
\end{align}
Here we construct the above matrix elements as
$\left\langle p\left|\left(\bar{u}\Gamma u-
\bar{d}\Gamma d\right)\right|p\right\rangle$, in which case
the function
$g_V(q^2)$ is also known as the Dirac form factor $F_1^p(q^2)-F_1^n(q^2)$ and
$\tilde{g}_T(q^2)$ as the Pauli form factor $F_2^p(q^2)-F_2^n(q^2)$.
Note that $\tilde{g}_T=\kappa_{u-d}\approx \kappa_p-\kappa_n$ determines
the difference
between the anomalous magnetic moments of the proton and the neutron
($\mu_p=1+\kappa_p$, $\mu_n=\kappa_n$),
$g_T=\langle 1\rangle_{\delta u-\delta d}$ is the first Mellin moment of
the isovector transversity distribution function and
$g_A=\langle 1\rangle_{\Delta u-\Delta d}$ that of the spin distribution
function.

With the exceptions of $g_P^*$ (defined below) and $\tilde{g}_T$
which require extrapolations in $q^2$,
all couplings can directly be accessed in the forward limit:
$g_S=g_S(0)$, $g_V=g_V(0)$, $g_A=g_A(0)$ and
$g_T=g_T(0)$. The determination of the pseudoscalar, axial and tensor
couplings requires polarized nucleon states.
We remark that $g_V$, $g_A$, $\tilde{g}_T$
and $g_P^*$
are scale independent while $g_T$, $g_P$ and $g_S$
carry anomalous dimensions. In these cases our
results will refer to the $\MS$-scheme at a renormalization
scale $\mu=2\,$GeV. Also note that the couplings $g_P$ and $g_S$ share the
negative anomalous dimension of the quark mass $m_{ud}$ so that
combinations $g_Sm_{ud}$ or $g_Pm_{ud}$ are scale independent.

The conservation of the isovector axial current (PCAC) implies
the relation~\cite{Adler:1966gd,Bernard:2001rs,Fuchs:2003vw}
\begin{equation}
\label{eq:mud}
m_{ud}g_P(q^2)=m_Ng_A(q^2)+\frac{q^2}{4m_N}\tilde{g}_P(q^2)\,.
\end{equation}
The right hand side of this expression can be extrapolated
to $q^2=0$, giving
\begin{equation}
\label{eq:gtr}
m_{ud}g_P=m_Ng_A=F_{\pi}g_{\pi NN}\left[1+\mathcal{O}(m_{\pi})^2\right]\,,
\end{equation}
where the second equality is the Goldberger-Treiman
relation~\cite{Goldberger:1958vp}, $F_{\pi}\approx 92\,$MeV denotes the pion
decay constant and $g_{\pi NN}$
the pion-nucleon-nucleon coupling. The chiral
perturbation theory corrections to
this relation due to the non-vanishing pion mass
are discussed
in Refs.~\cite{Gasser:1987rb,Bernard:1995dp,Fearing:1997dp,Fettes:1998ud}.
We will use the first equality in Eq.~(\ref{eq:gtr}) to determine $g_P$.

Equation~(\ref{eq:mud}) implies 
$\tilde{g}_P(q^2)=-4m_N^2g_A(q^2)/q^2$
at zero quark mass, which suggests
$\tilde{g}_P(q^2)$ is governed by a pion pole at small $q^2$
and $m_{\pi}$,
\begin{equation}
\label{eq:gp1}
\tilde{g}_P(q^2)= \frac{4c_N^2}{m_{\pi}^2-q^2}g_A(q^2)+\cdots\,,
\end{equation}
where the ellipses refer to corrections that are regular at $q^2<m_{\pi}^2$
or, equivalently, at $Q^2>-m_{\pi}^2$ and $c_N$ approaches the
nucleon mass as $m_{\pi}\rightarrow 0$.
Finally, the
induced pseudoscalar coupling
\begin{equation}
\label{eq:muon}
g^*_P=\frac{m_{\mu}}{m_N}\tilde{g}_P(-0.88\,m_{\mu}^2)
\end{equation}
quantifies the muon capture
process~\cite{Goldberger:1958vp,Bernard:1994wn,Bernard:1998gv}
$\mu^-p\rightarrow\nu_\mu n$, where the
scale $Q^2=0.88\,m_{\mu}^2$ corresponds to the kinematic threshold
and $m_{\mu}$ denotes the muon mass.

Responding to the phenomenological demand, several groups 
have recently determined $g_A$~\cite{Lin:2011sa,Dinter:2011sg,Capitani:2012gj,Green:2012ud,Lin:2012nv,Owen:2012ts,Horsley:2013ayv,Ohta:2013qda,Jager:2013kha,Alexandrou:2013jsa},
$g_T$~\cite{Aoki:2010xg,Alexandrou:2013wka},
$g_S$ and $g_T$~\cite{Green:2012ej}, $g_A$ and the induced pseudoscalar
form factor~\cite{Bratt:2010jn,Alexandrou:2010hf,Alexandrou:2013joa,Junnarkar:2014jxa},
$g_A$, $g_P$ and $g_P^*$~\cite{Lin:2008uz,Yamazaki:2009zq}
or $g_A$, $g_S$ and $g_T$~\cite{Bhattacharya:2013ehc,Alexandrou:2014wca}
or the related
form factors in lattice simulations. $g_V$ and $\tilde{g}_T$
are frequently determined in calculations of the electromagnetic
form factors~\cite{Yamazaki:2009zq,Syritsyn:2009mx,Bratt:2010jn,Collins:2011mk,Alexandrou:2011db,Alexandrou:2013joa,Jager:2013kha,Bhattacharya:2013ehc,Shanahan:2014uka,Green:2014xba,vonHippel:2014hla,Syritsyn:2014xwa},
also see Refs.~\cite{Syritsyn:2014saa,Alexandrou:2013asa,Brambilla:2014aaa,Constantinou:2014tga,Green:2014vxa}
for recent reviews. Here we compute the complete set of
isovector couplings down to a nearly physical quark mass.
We note that a
preliminary analysis on $g_A$, $g_S$ and $g_T$ using a sub-set of our ensembles
appeared in Ref.~\cite{Bali:2013nla}.

This article is organized as follows. In Sec.~\ref{latsetup}
we introduce our gauge ensembles and the analysis methods
used. Then in Sec.~\ref{sec:renorm} we check the non-perturbative
renormalization by computing $g_V$ and also
present results on $g_A$, which serves as a benchmark
quantity. In the latter case we find significant finite size effects.
These are addressed in Sec.~\ref{fse},
where we also investigate the volume dependence of
the pion mass $m_{\pi}$ and the pion decay constant $F_{\pi}$.
In Sec.~\ref{results} we present results on the remaining
couplings $g_S$, $g_T$, $\tilde{g}_T$, $g_P$, $g_P^*$ and
$g_{\pi NN}$. We summarize our findings in Sec.~\ref{conc}.

\section{Simulation details}
\label{latsetup}
\subsection{Lattice set-up}
We analyse several gauge ensembles that were generated
employing $N_{\mathrm{f}}=2$ non-perturbatively
improved Sheikholeslami-Wohlert (NPI Wilson-clover) fermions, using the
Wilson gauge action by the RQCD and QCDSF collaborations.
Three lattice spacings were realized,
corresponding to $a\approx 0.081\,$fm
($\beta=5.20$), $a\approx 0.071\,$fm ($\beta=5.29$)
and $a\approx 0.060\,$fm ($\beta=5.40$), where the lattice spacing was
set using the value $r_0= 0.5\,$fm
at vanishing quark mass, obtained
by extrapolating the nucleon mass
to the physical point~\cite{Bali:2012qs}.
This is consistent with determinations from the
$\Omega$ baryon mass~\cite{Capitani:2011fg}
or the kaon decay constant~\cite{Fritzsch:2012wq}.
With the exceptions of $\tilde{g}_T$ and $g_P^*$
we implement full order-$a$ improvement such that
our leading lattice spacing effects are
of $\mathcal{O}(a^2)$. We vary $a^2$ by
a factor of about 1.8. However, not all volumes and
quark masses are realized at all three lattice spacings.
\begin{table*}
\caption{\label{tab_1}
Details of the ensembles used in this analysis. $N(n)$
indicates the number of configurations $N$ and the number of
measurements per configuration $n$.
$N_{\mathrm{sm}}$ refers to the number of Wuppertal smearing iterations
and $t_{\mathrm{f}}$ to the sink-source time differences
realized. For small $t_{\mathrm{f}}$-values
the numbers of measurements per configuration $n$ were reduced
(indicated in brackets after the respective $t_{\mathrm{f}}/a$ entries).
Note that the pion and nucleon masses displayed were obtained
on the respective ensembles and are not extrapolated to their
infinite volume limits.
The two errors of $am_{\pi}$ and $am_{N}$ are statistical
and from varying the fit range, respectively. The error
of the pion mass in physical units includes both sources of uncertainty.}
\begin{center}
\begin{ruledtabular}
\begin{tabular}{cccccccccccc}
Ensemble& $\beta$ &  $a$ [fm] & $\kappa$     &   $V$   & $am_{\pi}$ & $m_{\pi}$ [GeV] & $am_N$&  $Lm_{\pi}$  &   $N(n)$&  $N_{\mathrm{sm}}$ & $t_{\mathrm{f}}/a$  \\
\hline
I&5.20 &  0.081& 0.13596 &  $32^3\times 64$ &  0.11516(73)(11) & 0.2795(18)	&  0.4480(31)(06)&  3.69 &  $1986(4)$ &   300 &    13  \\
\hline
II&5.29 &  0.071 & 
0.13620 &   $24^3\times 48$  & 0.15449(69)(26)    & 0.4264(20)  & 0.4641(53)(05) &   3.71   &   $1999(2)$   &  300  &   15   \\
III&&&0.13620  &   $32^3\times 64$  &  0.15298(43)(16) &   0.4222(13)   &  0.4486(22)(20) &    4.90   &   $1998(2)$   &  300  &   15,17 \\
IV&&&0.13632 &    $32^3\times 64$ & 0.10675(51)(08)  & 0.2946(14) &  0.3855(39)(23) &   3.42   &   $2023(2)$   &  400   &  7(1),9(1),11(1), \\
&&&&    &          &         &             &            &     &  & 13,15,17  \\
V&&&&   $40^3\times 64$  &  0.10465(37)(08) &  0.2888(11)    &   0.3881(32)(12) &   4.19   &   $2025(2)$   &  400  &   15  \\
VI&&&&   $64^3\times 64$  &  0.10487(24)(04)  & 0.2895(07)    &  0.3856(19)(05) &   6.71   &   $1232(2)$   &  400  &   15 \\
VII&&&0.13640 &    $48^3\times 64$ &  0.05786(51)(21)   & 0.1597(15)  & 0.3484(69)(21) &     2.78   &   $3442(2)$   &  400   &  15  \\
VIII&&&&   $64^3\times 64$ &  0.05425(40)(28) & 0.1497(13)   &  0.3398(61)(18)&   3.47   &   $1593(3)$   &  400   &  9(1), 12(2), 15\\
\hline
IX&5.40 & 0.060 &  
0.13640  &   $32^3\times 64$  &  0.15020(53)(06) &  0.4897(17)     &  0.3962(33)(06) & 4.81   &   $1123(2)$   &  400  &   17 \\
X&&&0.13647  &   $32^3\times 64$  &  0.13073(55)(28) & 0.4262(20)	& 0.3836(29)(14) & 4.18         &   $1999(2)$&     450 &    17  \\       
XI&&&0.13660  &   $48^3\times 64$  &  0.07959(25)(09)  & 0.2595(09)  & 0.3070(26)(43)&  3.82   &   $2177(2)$   &  600  &   17
\end{tabular}
\end{ruledtabular}
\end{center}
\end{table*}

\begin{figure}[t]
\centerline{
\includegraphics[width=.48\textwidth,clip=]{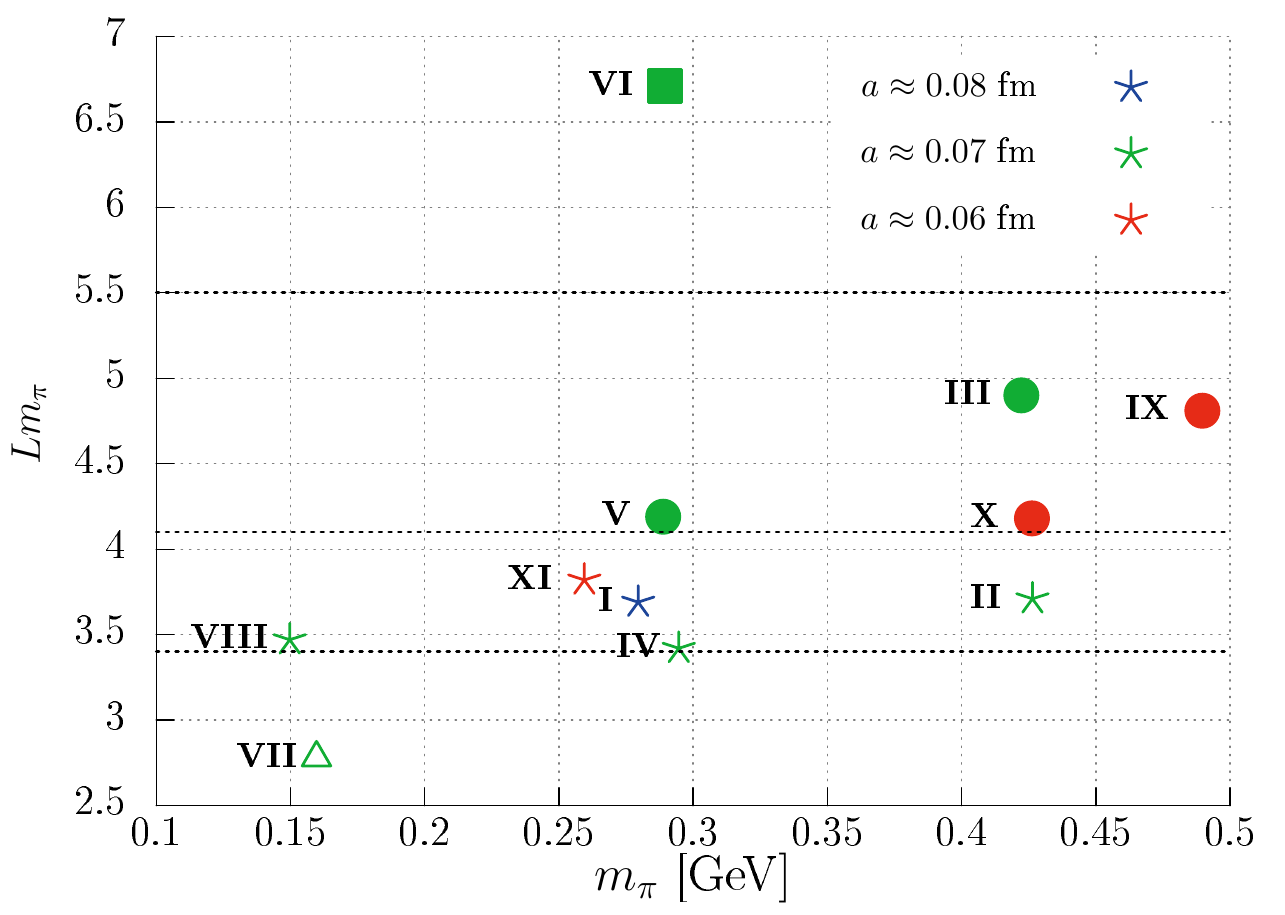}
}
\caption{Overview of the ensembles listed in Table~\protect\ref{tab_1}.
Colours encode the lattice spacings and symbols the lattice extents.
The colour and symbol labelling defined here will be used throughout
in Secs.~\protect\ref{sec:renorm} -- \protect\ref{results}.
The horizontal lines separate different volume ranges.}
\label{fig:overview}
\end{figure}

The analysed ensembles are listed in Table~\ref{tab_1} and illustrated
in Fig.~\ref{fig:overview}, see also
Ref.~\cite{Bali:2014gha}.  Our largest pion mass (ensemble IX)
corresponds to $m_{\pi}\approx 490\,$MeV. Around
$m_{\pi}\approx 425\,$MeV two lattice spacings and two different
spatial lattice extents $L$ are available.
Within the window $260\,\mathrm{MeV}\lesssim
m_{\pi}<290\,$MeV we cover three lattice spacings
and several volumes up to $Lm_{\pi}\approx 6.7$. The smallest
mass $m_{\pi}\approx
150\,$MeV was simulated at only one
lattice spacing ($a\approx 0.071$) but for two volumes
($Lm_{\pi}\approx 2.8$ and 3.5). In
Table~\ref{tab_1} we also list the nucleon masses in
lattice units. Note that, with the exception of ensemble IX, all
masses agree within one to two standard deviations with our previous
analysis~\cite{Bali:2012qs}, where in some cases we employed an
inferior quark smearing.

To improve the overlap of our nucleon interpolators with the
physical ground state, we follow Ref.~\cite{Bali:2005fu} and
employ Wuppertal (Gauss) smearing~\cite{Gusken:1989ad} of the
quark fields
\begin{equation}
\label{eq:wuppertal}
\phi^{(n)}_x=\frac{1}{1+6\delta}\left(\phi^{(n-1)}_x+
\delta\sum_{j=\pm 1}^{\pm 3}U_{x,j}\phi^{(n-1)}_{x+a\hat{\boldsymbol{\jmath}}}\right)\,,
\end{equation}
where we replace the spatial links $U_{x,j}$ by
APE-smeared~\cite{Falcioni:1984ei} gauge links
\begin{equation}
\label{eq:smear}
U_{x,i}^{(n)}= P_{\mathrm{SU}(3)}\!\left(\!\alpha\,U_{x,i}^{(n-1)}+\sum_{|j|\neq i}
U_{x,j}^{(n-1)}U^{(n-1)}_{x+a\hat{\boldsymbol{\jmath}},i}U^{(n-1)\dagger}_{x+a\hat{\boldsymbol{\imath}},j}\!\right)
\end{equation}
with $i\in\{1,2,3\}, j\in\{\pm 1,\pm 2,\pm 3\}$.
$P_{\mathrm{SU}(3)}$ denotes a projection into the $\mathrm{SU}(3)$ group
and the sum is over the four spatial ``staples'', surrounding $U_{x,i}$.
We employ 25 such gauge covariant smearing iterations and use 
the weight factor $\alpha=2.5$.
Within the Wuppertal smearing we set $\delta=0.25$
and adjust the number of iterations to optimize the
quality of the effective mass plateaus of smeared-smeared
nucleon two-point functions.

We label the nucleon source time as $t_{\mathrm{i}}=0$ and
the sink time as $t_{\mathrm{f}}$. The currents are inserted
at times $t\in[0,t_{\mathrm{f}}]$ and the relevant matrix
elements can be extracted from data within
the range $t\in[\delta t,t_{\mathrm{f}}-\delta t]$
where $\delta t\geq 2a$, due to the clover term in the action that couples
adjacent time slices. Using the sequential source method\footnote{
We also explored stochastic methods~\protect\cite{Bali:2013gxx},
see also Refs.~\protect\cite{Evans:2010tg,Alexandrou:2013xon}.}~\cite{Maiani:1987by},
all values of $t$ can be realized, essentially without overhead.
However, each $t_{\mathrm{f}}$-value requires additional computations of
sequential propagators, adding to the cost. On some of our ensembles
we vary this distance too, since this may be necessary to parameterize
and eliminate excited state contributions. The $t_{\mathrm{f}}$-values used,
the numbers of gauge configurations
$N$ and measurements per configuration $n$ are also included in
Table~\ref{tab_1}.
The statistical noise decreases with smaller Euclidean time
distances between source and sink, which means we can
reduce the number of
three-point function measurements in some cases (indicated in
brackets after the respective $t_{\mathrm{f}}/a$ entries).

\begin{figure}[t]
\centerline{
\includegraphics[width=.48\textwidth,clip=]{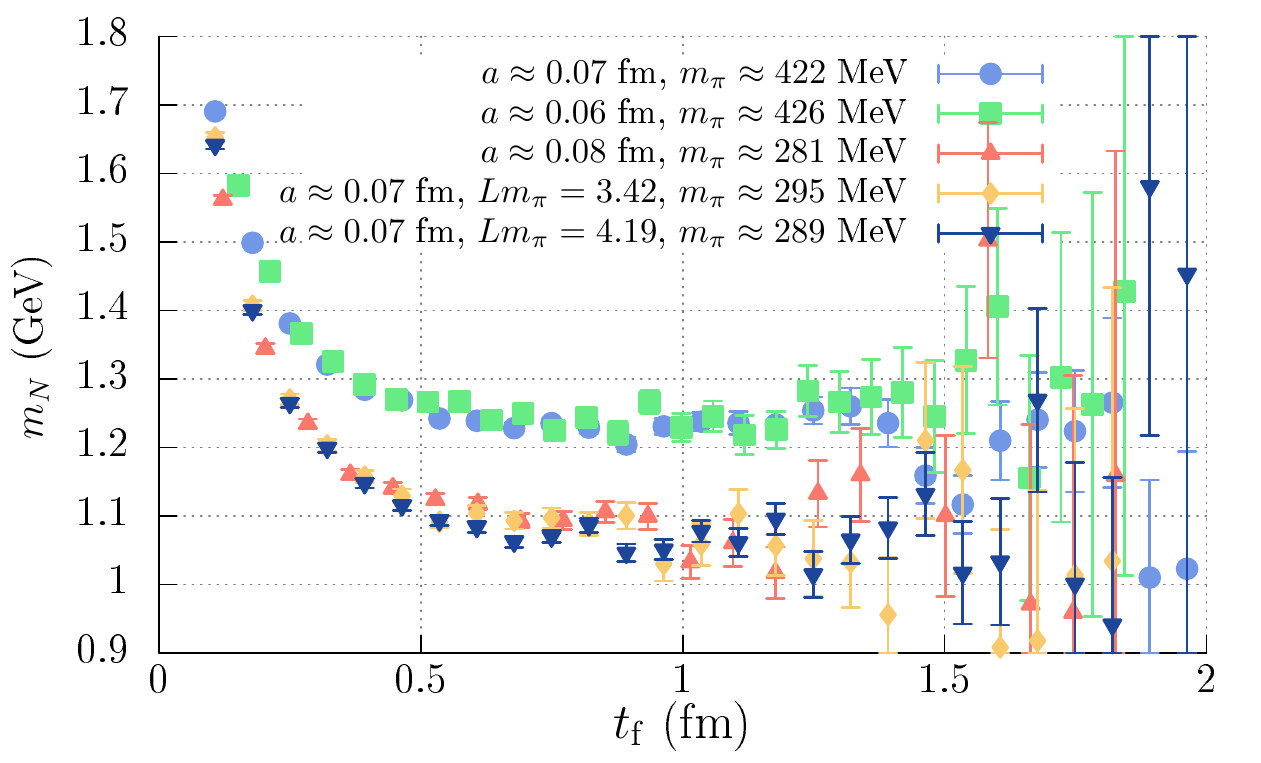}
}
\caption{Effective nucleon masses Eq.~(\protect\ref{eq:effective})
for five of our ensembles,
computed from smeared-smeared two-point functions $C_{\mathrm{2pt}}(t_{\mathrm{f}})$.}
\label{fig:smearing}
\end{figure}
\begin{figure}[t]
\centerline{
\includegraphics[width=.48\textwidth,clip=]{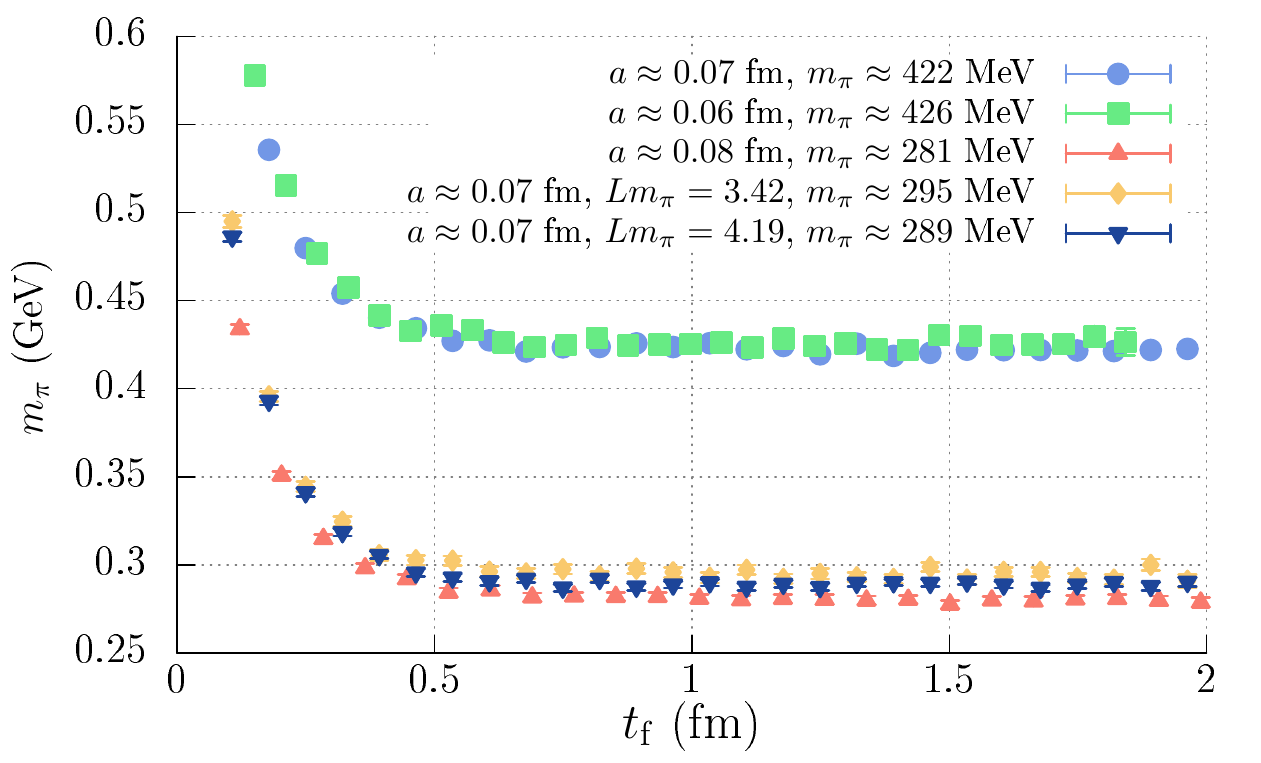}
}
\caption{The same as Fig.~\protect\ref{fig:smearing} for the
pion effective mass.}
\label{fig:smearing2}
\end{figure}

Naively, one would expect the optimal number of smearing
iterations $N_{\mathrm{sm}}$ to somewhat increase with decreasing
quark mass and, at a fixed mass, to scale with $1/a^2$,
maintaining a smearing radius that is constant in physical units.
As can be read off from the table, we approximately follow
this rule. In Fig.~\ref{fig:smearing} we compare our effective
nucleon masses
\begin{equation}
\label{eq:effective}
m_N(t_{\mathrm{f}}+a/2)=a^{-1}\ln
\left[\frac{C_{\mathrm{2pt}}(t_{\mathrm{f}})}{C_{\mathrm{2pt}}(t_{\mathrm{f}}+a)}\right]
\end{equation}
in physical units between ensembles III and X 
as well as between ensembles I, IV and V, see Fig.~\ref{fig:overview}.
These two groups of ensembles correspond to similar pion masses
but differ in terms of the lattice spacing.
Using our optimized smearing functions in the construction
of the nucleon interpolators, we do not detect any significant
lattice spacing dependence of the shapes of the resulting effective mass
curves. In Fig.~\ref{fig:smearing2} the same comparison is made
for smeared-smeared pion effective masses. Again, the shapes within
each group of ensembles are very similar while obviously
in this case we can resolve the small differences between the
lower pion masses.

Our nucleon sources were placed at different time slices
and spatial positions from configuration to configuration
to reduce autocorrelations.
Remaining autocorrelations were accounted for by binning
subsequent configurations within the jackknife error analysis
and varying the bin sizes until they were
bigger than four times the respective
estimated integrated autocorrelation times.

Recently, many groups investigated the issue of excited state
contamination of ground
state signals of three-point functions and, indeed, by
applying a more careful analysis, varying $t_{\mathrm{f}}$~\cite{Lin:2008uz,Capitani:2010sg,Dinter:2011sg,Capitani:2012gj,Green:2012ud,Lin:2012nv,Green:2012ej,Ohta:2013qda,Bhattacharya:2013ehc,Jager:2013kha,Bali:2014gha,Alexandrou:2014wca,Junnarkar:2014jxa},
using a variational approach~\cite{Owen:2012ts}
and/or by optimizing the ground state overlap of the nucleon
interpolator~\cite{Lin:2012nv,Bali:2014gha} significant effects were
detected in many matrix elements. Hence, for three of our
ensembles, covering the pion masses
150\,MeV (VIII), 290\,MeV (IV) and 425\,MeV (III),
we vary the source-sink distance $t_{\mathrm{f}}$ in addition to
the position of the current $t$, see Table~\ref{tab_1}.
Based on these results and our observation of very similar shapes
as a function of time of the effective masses computed from our
nucleon two-point functions
(see Fig.~\ref{fig:smearing}), for the remaining ensembles we
fix $t_{\mathrm{f}}\gtrsim 1\,$fm.

\subsection{Excited state analysis}
\label{sec:fits}
The spectral decompositions
for two- and three-point functions read
\begin{align}\label{eq:fit1}
C_{\mathrm{2pt}}(t_{\mathrm{f}})&=A_0e^{-m_Nt_{\mathrm{f}}}\left(
1+A_1e^{-\Delta m_Nt_{\mathrm{f}}}+\cdots\right)\,,\\
C_{\mathrm{3pt}}(t,t_{\mathrm{f}})&=A_0e^{-m_Nt_{\mathrm{f}}}\nonumber\\\nonumber
&\times
\left[B_0+B_{01}e^{-\Delta m_N {t_{\mathrm{f}}}/2}\cosh(\Delta m_N t)\right.\nonumber\\\label{eq:fit2}
&\quad\left.+B_1e^{-\Delta m_Nt_{\mathrm{f}}}+\cdots\right]\,,
\end{align}
where $\Delta m_N=m_{N'}-m_N$ denotes the mass gap between the nucleon
ground state and its first excitation and the ellipses denote
contributions from higher excited states. The coefficients
$A_0$, $A_1$, $B_0$, $B_{01}$ and $B_1$
are real if the current is self-adjoint (or anti-self-adjoint)
and the same
interpolator (i.e.\ smearing) is used at the source and the sink.
Above we assumed the temporal lattice extent to be much bigger
than $t_{\mathrm{f}}$ which holds in our case.

For a current $J=\bar{u}\Gamma d$, a nucleon interpolator
$\Phi$, a nucleon state $|N\rangle$ (and first
excitation $|N'\rangle$) and a vacuum state
$|0\rangle$ the coefficients read\footnote{In our normalization we
assume $|N'\rangle$ to be a one-particle state. However,
the precise nature of $|N'\rangle$ does not have any impact on the
discussion below nor does it affect any of the
arguments or the analysis.}
\begin{align}
A_0=\frac{|\langle 0|\Phi|N\rangle|^2}{2m_N}\,,\quad
A_1=\frac{|\langle 0|\Phi|N'\rangle|^2}
{2m_{N'}A_0}\,,\\
B_0=\frac{\langle N|J|N\rangle}{2m_N}\,,\quad
B_1=A_1\frac{\langle N'|J|N'\rangle}{2m_{N'}}\,,\label{eq:b1}\\
B_{01}=\frac{2\,\mathrm{Re}\left(\langle 0|\Phi|N\rangle\langle
N|J|N'\rangle\langle N'|\Phi^{\dagger}|0\rangle\right)}{4m_Nm_N' A_0}\,.
\end{align}
If for instance the transition matrix element $\langle N|J|N'\rangle$
and therefore $B_{01}$ is small, this does not imply a small
coefficient $B_1$ and vice versa. Hence it is essential to employ
interpolators that minimize overlaps with
higher excitations (i.e.\ $|\langle 0|\Phi|N'\rangle|\ll
|\langle 0|\Phi|N\rangle|$ etc.) and
to choose $t_{\mathrm{f}}$ sufficiently large.

For two-point functions excited states are suppressed
by factors $e^{-\Delta m_N t_{\mathrm{f}}}$ while in the three-point functions
there exist contributions $\propto e^{-\Delta m_N t_{\mathrm{f}}/2}$. If the
ratio of the three-point function over the two-point function
is constant upon varying $t$, this indicates
a small $B_{01} e^{-\Delta m_N t_{\mathrm{f}}/2}$ term, but still
terms $(B_1-A_1)e^{-\Delta m_N t_{\mathrm{f}}}$ may be present that
can only be isolated if $t_{\mathrm{f}}$ is varied as well.
Up to such corrections the ratio
reads
\begin{equation}\label{eq:ratio}
R(t,t_{\mathrm{f}})\equiv\frac{C_{\mathrm{3pt}}(t,t_{\mathrm{f}})}{
C_{\mathrm{2pt}}(t_{\mathrm{f}})}=
\frac{\langle N|J|N\rangle}{2m_N}+\cdots\,,
\end{equation}
where $\langle N|J|N\rangle$ is the matrix element of interest.
Fitting this combination to a constant suffers from the obvious caveats
described above.

Recently, the summation method~\cite{Maiani:1987by}
\begin{equation}
\label{eq:summa}
\frac{a}{t_{\mathrm{f}}}\sum_{t=\delta t}^{t_{\mathrm{f}}-\delta t}
R(t,t_{\mathrm{f}})=\frac{\langle N|J|N\rangle}{2m_N}+c\frac{a}{t_{\mathrm{f}}}+
\mathcal{O}(e^{-\Delta m_Nt_{\mathrm{f}}})
\end{equation}
was advertized~\cite{Capitani:2010sg} as a more reliable alternative.
In this case corrections $\propto  e^{-\Delta m_Nt_{\mathrm{f}}/2}$ are removed,
but a $c/t_{\mathrm{f}}$ term is introduced, adding a not
necessarily small parameter $c$ to the fit function. We refrain from
quoting the corresponding results as direct fits to the known parametrization
Eqs.~(\ref{eq:fit1}) and (\ref{eq:fit2}) are cleaner theoretically
and utilize the whole functional dependence of the data
on $t$ and $t_{\mathrm{f}}$. Since the summation method appears to be
very popular, we discuss it in more detail in Sec.~\ref{sec:sum} below.

\begin{figure}[t]
\centerline{
\includegraphics[width=.48\textwidth,clip=]{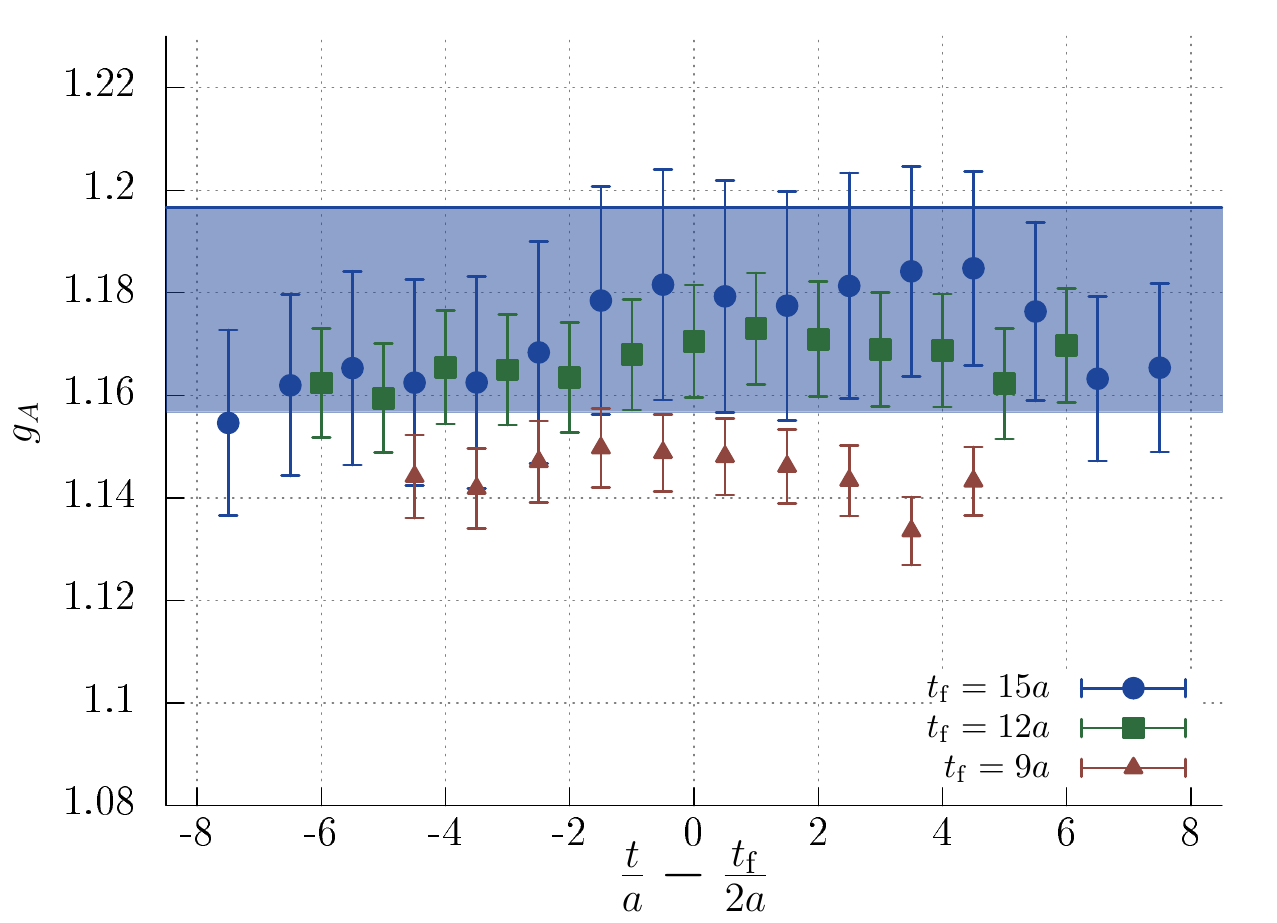}
}
\caption{The renormalized ratio
Eq.~(\protect\ref{eq:ratio})
for the example of $g_A$ obtained on ensemble VIII~($m_{\pi}\approx 150\,$MeV, $a\approx 0.071\,$fm) for three different values
of $t_{\mathrm{f}}$. The shaded region represents the
result of a constant fit in the range $t/a\in[4,11]$ to the
$t_{\mathrm{f}}=15a$ data.}
\label{fig_gab_fit}
\end{figure}

First we discuss $g_A$.
In Fig.~\ref{fig_gab_fit} we display the ratio Eq.~(\ref{eq:ratio})
of the renormalized
(see Sec.~\ref{sec:renorm} below) three-point over the two-point
function obtained from ensemble
VIII~($m_{\pi}\approx 150\,$MeV) at
$t_{\mathrm{f}}=15a\approx 1.07\,$fm, $t_{\mathrm{f}}=12a$ and
$t_{\mathrm{f}}=9a$. All three sets are compatible with constants,
however, the $t_{\mathrm{f}}=9a\approx 0.64\,$fm data are significantly
lower than the two other sets. This indicates a small $B_{01}$-coefficient
in Eq.~(\ref{eq:fit2}).
The effect of $B_1-A_1$ (or higher excitations)
becomes visible at $t_{\mathrm{f}}<1\,$fm. 
Whenever $B_{01}$ could not be resolved, such as in the case shown
in the figure, $g_A$ was obtained from a fit of the plateau to a constant.
Otherwise multi-exponential fits Eqs.~(\ref{eq:fit1}) and (\ref{eq:fit2})
were performed, where $B_1$ was set to zero for the ensembles
with only one $t_{\mathrm{f}}$-value.
These multi-exponential fits gave numbers compatible with those
obtained by fitting the $t_{\mathrm{f}}\gtrsim 1\,$fm ratios to constants
for $g_A$ as well as for all the other couplings discussed in this
article.

In all analyses presented in this article
the fit ranges were selected based on the goodness
of the correlated $\chi^2$-values and the stability of the results
upon reducing the fit range, i.e.\ increasing the minimal
distance between the current and the source-sink $\delta t$ or
reducing the number of $t_{\mathrm{f}}$-values entering the fit.
A systematic error was then estimated by varying the fit-range,
and the parametrization, e.g., allowing for
$B_1\neq 0$ in cases where this parameter was consistent with zero.

\begin{figure}[t]
\centerline{
\includegraphics[width=.48\textwidth,clip=]{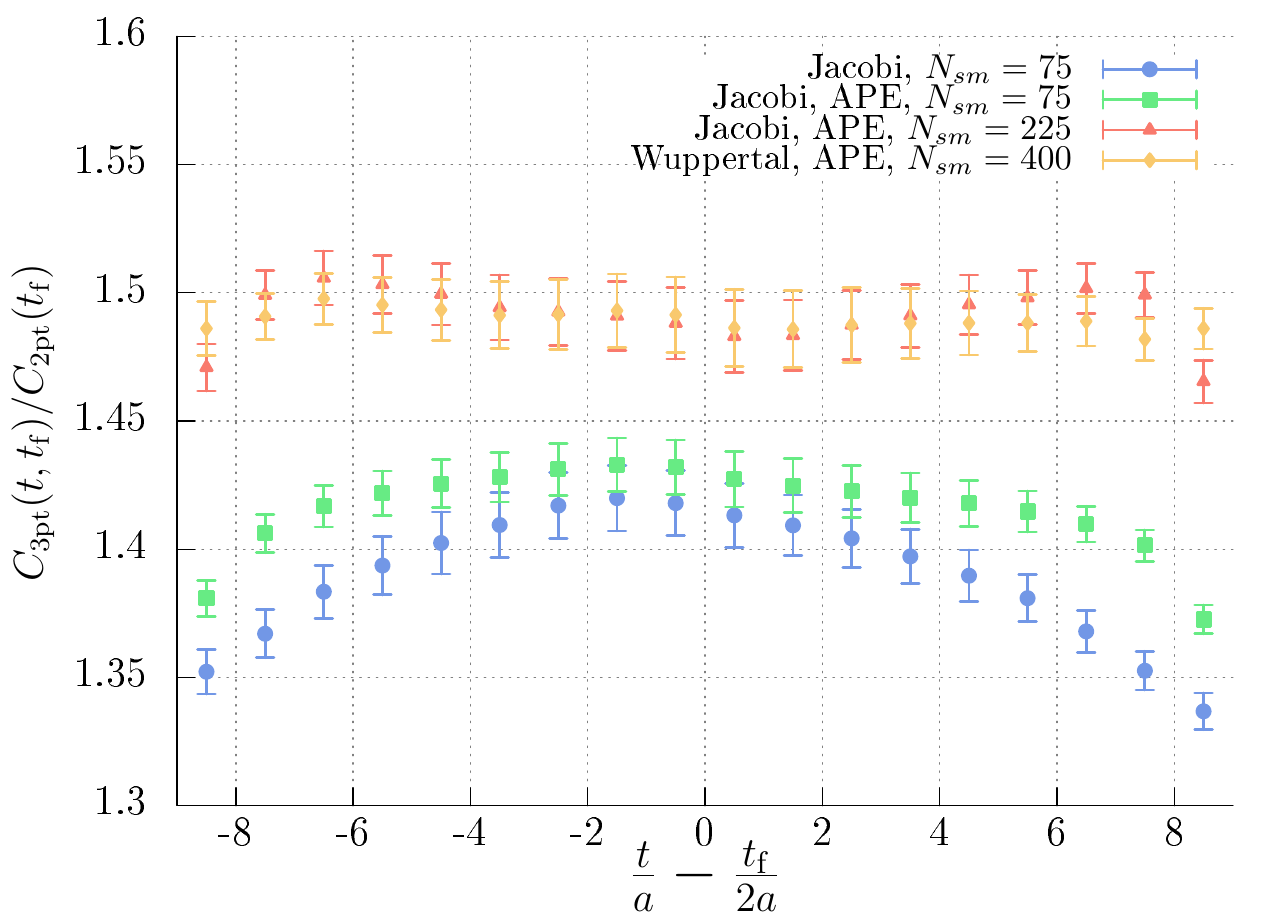}
}\caption{The ratio of the three- over the two-point function for
$g_A$ at $t_{\mathrm{f}}=17a$ on ensemble IX~($m_{\pi}\approx 490\,$MeV,
$a\approx 0.060\,$fm) with
different smearing methods.}
\label{fig_smear_gab}
\end{figure}

In some publications a dependence of the ratio of the axial
three-point over the two-point function on $t_{\mathrm{f}}$ and on $t$
is reported that is much stronger than what we observe,
see, e.g., Refs.~\cite{Green:2012ud,Jager:2013kha,Junnarkar:2014jxa} while
the results of, e.g., Ref.~\cite{Bhattacharya:2013ehc}
are quite similar to ours. This motivates us to
compare two different smearing methods found in
the literature on ensemble IX: Jacobi smearing~\cite{Allton:1993wc}
and Wuppertal smearing~\cite{Gusken:1989ad}.
With the optimized root
mean squared smearing radius\footnote{All three quarks
within the interpolator $\Phi^{\dagger}$, used to create a state
with the quantum numbers of the nucleon, are smeared applying
the same matrix $A$ to $\delta$-sources. For the case of Wuppertal
smearing this matrix $A$ with space and colour indices is iteratively defined
in Eq.~(\protect\ref{eq:wuppertal}). We compute a gauge
invariant smearing function $\psi(\mathrm{r})\geq 0$:
$\psi^2(\mathrm{r})
=\sum_{ab}|(A\delta^a)_{\mathrm{r},b}|^2$, where 
the $\delta$-source has only one non-vanishing entry,
at the spatial origin and of colour $a$. The RMS radius is computed in
the usual way:
$r^2_{\mathrm{RMS}}=[\sum_{\mathbf{n}}r^2\psi(\mathbf{n}a)]/[\sum_{\mathbf n}\psi(
\mathbf{n}a)]$,
where the sum extends over all (three-dimensional) lattice points
and $r^2=\sum_i \min[(an_i)^2,(an_i-L)^2]$, taking account of the
periodic boundary conditions. In principle one could also, by analogy with
quantum mechanics, define $r_{\mathrm{RMS}}$ with a weight factor
$\psi(\mathbf{r})^2$, rather than $\psi(\mathbf{r})$. Due to the
approximately 
Gaussian profile, this definition will result in a radius that is smaller
by a factor of about $\sqrt{2}$ than the numbers we quote.}
$r_{\mathrm{RMS}}\approx 0.58\,$fm both methods give
similar results, see the comparison between the
$N_{\mathrm{sm}}=225$ Jacobi and the $N_{\mathrm{sm}}=400$ Wuppertal
smearing in Fig.~\ref{fig_smear_gab}. In these cases
the parameter $B_{01}$ is statistically compatible with zero.
Without realizing additional $t_{\mathrm{f}}$-values we cannot
determine $B_1$ but, based on our detailed investigations
on ensembles III, IV and VIII, it is reasonable to assume that the
effect of this term is statistically insignificant at
$t_{\mathrm{f}}=17a\approx 1.03\,$fm.

For the Jacobi algorithm additionally we realize $N_{\mathrm{sm}}=75$,
reducing the smearing radius to $r_{\mathrm{RMS}}\approx 0.37\,$fm
and $r_{\mathrm{RMS}}\approx 0.34\,$fm with and
without APE smearing, respectively.
This results in some curvature due to the effect of 
excited states, i.e.\ the parameter
$B_{01}$ now significantly differs from zero.
Comparing the two $N_{\mathrm{sm}}=75$
results illustrates that
APE smearing the spatial gauge links is less important than
varying the number of smearing iterations. However, APE smearing
further increases the overlap with the physical ground state.

For $t_{\mathrm{f}}\rightarrow\infty$ and $t\approx t_{\mathrm{f}}/2$
obviously all four data sets must approach the same asymptotic
value. However, from the comparison shown in
Fig.~\ref{fig_smear_gab} it is clear that with the two
inferior smearing functions $t_{\mathrm{f}}$ needs to be chosen much larger ---
or at least additional source-sink distances need to be realized,
to enable a determination of the parameters $B_1$ and $B_{01}$ and a subsequent
extrapolation. Otherwise, in these cases
an incorrect result would be obtained: Clearly,
the minimal sensible value of $t_{\mathrm{f}}$
does not only depend on the statistical accuracy
but also on the quality of the interpolator.
For instance, an ideal interpolator $\Phi$ with 100\%
ground state overlap would, up to issues related to the
locality of the action,
eliminate the time-dependence altogether.

\begin{figure}[t]
\centerline{
\includegraphics[width=.48\textwidth,clip=]{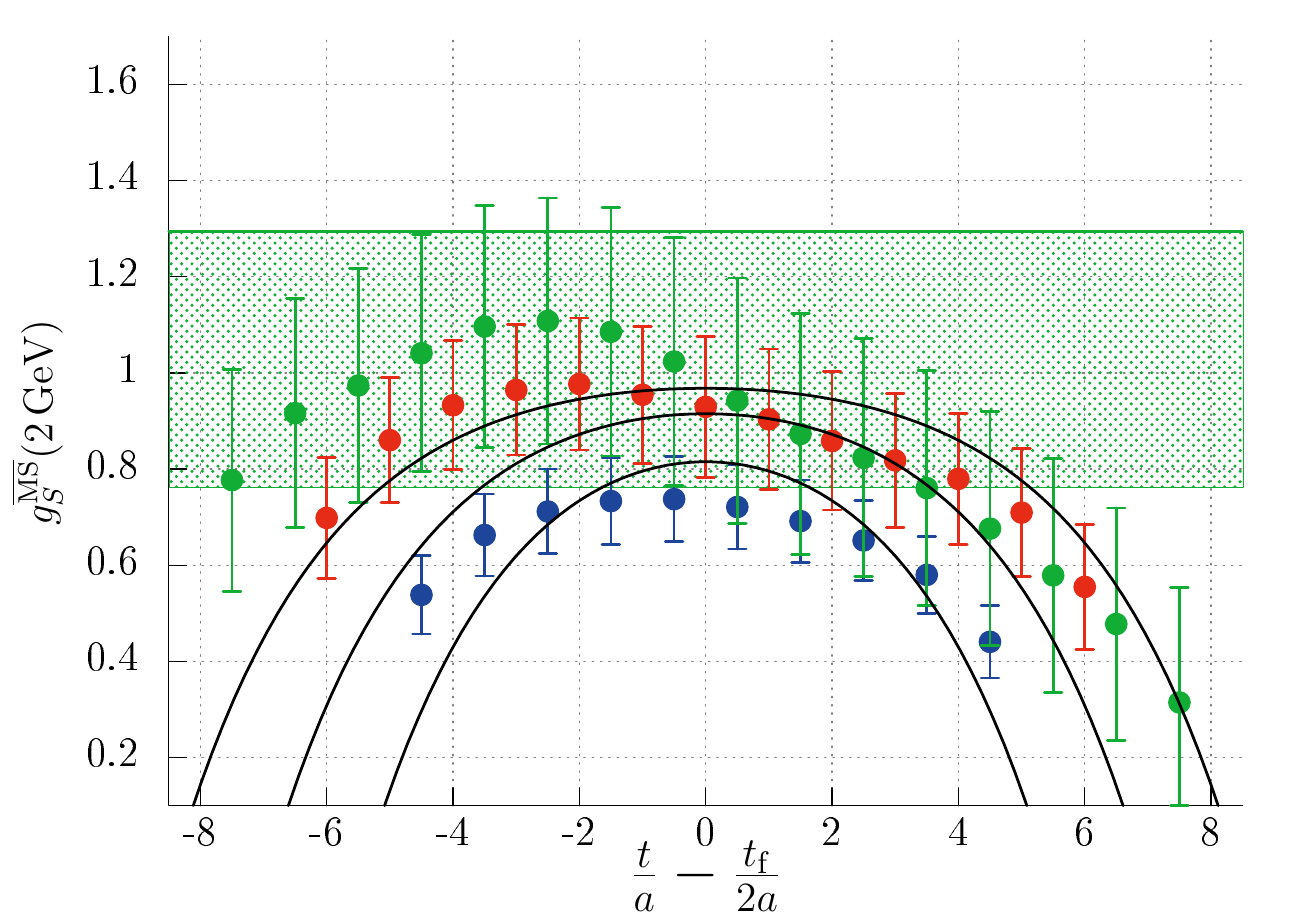}
}
\caption{The combination
$C_{\mathrm{3pt}}(t,t_{\mathrm{f}})/(A_0e^{-m_Nt_{\mathrm{f}}})$ with
$t_{\mathrm{f}}/a\in\{9, 12, 15\}$ on ensemble VIII~($m_{\pi}\approx 150\,$MeV, $a\approx 0.071\,$fm), multiplied by the appropriate
  renormalization factors to give $g_S^{\MS}(2\,\mathrm{GeV})$.
$A_0e^{-m_Nt_{\mathrm{f}}}$ corresponds to the
  ground state contribution to $C_{\mathrm{2pt}}(t_{\mathrm{f}})$ obtained
  from a simultaneous fit according to Eqs.~(\protect\ref{eq:fit1}) and (\protect\ref{eq:fit2})
to $C_{\mathrm{3pt}}$ and $C_{\mathrm{2pt}}$. The fit ranges were
  $t_{\mathrm{f}}/a\in[2,26]$ for $C_{\mathrm{2pt}}$ and $\delta t=2a$ for
  $C_{\mathrm{3pt}}$ where $B_1$ is set to zero.
Also shown are the resulting fit curves for
  each $t_{\mathrm{f}}$. The shaded region indicates the fitted value
  of $g_S^{\MS}(2\,\mathrm{GeV})$ and the
  corresponding statistical uncertainty.}
\label{fig_gs_fit}
\end{figure}

\begin{figure}[t]
\centerline{
\includegraphics[width=.48\textwidth,clip=]{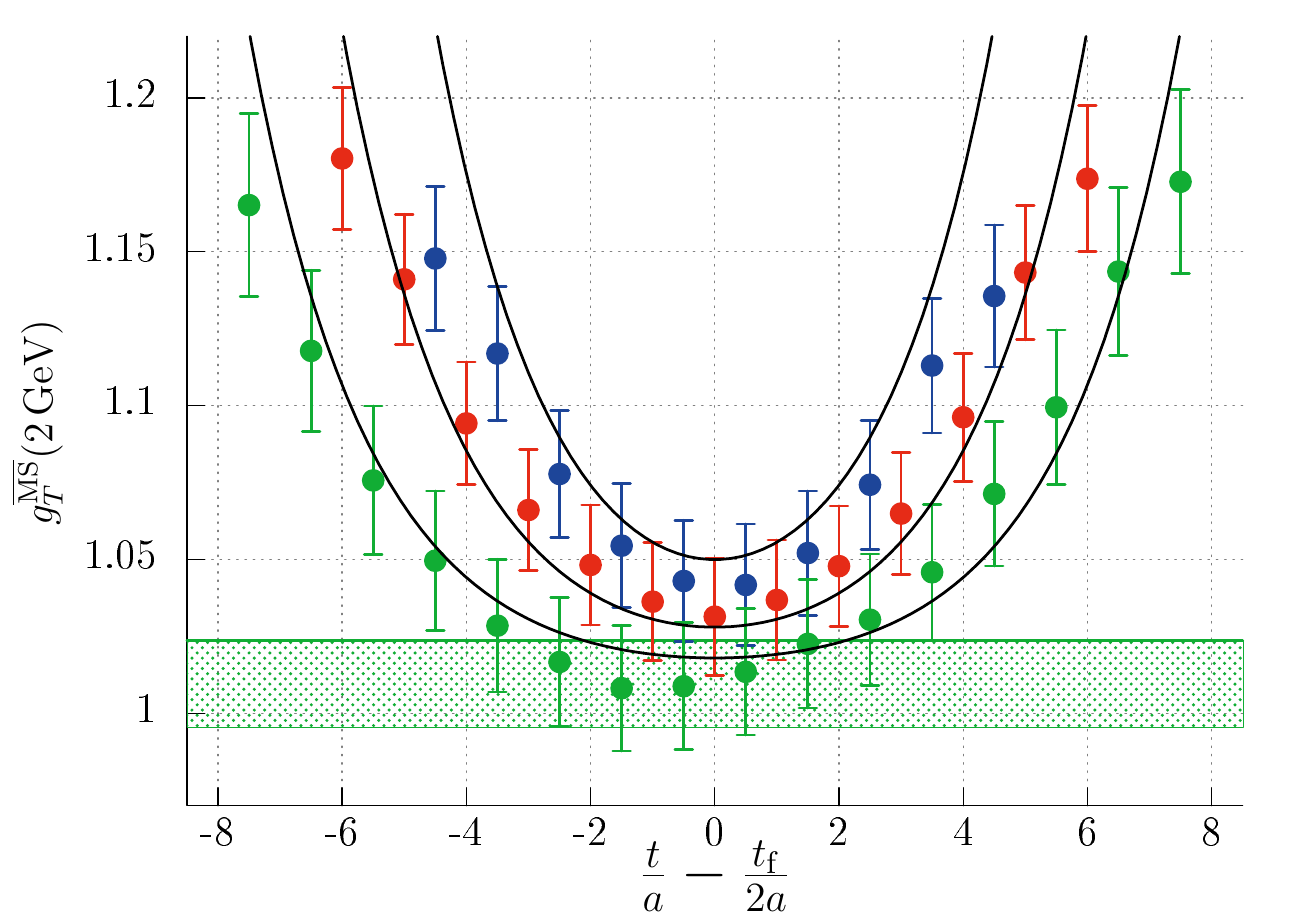}
}
\caption{The same as Fig.~\protect\ref{fig_gs_fit} for $g_T$.}
\label{fig_gt_fit}
\end{figure}

\begin{figure}[t]
\centerline{
\includegraphics[width=.48\textwidth,clip=]{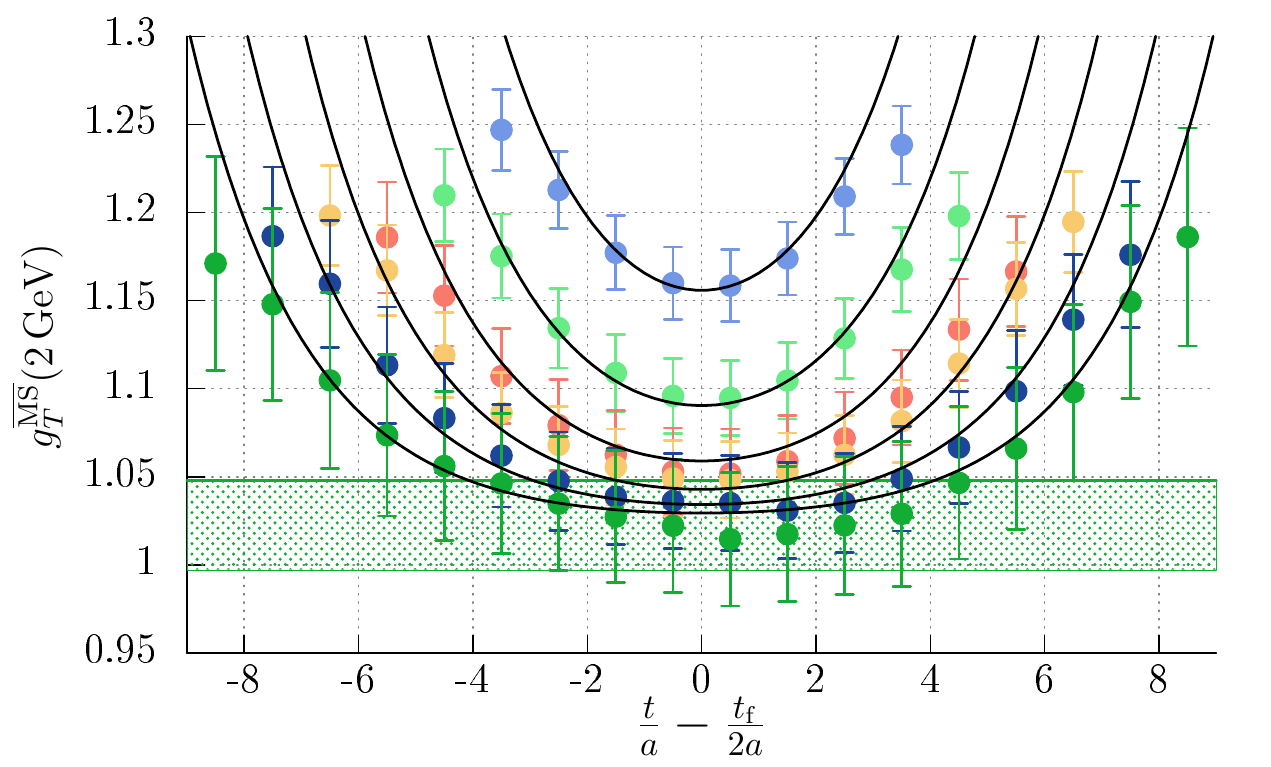}
}
\caption{The same as Fig.~\protect\ref{fig_gt_fit} on ensemble
IV ($m_{\pi}\approx 290\,$MeV, $a\approx 0.071\,$fm) and
$t_{\mathrm{f}}/a\in\{7, 9, 11, 13, 15, 17\}$.}
\label{fig_gt_fit2}
\end{figure}

\begin{figure}[t]
\centerline{
\includegraphics[width=.48\textwidth,clip=]{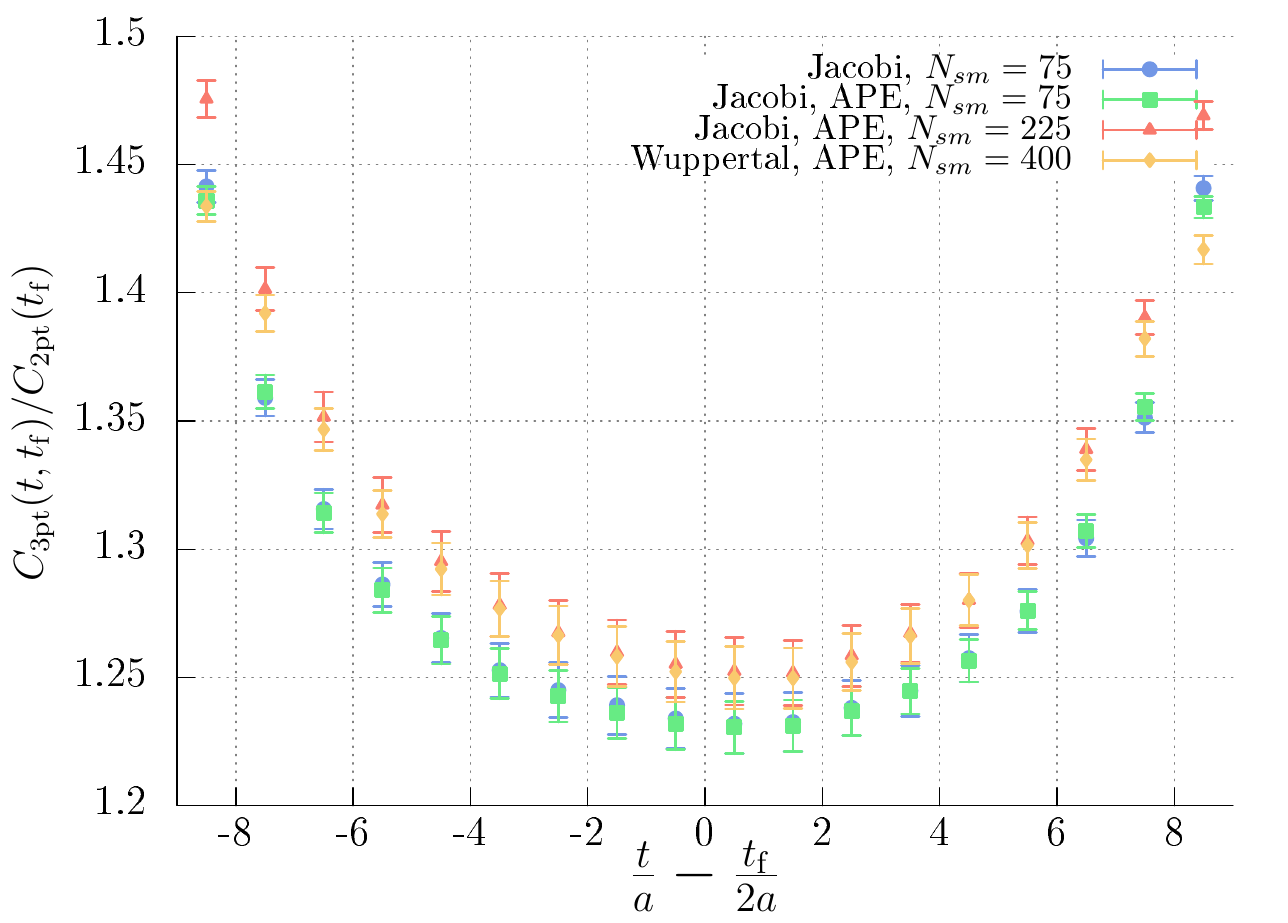}
}
\caption{
The same as Fig.~\protect\ref{fig_smear_gab} for
$g_T$.}
\label{fig_smear_gtb}
\end{figure}

\begin{figure}[t]
\centerline{
\includegraphics[width=.48\textwidth,clip=]{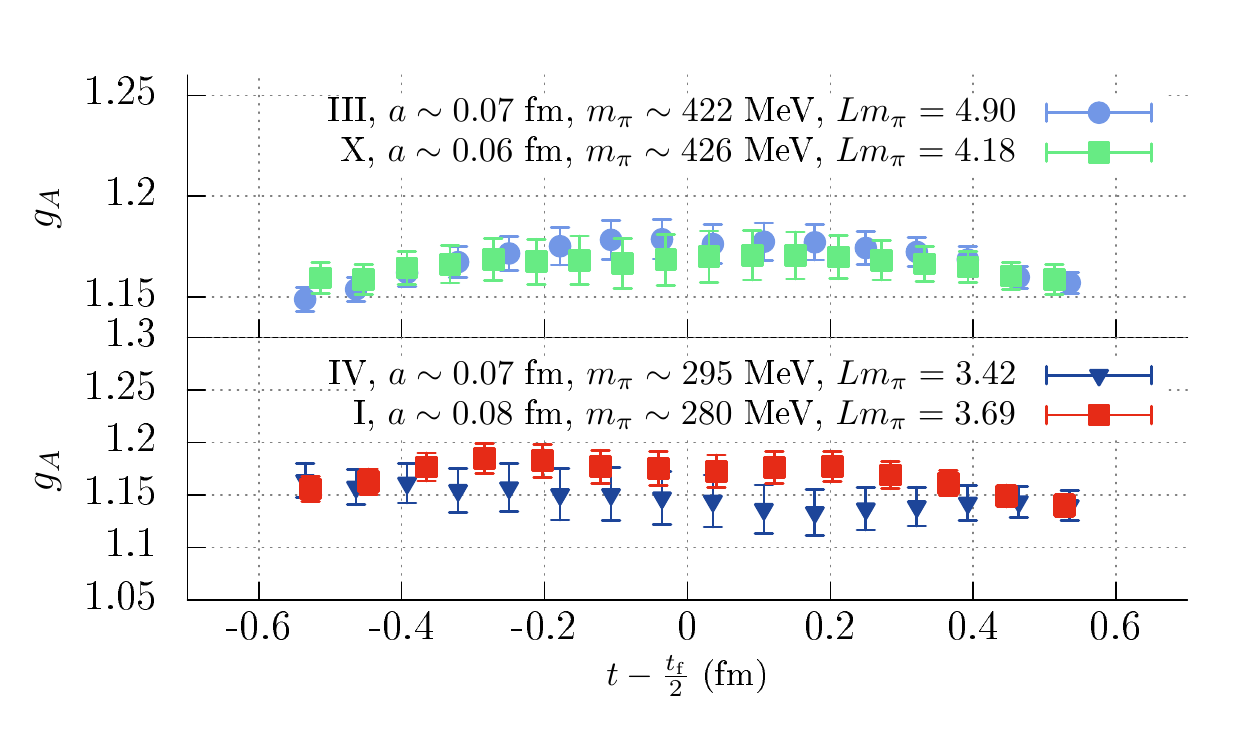}
}
\caption{Ratios of renormalized three- over two-point functions,
giving $g_A$ in the limit $0\ll t\ll t_{\mathrm{f}}$ for four of our ensembles.}
\label{fig:smearing3}
\end{figure}
\begin{figure}[t]
\centerline{
\includegraphics[width=.48\textwidth,clip=]{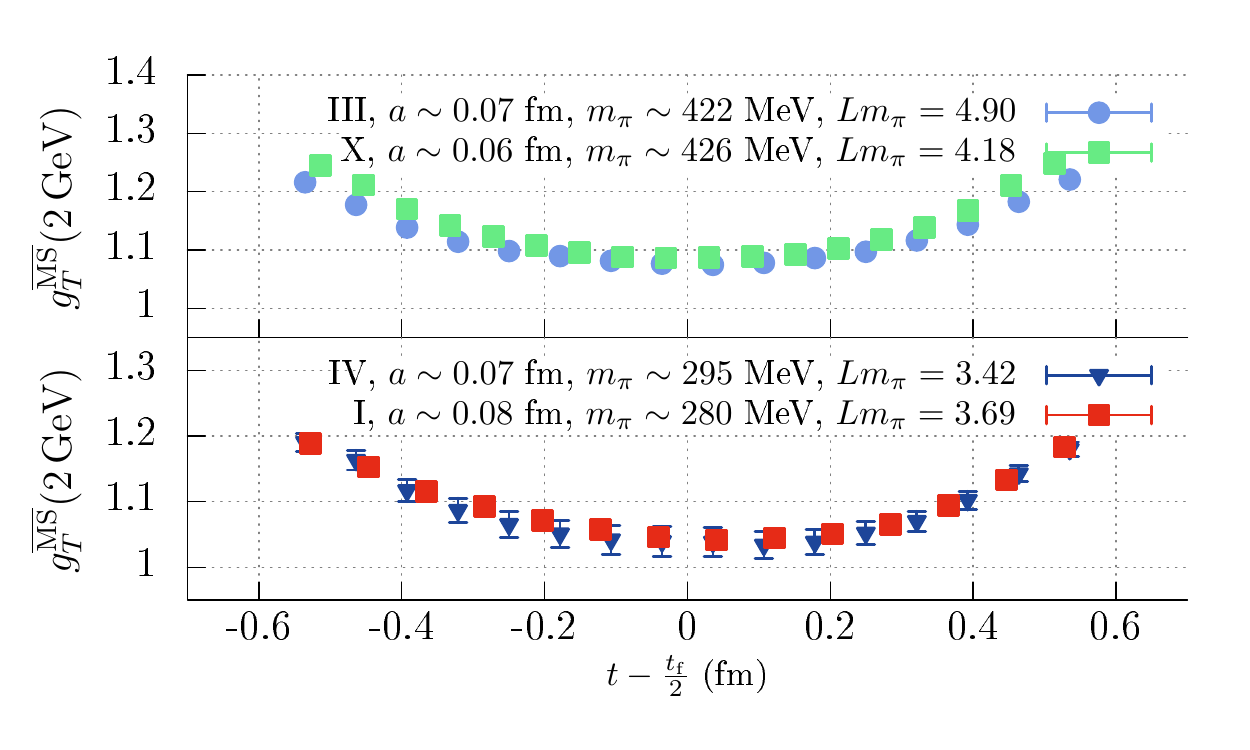}
}
\caption{The same as
Fig.~\protect\ref{fig:smearing3} for $g_T^{\MS}(2\,\mathrm{GeV})$.}
\label{fig:smearing4}
\end{figure}

In Fig.~\ref{fig_gs_fit} we show data for the renormalized scalar density 
for the same $m_{\pi}\approx 150\,$MeV ensemble
as in Fig.~\ref{fig_gab_fit}. In this case $B_{01}$
significantly differs from zero. We divide the three-point
functions by the asymptotic parametrization of
the two-point function $A_0e^{-m_Nt_{\mathrm{f}}}$, obtained
from the combined fit. The curves correspond
to the multi-exponential fit Eqs.~(\ref{eq:fit1}) and (\ref{eq:fit2})
with $\delta t=2a$. $B_1$ is compatible with zero.
The figure demonstrates that varying $t_{\mathrm{f}}$
helps to obtain a reliable result. However, it is also clear that
within statistical errors the $t_{\mathrm{f}}=15a>1\,$fm data
alone would have given the correct value.

Finally, we
discuss the tensor charge $g_T$ where the relative
errors are --- in contrast to $g_S$ --- not much bigger than for $g_A$ but
excited state contributions are clearly present, as is illustrated
in Figs.~\ref{fig_gt_fit} and \ref{fig_gt_fit2}
for the examples of $m_{\pi}\approx 150\,$MeV and $m_{\pi}\approx 290\,$MeV,
respectively. Again, the error bands shown are from multi-exponential
fits. In Fig.~\ref{fig_smear_gtb} we compare the different
smearing methods for the case of $g_T$. The effect
is visible, however, much less dramatic than for $g_A$
(see Fig.~\ref{fig_smear_gab}).
In the case of $g_T$ the smearing has only
a minor effect on the shape as a function of $t$
but still moves the ratio vertically.

We conclude this section by investigating the lattice spacing
dependence of ratios of renormalized three- over two-point functions.
This is important as we have only varied $t_{\mathrm{f}}$ on
three of our ensembles, albeit at three very different pion masses.
From these detailed investigations we concluded that --- within
the statistics that we have been able to accumulate and
with the smearing employed
--- a single value $t_{\mathrm{f}}\approx 1\,$fm
was sufficient to obtain the correct ground state results.
No lattice spacing effects are visible for effective masses,
see Figs.~\ref{fig:smearing} and \ref{fig:smearing2}.
However, in principle the situation may differ for three-point
functions. Therefore, we plot a comparison of the three-point
function, normalized with respect to the two-point function
for two different pion masses for the couplings with the
highest statistical
accuracy, $g_A$ and $g_T$, respectively,
in Figs.~\ref{fig:smearing3}
and \ref{fig:smearing4}; no significant dependence of the
shape on the lattice spacing can be recognized.

Similar excited state analyses to those detailed above
were carried out for all the couplings
on all the different ensembles displayed in Table~\ref{tab_1},
also shown in Fig.~\ref{fig:overview}.

\subsection{Comparison with the summation method}
\label{sec:sum}

\begin{figure}[t]
\centerline{
\includegraphics[width=.48\textwidth,clip=]{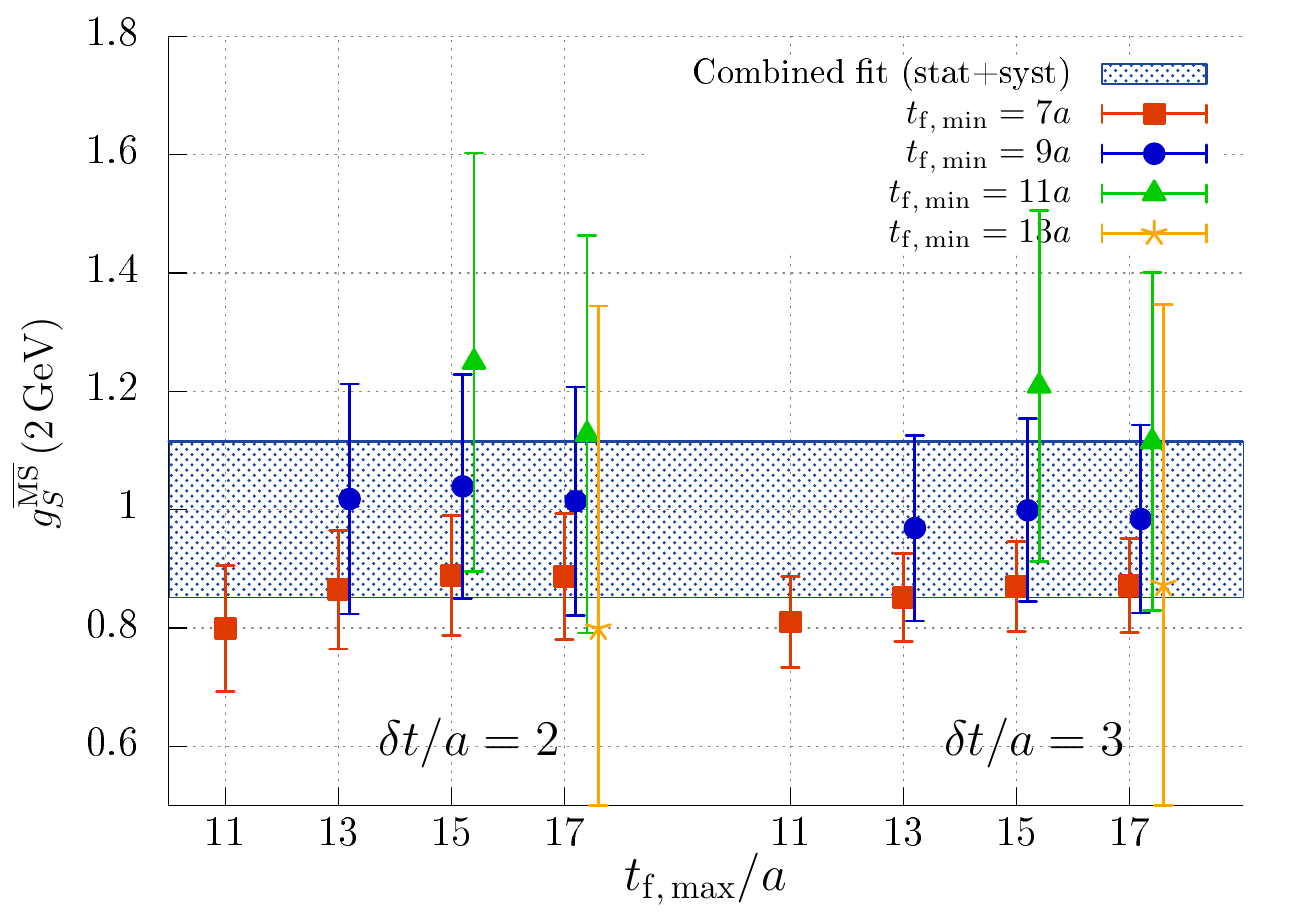}
}
\caption{Results on $g_S^{\MS}(2\,\mathrm{GeV})$ obtained with the
summation method Eq.~(\protect\ref{eq:summa2}) for different
fit ranges $t_{\mathrm{f}}\in [t_{\mathrm{f},\min},t_{\mathrm{f},\max}]$
and $\delta t/a\in\{2,3\}$ on
ensemble IV ($m_{\pi}\approx 290\,$MeV, $a\approx 0.071\,$fm).
The error band corresponds to the result obtained with the
fit method detailed in Sec.~\ref{sec:fits}, including our
assignment of systematic errors.
All data are normalized with respect to the $\MS$ scheme. The error
of the renormalization factor is smaller by more than one order of magnitude
than any of the statistical errors displayed and can be neglected.}
\label{fig:summation_gs}
\end{figure}

\begin{figure}[t]
\centerline{
\includegraphics[width=.48\textwidth,clip=]{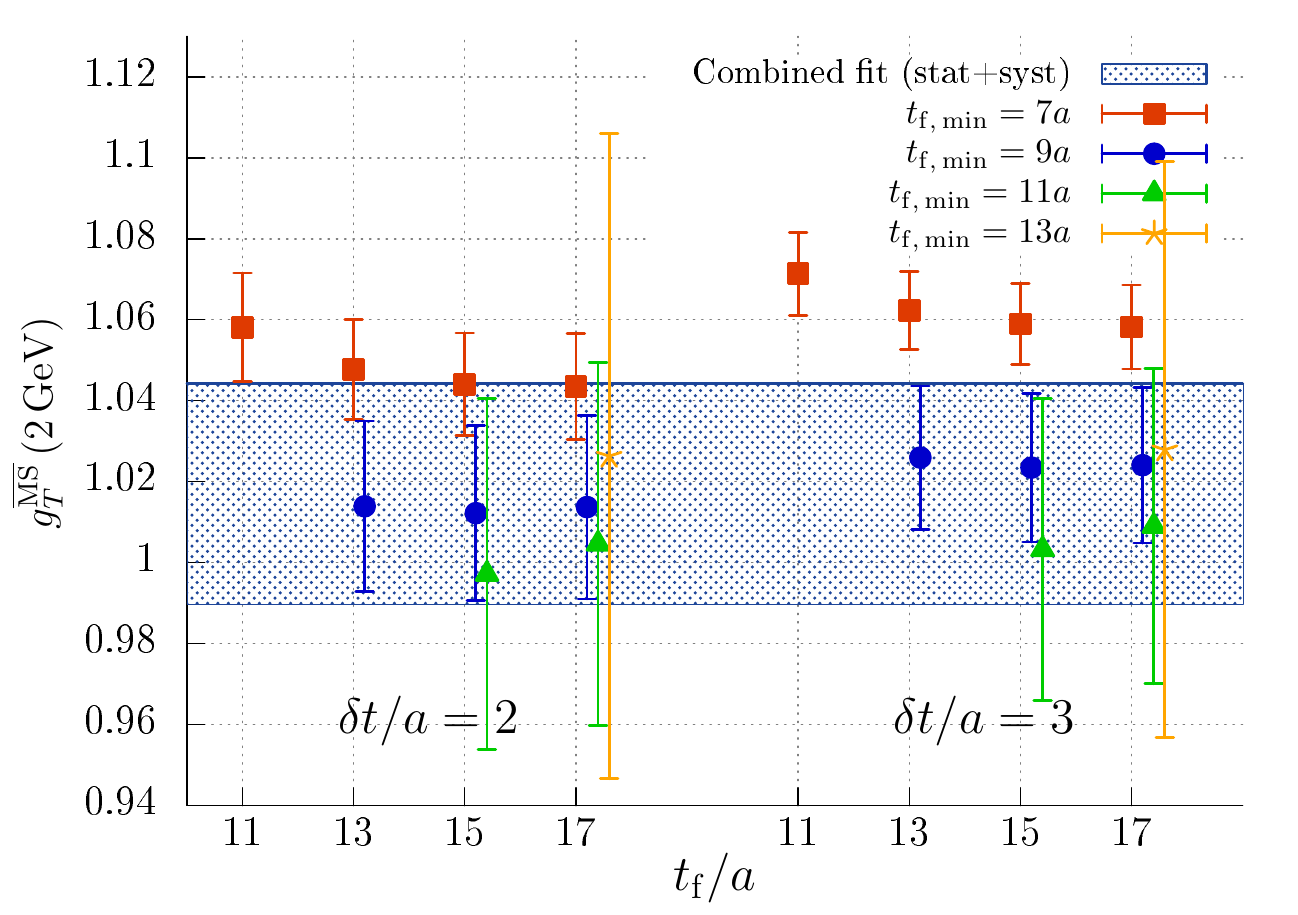}
}
\caption{The same as Fig.~\protect\ref{fig:summation_gs} for $g_T^{\MS}(2\,\mathrm{GeV})$.}
\label{fig:summation_gt}
\end{figure}

The summation method~\cite{Maiani:1987by}
has recently gained in popularity~\cite{Capitani:2010sg}.
Fitting ratios in $R(t,t_{\mathrm{f}})$ to a plateau in $t$,
see Eq.~(\ref{eq:ratio}), there are
corrections of order $\exp(-\Delta m_Nt_{\mathrm{f}}/2)$.
Instead, the summation method comprises of computing sums
\begin{equation}
\label{eq:summa2}
S(t_{\mathrm{f}},\delta t)=\sum_{t=\delta t}^{t_{\mathrm{f}}-\delta t}
R(t,t_{\mathrm{f}})=c(\delta t)+\frac{t_{\mathrm{f}}}{a}\left[\frac{\langle N|J|N\rangle}{2m_N}+\cdots\right]\,,
\end{equation}
see Eq.~(\protect\ref{eq:summa}),
and fitting these linearly in $t_{\mathrm{f}}$ within an
interval $t_{\mathrm{f}}\in[
t_{\mathrm{f},\min},t_{\mathrm{f},\max}]$.
It is easy to see from Eqs.~(\ref{eq:fit1}) and (\ref{eq:fit2})
that the corrections to the slope, and thereby to the desired
matrix element, in this case are only of order $\exp(-\Delta m_Nt_{\mathrm{f}})$.
Therefore, for $\delta t$ chosen sufficiently large and
$t_{\mathrm{f},\min}\geq t_{\mathrm{f},\max}/2$, the convergence of
the slope as a function of $t_{\mathrm{f},\max}$ towards the
asymptotic value is faster than the convergence
of results of plateau fits as a function of $t_{\mathrm{f}}$, 
at the price of introducing a second fit parameter $c$.
It is not clear why one would compare this procedure to simple plateau fits:
In that case, introducing for each $t_{\mathrm{f}}$-value
additional fit parameters
$c'=B_{01}\exp(-\Delta m_Nt_{\mathrm{f}}/2)$ and $m_N$, the
dependence on $\exp(-\Delta m_Nt_{\mathrm{f}}/2)$ can be removed too.
If more than one $t_{\mathrm{f}}$-value is available,
which is a pre-requisite of the summation method,
it is also not obvious why one should not attempt the
combined fit Eqs.~(\ref{eq:fit1}) and (\ref{eq:fit2}),
rather than transforming (and reducing) the available data
into sums $S(t_{\mathrm{f}},\delta t)$.

For $g_A$, with our interpolator, differences between
plateau fits, our combined fit and the summation method cannot be
resolved statistically as all $R(t,t_{\mathrm{f}})$ data for
different $t_{\mathrm{f}}$ and $t\approx t_{\mathrm{f}}/2$ 
basically agree within errors. For examples of these ratios, see
Figs.~\ref{fig_gab_fit} and \ref{fig:smearing3}
and the $N_{\mathrm{sm}}=400$ ratio shown in Fig.~\ref{fig_smear_gab}.
In Fig.~\ref{fig:summation_gs} we compare the result of our combined fit
(including a systematic error from varying the fit range and parametrization)
to results of the summation method Eq.~(\ref{eq:summa2}) for the
example of $g_S^{\MS}(2\,\mathrm{GeV})$ on ensemble IV.
We employ two different minimal distances $\delta t$
of the summation region in $t$ from the source
and sink positions 
and fit to different intervals
$t_{\mathrm{f}}\in[t_{\mathrm{f},\min},t_{\mathrm{f},\max}]$.
Indeed, the summation method converges towards the asymptotic result
and the convergence rate improves for larger values of $t_{\mathrm{f},\min}$.
The same can be seen in Fig.~\ref{fig:summation_gt} for
the tensor coupling $g_T^{\MS}(2\,\mathrm{GeV})$.

The form factors $\tilde{g}_T(Q^2)$ and $\tilde{g}_P(Q^2)$
at different virtualities $Q^2$ show a similar behaviour.
For the example
of the second Mellin moment of the isovector
spin-independent structure function $\langle x\rangle_{u-d}$,
a comparison between the methods
was presented in Ref.~\cite{Bali:2014gha}.
Also in that case we found agreement between the results of the two
methods within the respective $\delta t$- and $t_{\mathrm{f}}$-windows
of applicability, however, the combined fits utilize more information
than the summation method.

\section{$g_V$, $g_A$ and the renormalization}
\label{sec:renorm}

\begin{table*}
\caption{\label{tab_2} Values of the pion mass, the PCAC lattice quark mass
Eq.~(\protect\ref{eq:mpcac}),
the un-renormalized pion decay constant $F_{\pi}^{\lat}$ and the couplings
$g_V^{\lat}$, $g_A^{\lat}$, $g_S^{\lat}$ and $g_T^{\lat}$. The errors are
statistical and systematic (from varying the fit range and parametrization),
respectively.}
\begin{center}
\begin{ruledtabular}
\begin{tabular}{cccccccc}
Ensemble&  $am_{\pi}$ & $a\tilde{m}$ & $aF_{\pi}^{\lat}$ & $g_V^{\lat}$ & $g_A^{\lat}$ & $g_S^{\lat}$ & $g_T^{\lat}$ \\\hline
I    & 0.11516(73)(11)    & 0.003676(38)(10)   & 0.05056(18)(07)  &1.3714(24)(03) & 1.566(23)(14) & 1.59(17)(05)  & 1.239(19)(16)\\
II   & 0.15449(69)(26)    & 0.007987(44)(06)   & 0.04841(43)(05)  &1.3461(87)(04) & 1.473(31)(04) & 1.15(19)(03) & 1.275(35)(07)\\
III  & 0.15298(43)(16)    & 0.007964(32)(10)   & 0.04943(28)(03)  &1.3387(17)(01) & 1.550(15)(09) & 1.35(07)(03) & 1.264(14)(11)  \\
IV   & 0.10675(51)(08)    & 0.003794(27)(06)   & 0.04416(37)(05)  &1.3539(57)(05) & 1.491(30)(02) & 1.58(18)(11) & 1.188(30)(11)\\
V    & 0.10465(37)(08)    & 0.003734(21)(04)   & 0.04449(12)(04)  &1.3473(30)(05) & 1.600(19)(09) & 1.49(14)(03) & 1.267(20)(05)\\
VI   & 0.10487(24)(04)    & 0.003749(16)(08)   & 0.04490(12)(04)  &1.3445(14)(04) & 1.585(17)(05) & 1.51(09)(02) & 1.221(17)(04)\\
VII  & 0.05786(51)(21)    & 0.001129(18)(04)   & 0.04048(48)(13)  &1.3395(120)(04)& 1.521(28)(02) & 1.48(38)(05) & 1.196(27)(20) \\
VIII & 0.05425(40)(28)    & 0.000985(17)(08)   & 0.04029(30)(34)  &1.3440(110)(17)& 1.540(26)(03) & 1.68(28)(13) & 1.181(17)(07)\\
IX   & 0.15020(53)(06)    & 0.009323(21)(13)   & 0.04351(33)(03)  &1.3141(15)(02) & 1.489(14)(00) & 1.57(07)(03) & 1.201(22)(10) \\
X    & 0.13073(55)(28)    & 0.007005(23)(04)   & 0.04152(27)(03)  &1.3190(23)(04) & 1.492(15)(00) & 1.42(10)(01)& 1.249(20)(05) \\
XI   & 0.07959(25)(09)    & 0.002633(13)(04)   & 0.03651(33)(04)  &1.3233(50)(06) & 1.540(19)(09) & 1.51(15)(02) & 1.179(17)(18)
\end{tabular}
\end{ruledtabular}
\end{center}
\end{table*}

Following the procedure outlined in Sec.~\ref{sec:fits},
we obtain the un-renormalized values $g_A^{\lat}$, $g_S^{\lat}$, $g_P^{\lat}$ and $g_T^{\lat}$
listed in Table~\ref{tab_2}. The induced couplings
$\tilde{g}_T$ and $g_P^*$ require an
extrapolation of non-forward three-point functions in the
virtuality $Q^2$ and will be discussed in detail together with
$g_S$ and $g_T$ in Sec.~\ref{results} below. Here we
concentrate on $g_V$ and $g_A$. 
We also list the pion masses and PCAC lattice quark masses, obtained from
the axial Ward identity
\begin{equation}
\label{eq:mpcac}
\tilde{m}=
\frac{\partial_4\langle 0|A_4|\pi\rangle}
{2\langle 0|P|\pi\rangle}\left[1+am(b_A-b_P)\right]\,,
\end{equation}
where $|\pi\rangle$ is the physical pion state created
by an interpolator of spin/flavour structure
$(\bar{u}\gamma_5d)^{\dagger}$, $\partial_{\mu}$ denotes
the symmetrized lattice derivative, $P=\bar{u}\gamma_5d$ is
the local pseudoscalar density and
$A_{\mu}=\bar{u}\gamma_4\gamma_5d +ac_A\partial_{\mu}P$ is the
non-perturbatively improved axial current ($P$ is automatically
order-$a$ improved). $c_A$ was obtained in Ref.~\cite{DellaMorte:2005se},
the improvement factor $b_A-b_P$ is explained below
and $m$ denotes the lattice vector quark mass defined through
\begin{equation}
m=\frac{1}{2a}\left(\frac{1}{\kappa}-\frac{1}{\kappa_{\mathrm{crit}}}\right)\,,
\label{eq:quarkmass}
\end{equation}
where $\kappa_{\mathrm{crit}}$ is the value of the hopping parameter
where the PCAC mass vanishes. The lattice quark masses $m$
can easily be computed from the $\kappa$-values given in Table~\ref{tab_1}
and the critical hopping parameter values listed in Table~\ref{tab_3}.
The PCAC quark masses $\tilde{m}$ (listed in Table~\ref{tab_2})
can be translated into the $\MS$ scheme at
2\,GeV, upon multiplication with $Z_A/Z_P$ (see below).
The pion decay constant is obtained through
\begin{equation}
F_{\pi}^{\lat}=\frac{\langle 0|A_4|\pi\rangle}{\sqrt{2}\,m_{\pi}}\,,
\end{equation}
where we use the normalization that corresponds
to the experimental value $F_{\pi}=Z_A(1+amb_A)F_{\pi}^{\lat}\approx 91\,$MeV.

\begin{table*}
\caption{The critical hopping parameters $\kappa_{\mathrm{crit}}$, $m=0$ plaquette
values $P$ and 
renormalization constants~\protect\cite{Gockeler:2010yr} of the
lattice currents relative to the $\MS$-scheme at $\mu=2\,$GeV.
The errors given include systematics.\label{tab_3}}
\begin{center}
\begin{ruledtabular}
\begin{tabular}{cccccccc}
$\beta$ & $\kappa_{\mathrm{crit}}$ & $P$&$Z_A$ & $Z_V$ & $Z_S^{\MS}(2\,\mathrm{GeV})$ & $Z_P^{\MS}(2\,\mathrm{GeV})$ & $Z_T^{\MS}(2\,\mathrm{GeV})$\\\hline
5.20 &0.1360546(39)&0.53861& 0.7532(16)   & 0.7219(47)  & 0.6196(54)  & 0.464(12) &  0.8356(15) \\
5.29 &0.1364281(12)&0.54988& 0.76487(64)  & 0.7365(48)  & 0.6153(25)  & 0.476(13) &  0.8530(25) \\
5.40 &0.1366793(11)&0.56250& 0.77756(33)  & 0.7506(43)  & 0.6117(19)  & 0.498(09) &  0.8715(14)
\end{tabular}
\end{ruledtabular}
\end{center}
\end{table*}

The lattice couplings extracted
from the respective matrix elements
need to be renormalized too:
\begin{equation}
g_X=Z_X(1+amb_X)g_X^{\lat}\,,
\end{equation}
where $X\in\{S,P,V,A,T\}$.
The renormalization factors $Z_X$ and the improvement coefficients $b_X$
depend on the inverse lattice coupling $\beta$. 
No anomalous dimension is encountered for
$g_V$ and $\tilde{g}_T$ due to baryon number conservation
and for $g_A$ and $g^*_P$ due to the PCAC relation.
In the other cases we quote the
values in the $\MS$ scheme at a scale $\mu=2\,$GeV.
As detailed in Ref.~\cite{Gockeler:2010yr}, the renormalization
factors are first determined non-perturbatively in the
RI'MOM scheme, using the Roma-Southampton method~\cite{Martinelli:1994ty},
and then converted perturbatively at
three-loop order to the $\MS$-scheme.
The improvement factors $amb_X$
were computed in Ref.~\cite{Sint:1997jx}
($X\in\{S,P,V,A\}$) to one loop
and confirmed in Refs.~\cite{Taniguchi:1998pf,Capitani:2000xi}, where
$b_T$ is given as well.
These are very close to unity, due to the smallness of $am$, and 
can be taken into account perturbatively:
\begin{align}
b_A&=1+0.15219(5)g^2\,,\quad
b_V=1+0.15323(5)g^2\,,\nonumber\\\nonumber
b_P&=1+0.15312(3)g^2\,,\quad
b_S=1+0.19245(5)g^2\,,\\
b_T&=1+0.1392 (1)g^2\,.
\label{eq:oneloop}
\end{align}
In this context we use the ``improved'' coupling
$g^2\equiv -3\ln P=
6/\beta+\mathcal{O}(g^4)$, where $P$ denotes the average
plaquette with the normalization $P=1$ at $\beta=\infty$.
The corresponding chirally extrapolated values of $P$ are
displayed in Table~\ref{tab_3}.
Note that $b_m=-b_S/2$ as well as the combination
$b_A-b_P\approx 0$ were determined
non-perturbatively~\cite{Fritzsch:2010aw} and
for $b_S$ we use the interpolating formula of this reference
\begin{equation}
b_S=\left(1+0.19246g^2\right)\frac{1-0.3737g^{10}}{1-0.5181g^4}\,,
\end{equation}
instead of the one-loop expression given in Eq.~(\ref{eq:oneloop}).

For convenience we list, in addition to the
critical hopping parameter values, the renormalization
factors $Z_X$ between the lattice and the $\MS$ schemes
determined in Ref.~\cite{Gockeler:2010yr}
(and slightly updated here) in Table~\ref{tab_3}.
Note that our $Z_A$-value at $\beta=5.2$
is by about 2\% smaller than that obtained in Ref.~\cite{Fritzsch:2012wq}
from the Schr\"odinger functional. This is indicative of the
$\mathcal{O}(a^2)$ difference
between cut-off effects of the two
methods. This disagreement indeed
reduces with increasing $\beta$~\cite{DellaMorte:2008xb}.
Also note that the ratios $Z=Z_P/(Z_SZ_A)$ are consistent
with the parametrization obtained from
the dependence of the
PCAC quark mass on a valence quark hopping
parameter by the ALPHA Collaboration~\cite{Fritzsch:2010aw}.

\begin{figure}[t]
\centerline{
\includegraphics[width=.48\textwidth,clip=]{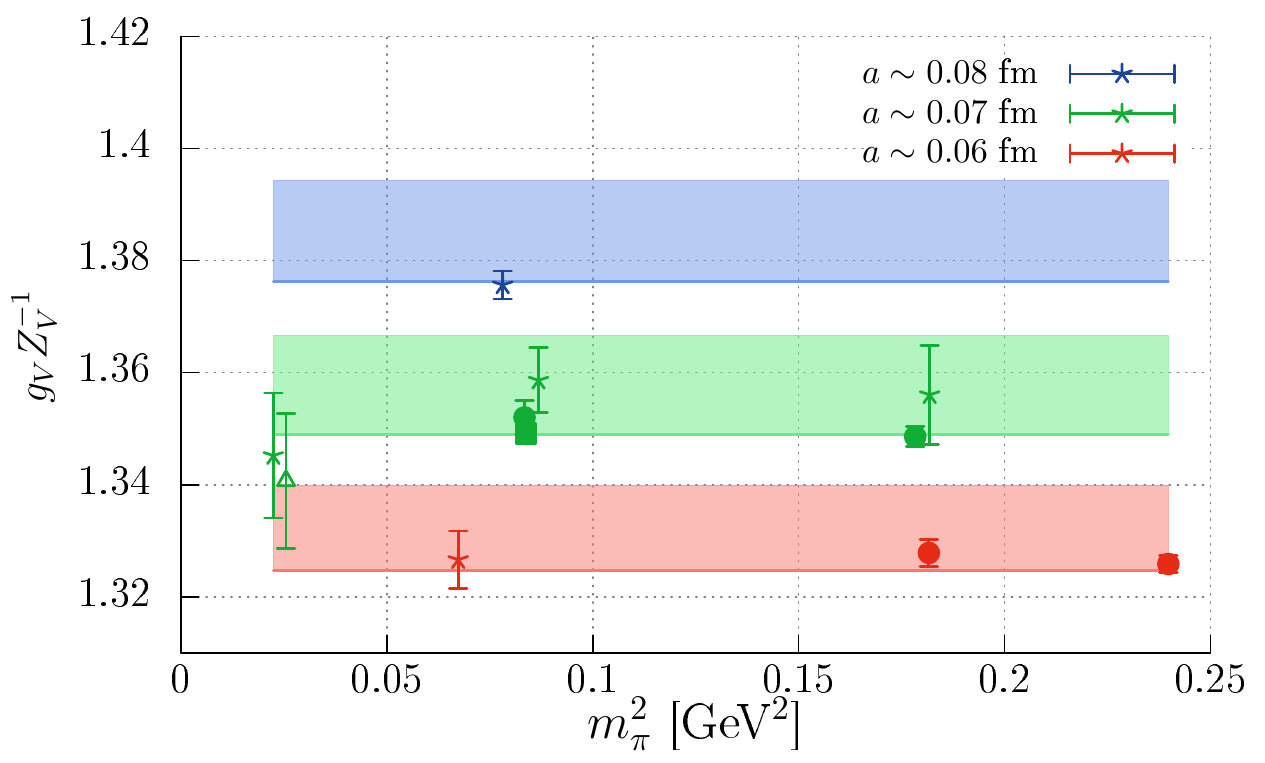}
}
\caption{$g_V/Z_V\equiv g_V^{\lat}(1+amb_V)$ as a function of $m_{\pi}^2$ for
all ensembles. Symbols are as in Fig.~\protect\ref{fig:overview}.
Shown as solid bands are the $1/Z_V$-values determined
non-perturbatively~\protect\cite{Gockeler:2010yr} (updated in
Table~\protect\ref{tab_3})
for the three $\beta$-values.}
\label{fig_gv_zeromom}
\end{figure}

For all ensembles,
in Fig.~\ref{fig_gv_zeromom} we compare 
the $g_V^{\lat}$-values, multiplied by the improvement terms
$[1+amb_V(\beta)]$, to the corresponding
renormalization factors $1/Z_V(\beta)$ of Table~\ref{tab_3} to
confirm the relation $g_V=Z_Vg_V^{\lat}[1+amb_V+\mathcal{O}(a^2)]$.
We find perfect agreement within errors.
The non-perturbative determination of $Z_A$ is very similar to
that of $Z_V$. Therefore, based on this independent validation
of $g_V=1$, we would not expect
any problems related to the renormalization of $g_A$ either.

\begin{figure}[t]
\centerline{
\includegraphics[width=.48\textwidth,clip=]{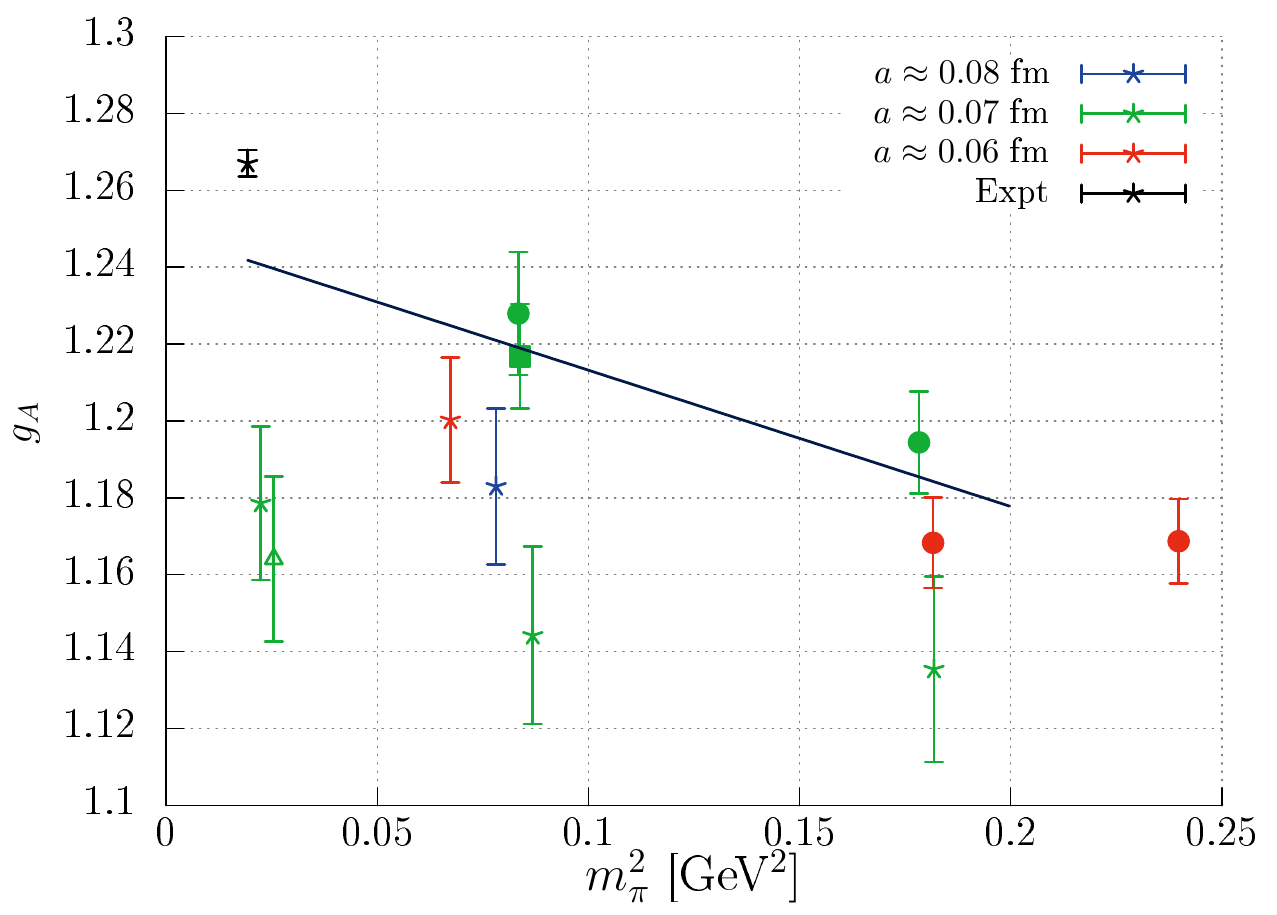}
}
\caption{$g_A$ as a function of $m_{\pi}^2$ for all ensembles.
Symbols are as in Fig.~\protect\ref{fig:overview}: the square
corresponds to $Lm_{\pi}\approx 6.7$, circles to $Lm_{\pi}>4.1$, stars
to $Lm_{\pi}\in[3.4,4.1]$ and the triangle to $Lm_{\pi}\approx 2.8$.
The line drawn to guide the eye
represents the result of a linear fit to the four
$m_{\pi}<430\,$MeV points with $Lm_{\pi}>4.1$.}
\label{fig_ga}
\end{figure}

In Fig.~\ref{fig_ga} we show the renormalized axial coupling
as a function of the squared pion mass for all ensembles.
The different symbols encode the linear lattice extents $Lm_{\pi}$
and the colours the lattice spacings, see Fig.~\ref{fig:overview}.
Finite lattice spacing effects cannot be resolved within our errors.
Comparing volumes similar in units of $m_{\pi}$,
$g_A$ increases with decreasing
pion mass. It also increases, enlarging the volume at a fixed
pion mass:
by about 5\% increasing
$Lm_{\pi}$ from 3.7 to 4.9 at $m_{\pi}\approx 425\,$MeV and
by about 6\% going from $Lm_{\pi}\approx 3.4$
to 4.2 at $m_{\pi}\approx 290\,$MeV.
When further pushing $Lm_{\pi}$ from 4.2 to 6.7, $g_A$
remains constant within a combined error of 1.7\%.
At the near-physical
pion mass the larger volume has an extent $Lm_{\pi}\approx 3.5$ only,
possibly explaining the underestimation of the experimental value by
about 7\%. Unfortunately, at this pion mass, we do not have a volume
with $Lm_{\pi}> 4.1$ at our disposal which would have required
simulating a spatial box of $80^3$ points. There is little
effect, however, moving
from $Lm_{\pi}\approx 3.5$ down to $Lm_{\pi}\approx 2.8$. One should
not over-interpret this though as it is conceivable that
the volume dependence
could be small within some range of volumes, due to
other effects competing with $N\pi$ and $\Delta \pi$ 
loop corrections. Naively, one would
expect volume effects mediated by pion exchange to be
proportional to $m_{\pi}^2$ when keeping the lattice extent 
fixed in terms of the pion Compton wave length. Comparing
the 290\,MeV pion mass points to the 425\,MeV points, there is no
indication though for the change being larger in the latter case,
suggesting a more complex behaviour --- at least for $Lm_{\pi}<4$.

Fitting the $Lm_{\pi}>4.1$ values of
$g_A(m_{\pi}^2)$ alone for $m_{\pi}<430\,$MeV
as a linear function of $m_{\pi}^2$ gives the line
drawn in Fig.~\ref{fig_ga}, illustrating
the remarks made above. The line suggests consistency
with experiment. At the physical point it
reads $g_A=1.242(15)$, two standard deviations
below the known value. However, clearly, with few
ensembles at small quark masses and $Lm_{\pi}>4$, we cannot
at present perform such an extrapolation with any confidence,
in particular as the slope is expected to change
its sign towards very small pion masses, see, e.g.,
Ref.~\cite{Hemmert:2003cb}
as well as Sec.~\ref{fse} below.

\begin{figure}[t]
\centerline{
\includegraphics[width=.48\textwidth,clip=]{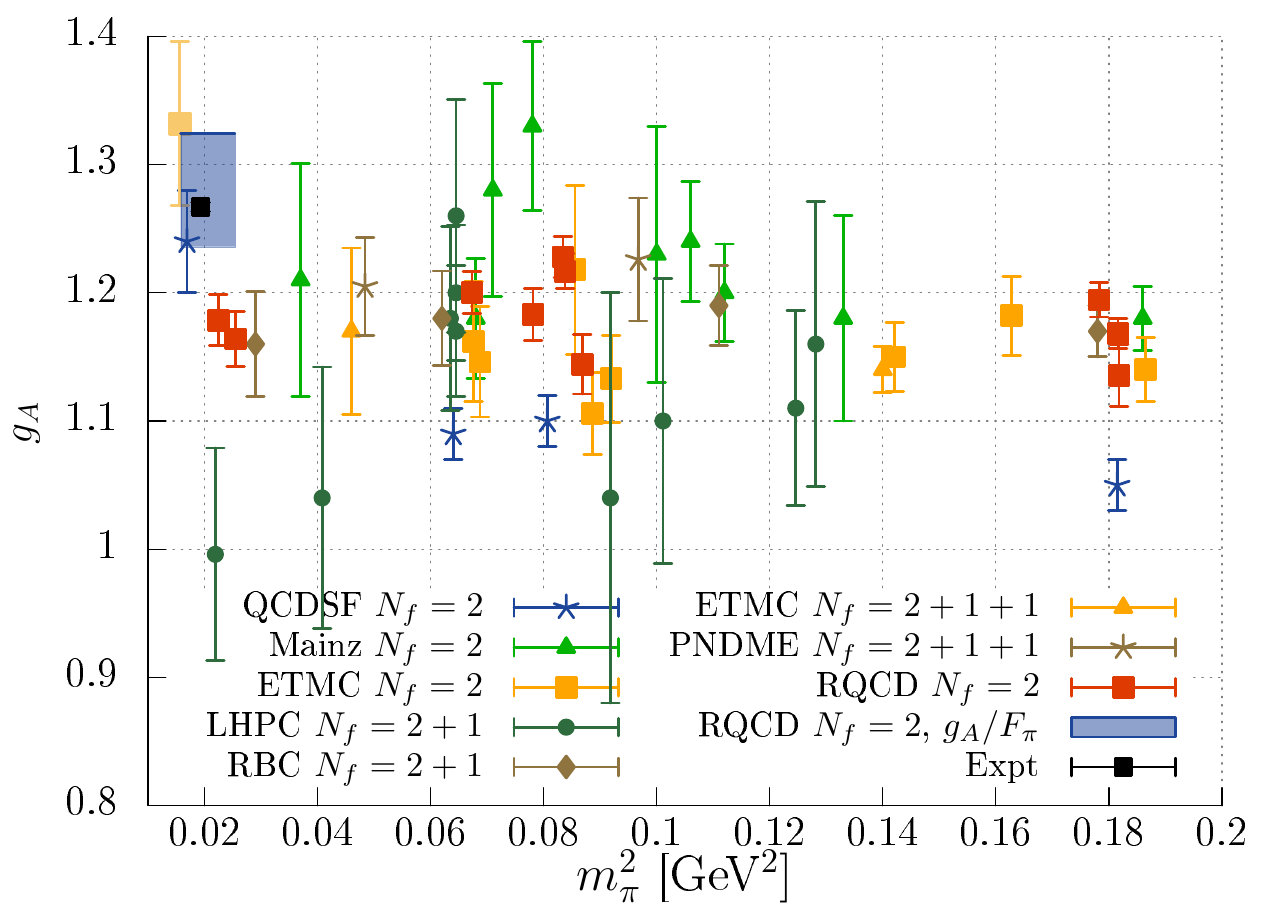}
}
\caption{$g_A$ as a function of $m_{\pi}^2$: our results (RQCD,
non-perturbatively improved (NPI) Wilson-clover)
in comparison
with other results (fermion action used in brackets). $N_{\mathrm{f}}=2$:
QCDSF~\protect\cite{Horsley:2013ayv} (NPI Wilson-clover),
Mainz\textsuperscript{\protect\ref{fn}}~\protect\cite{Jager:2013kha}
(NPI Wilson-clover),
ETMC~\protect\cite{Alexandrou:2013jsa} (twisted mass).
$N_{\mathrm{f}}=2+1$: LHPC~\protect\cite{Green:2012ud}
(HEX-smeared Wilson-clover),
RBC/UKQCD~\protect\cite{Ohta:2013qda} (domain wall).
$N_{\mathrm{f}}=2+1+1$: ETMC~\protect\cite{Alexandrou:2013joa} (twisted mass),
PNDME~\protect\cite{Bhattacharya:2013ehc} (Wilson-clover on a
HISQ staggered sea). Also indicated as a shaded area is the result from
extrapolating our $g_A/F_{\pi}$ data to the physical point,
see Sec.~\protect\ref{fse}.}
\label{fig_ga2}
\end{figure}

Prior to investigating the finite volume behaviour in more detail in the
next section, in Fig.~\ref{fig_ga2} we put our $N_{\mathrm{f}}=2$ results on $g_A$
in perspective, comparing these to recent determinations obtained by other
collaborations, namely QCDSF~\cite{Horsley:2013ayv},
the Mainz group\footnote{\label{fn}For each of the ensembles studied by
the Mainz group two results are given in their article, obtained from
plateau fits and
from the summation method.
We include the summation results since this appears to be their
preferred method.}~\cite{Jager:2013kha}
and ETMC~\cite{Alexandrou:2013jsa} for $N_{\mathrm{f}}=2$,
LHPC~\cite{Green:2012ud} and
RBC/UKQCD~\cite{Ohta:2013qda}
for $N_{\mathrm{f}}=2+1$ as well as ETMC~\cite{Alexandrou:2013joa} and
PNDME~\cite{Bhattacharya:2013ehc} 
for $N_{\mathrm{f}}=2+1+1$.
Most errors displayed are larger than ours, which
include the systematics from the renormalization factors,
varying fit ranges and parametrizations. This precision is in
particular due to our large numbers of measurements and the effort
that went into
the optimization of the nucleon interpolators.
We also indicate in the figure as a shaded area the
result of a chiral extrapolation of our data on the ratio
$g_A/F_{\pi}$, which we expect to be less affected
by finite volume effects, see Sec.~\ref{fse}.

Note that the recent QCDSF study~\cite{Horsley:2013ayv}
utilizes a smearing different from ours
for $m_{\pi}>250\,$MeV but has significant overlap in terms
of the gauge ensembles and the values of $Z_A$ used.
These results also carry quite small errors, 
however, their $g_A$-values are systematically lower, suggesting in these
cases that smearing could be an issue, see Fig.~\ref{fig_smear_gab}.
The left-most point of that study, that they associate with
$m_{\pi}\approx 130\,$MeV,
was obtained using the same smearing that we employ
on a sub-set of ensemble VII [$m_{\pi}(L)\approx 160\,$MeV,
$Lm_{\pi}\approx 2.8$, $m_{\pi}(\infty)\approx 149.5\,$MeV].
Their result at this point (leftmost circle)
is compatible within errors
not only with experiment but also
with our corresponding high statistics result (second red square
from the left).

Within errors all recent determinations (with the exception
of $m_{\pi}>250\,$MeV QCDSF results) are consistent with
our data. In particular, differences
between including the strange or even the charm quark or
ignoring these vacuum polarization effects
are not obvious. Moreover, in all studies
the $g_A$-values appear to be constant or increasing with decreasing
pion mass and,
where this could be resolved, correlated with the lattice size.
In none of the simulations could
any significant lattice spacing effects be detected.

\section{Finite size effects and the axial charge $g_A$}
\label{fse}
Above we have seen a noticeable dependence of $g_A$ on the
lattice volume for $Lm_{\pi}<4.1$.
Chiral perturbation theory not only predicts the functional
form of the pion mass dependence of hadronic observables
but also their finite volume
effects, as long as $m_{\pi}$ is small enough and
$\lambda=Lm_{\pi}$ sufficiently large. 
To leading non-trivial order~\cite{Gasser:1986vb,Gasser:1987zq}, the finite
size effects on the pion mass read
\begin{align}
\label{eq:fsepi}
\frac{m_{\pi}(L)-m_{\pi}}{m_{\pi}}
&=\frac{2}{N_{\mathrm{f}}}h(Lm_{\pi},m_{\pi})\,,\\
h(\lambda,m_{\pi})&=\frac{m_{\pi}^2}{16\pi^2 F^2}
\sum_{\mathbf{n}\neq{\bf 0}}\frac{K_1(\lambda
|\mathbf{n}|)}{\lambda|\mathbf{n}|}\,,
\label{eq:fsepi2}
\end{align}
where $F$ is the pion decay constant in the chiral limit,
$m_{\pi}=m_{\pi}(\infty)$ is the infinite volume pion mass,
$\mathbf{n}\in\mathbb{Z}^3$ are integer component vectors
and $K_1(x)$ is the modified Bessel function of the second kind.

The only parameter appearing in Eq.~(\ref{eq:fsepi}), apart
from $F=85.8(6)\,$MeV~\cite{Aoki:2013ldr,Agashe:2014kda},
is the infinite volume pion mass. Going beyond this order
of chiral perturbation theory~\cite{Colangelo:2003hf,Colangelo:2005gd},
several low-energy constants (LECs) are encountered, namely
$\bar{\ell}_i$, $i=1,2,3,4$ at $\mathcal{O}(p^4)$
and $\tilde{r}_i(m_{\rho})$, $i=1,2,\ldots,6$
at $\mathcal{O}(p^6)$ (next-to-next-to-leading order, NNLO).
We use the parametrization with NNLO chiral perturbation theory input
of Ref.~\cite{Colangelo:2005gd} to investigate finite volume effects
of the pion mass, setting $F=86\,$MeV and using the FLAG
values~\cite{Aoki:2013ldr} $\bar{\ell}_3=3.41(41)$,
$\bar{\ell}_4=4.62(22)$ for these two LECs.
For $\bar{\ell}_1$, $\bar{\ell}_2$ and
$\tilde{r}_i$ we take the central values given in
Ref.~\cite{Colangelo:2001df} that were also used in
Ref.~\cite{Colangelo:2005gd}.

\begin{figure}[t]
\centerline{
\includegraphics[width=.48\textwidth,clip=]{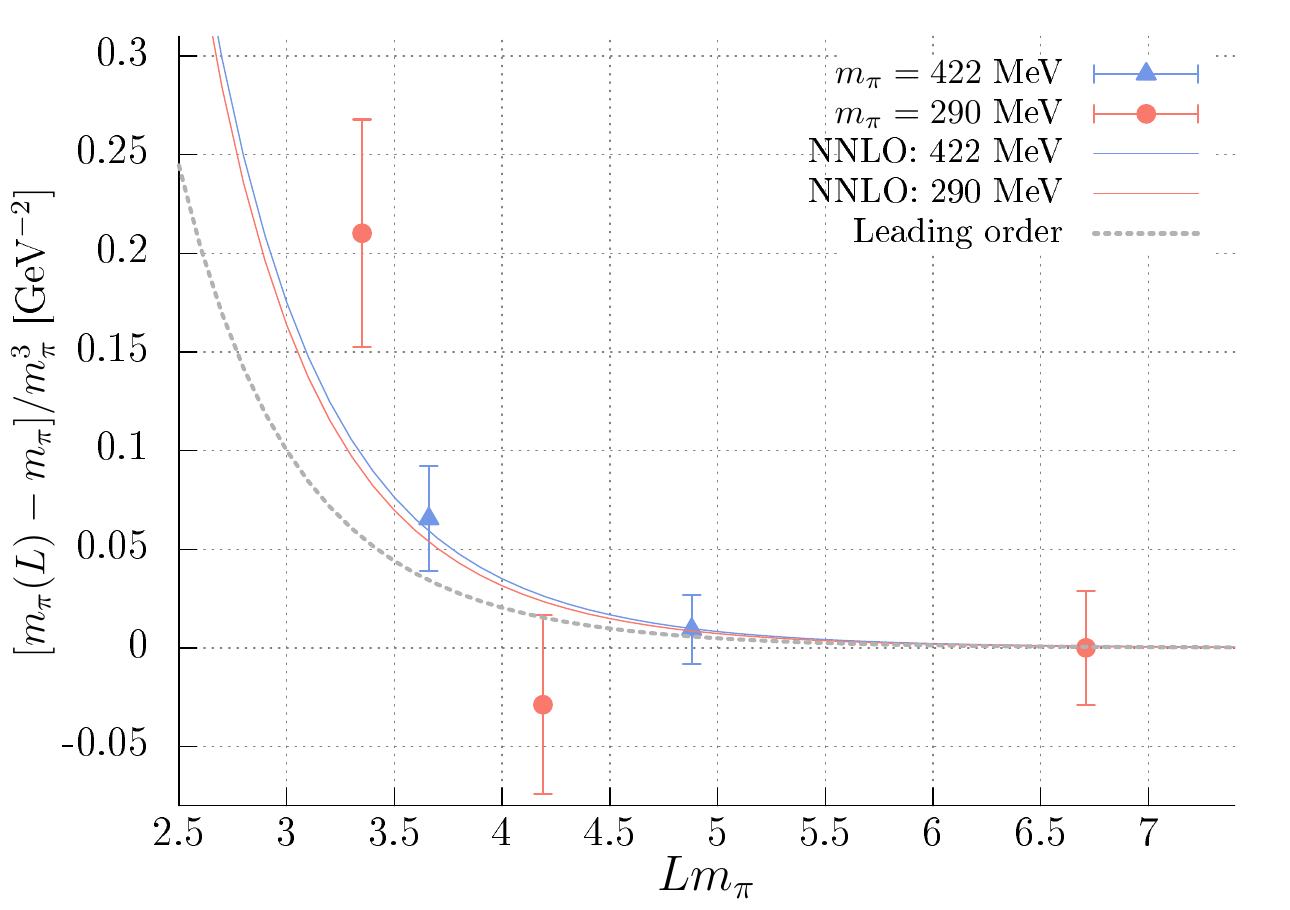}
}
\caption{The combination $[m_{\pi}(L)-m_{\pi}]/m_{\pi}^3$ as
a function of the linear lattice extent, in comparison with
the leading order~\protect\cite{Gasser:1986vb} [Eq.~(\protect\ref{eq:fsepi})]
and NNLO~\protect\cite{Colangelo:2005gd} chiral perturbation theory expectations.}
\label{fig_fse1}
\end{figure}

We are now in a position to estimate the infinite volume pion
masses. We do this by matching the NNLO finite size
formula~\cite{Colangelo:2005gd} in each case to the pion mass obtained
on the largest available volume.  Extrapolating this to infinite
volume lowers the central value of the pion mass on ensemble III from
422.2\,MeV by half a standard deviation to 421.5\,MeV, that on ensemble
VI (289.5\,MeV) by 0.02\,MeV and that on ensemble VIII from 149.7\,MeV by
one sixth of a standard deviation to 149.5\,MeV. Having eliminated the
free parameter by this matching, we can compare the combination
$[m_{\pi}(L)-m_{\pi}]/m_{\pi}^3$ to the leading order chiral
expectation $h(\lambda,m_{\pi})/m_{\pi}^2$, see
Eqs.~(\ref{eq:fsepi}) and (\ref{eq:fsepi2}), and the NNLO formula of
Ref.~\cite{Colangelo:2005gd}.  This comparison is shown in
Fig.~\ref{fig_fse1}.  Note that we omit the $m_{\pi}\approx 150\,$MeV
data from the figure.  In this case
$[m_{\pi}(3.42\,\mathrm{fm})-m_{\pi}]/m_{\pi}^3\approx
3\,\mathrm{GeV}^{-2}$, well off the scale of the figure, while the
leading order prediction Eq.~(\ref{eq:fsepi}) amounts to
$0.20\,\mathrm{GeV}^{-2}$ and the NNLO
prediction~\cite{Colangelo:2005gd} to $0.27\,\mathrm{GeV}^{-2}$. On
one hand the expansion seems to break down around $Lm_{\pi}\approx
3.5$ where the differences between the leading order and NNLO curves become
large. Already the leftmost point shown
in the figure appears to deviate from the predictions.  On the other
hand, in the safe $Lm_{\pi}>4$ region, the exponentially small finite
size effects cannot be resolved within the precision of the lattice
data.

In Refs.~\cite{Gasser:1986vb,Gasser:1987zq} the leading order
finite size expression of the pion decay constant is given
too:
\begin{equation}
\label{eq:fsefpi}
\frac{F_{\pi}(L)-F_{\pi}}{F_{\pi}}
=-2N_{\mathrm{f}}h(Lm_{\pi},m_{\pi})\,.
\end{equation}
The leading order finite volume effect of the axial charge
in SU(2) chiral perturbation theory contains the same $h(\lambda,m_{\pi})$ 
term~\cite{Beane:2004rf,Procura:2006bj,Khan:2006de}:
\begin{equation}
\label{eq:fsega}
\frac{g_A(L)-g_A(\infty)}{g_A^0}
=-4h(Lm_{\pi},m_{\pi})+D(L,m_{\pi},\Delta_0)\,,
\end{equation}
where $g_A^0=g_A(\infty)$ at $m_{\pi}=0$ and we
have suppressed the pion mass dependence of $g_A(L)$.
The correction $D(L,m_{\pi},\Delta_0)$ has been computed 
taking into account
also transitions between the nucleon and the $\Delta(1232)$ resonance
in Ref.~\cite{Khan:2006de}, using the
small scale expansion (SSE) technique~\cite{Hemmert:1997ye}.
Consequently, it depends on the mass difference
$\Delta_0$ between the nucleon and the real
part of the $\Delta$ pole as well as on the
squares of the pion-nucleon-nucleon and pion-nucleon-$\Delta$ couplings
and the ratio of the $\Delta$ axial charge over the
nucleon axial charge $g_1^0/g_A^0$.
In the chiral limit the pion-nucleon-nucleon
and pion-nucleon-$\Delta$ couplings can be re-expressed
in terms of $g_A^0$, see Eq.~(\ref{eq:gtr}),
and the axial transition charge $c_A^0$, respectively.
In the $\mathrm{SU}(2N_{\mathrm{f}})$ quark model $g_1^0/g_A^0=9/5$.
Note that, although this may not be obvious immediately,
the result of Ref.~\cite{Beane:2004rf} is identical
to the expression of Ref.~\cite{Khan:2006de}
in terms of the volume-dependence Eq.~(\ref{eq:fsega}).

\begin{figure}[t]
\centerline{
\includegraphics[width=.48\textwidth,clip=]{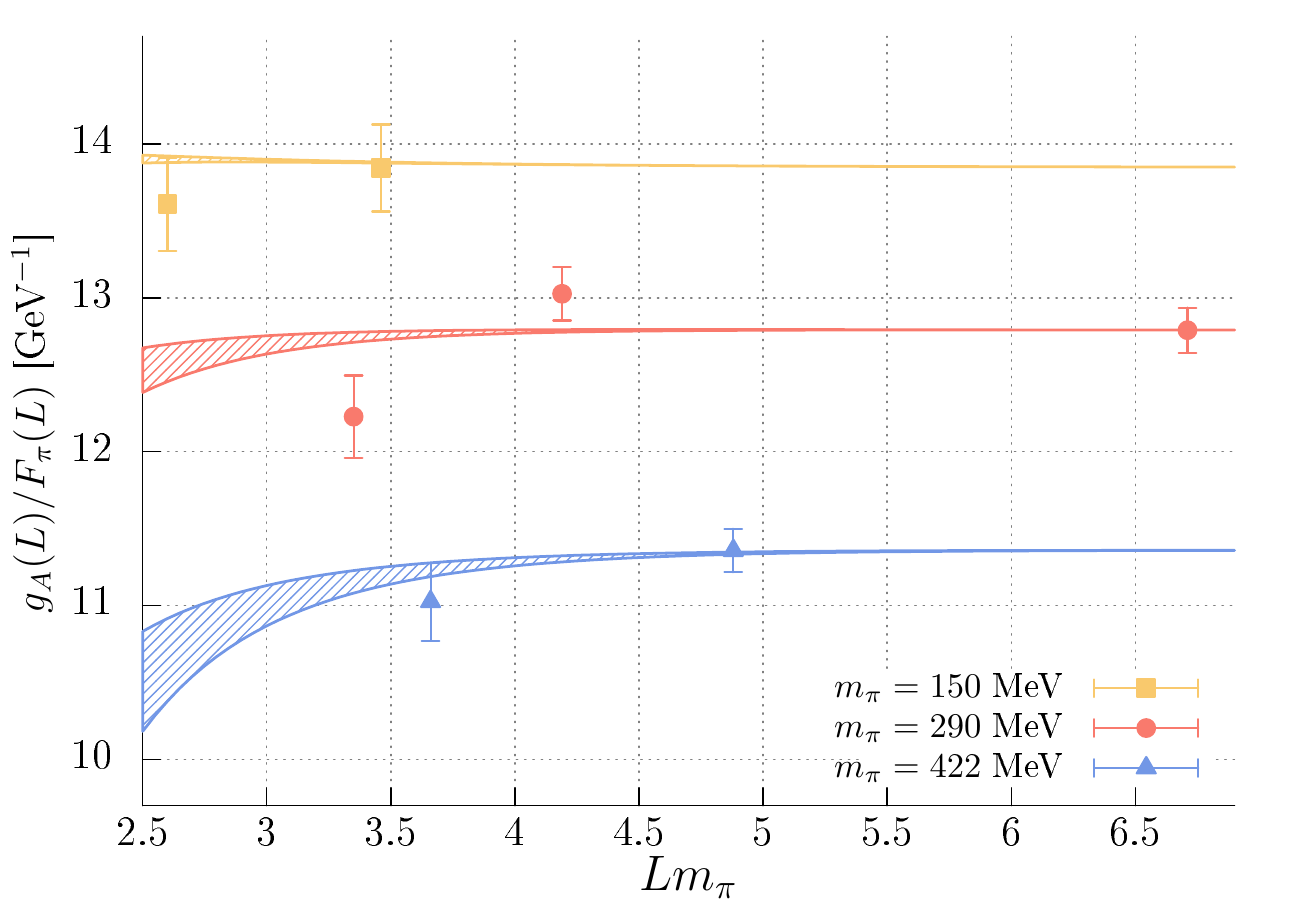}
}
\caption{The ratio $g_A(L)/F_{\pi}(L)$ as a function of the linear
lattice extent for three different pion masses.
The error bands are the predictions of Eq.~(\protect\ref{eq:fseratio}),
multiplied by constants $g_A(\infty)/F_{\pi}$
to match the three data sets.
The widths of the error bands are from varying the ratio
$g_A^0/g_A(\infty)\in[0.9,1.1]$.}
\label{fig_fse2}
\end{figure}

In Ref.~\cite{Horsley:2013ayv} an approximate cancellation between
different contributions to
$D(L,m_{\pi},\Delta_0)$ over a large range of
$L$- and $m_{\pi}$-values was observed, which motivated the
authors to study the ratio $g_A/F_{\pi}$.
From Eqs.~(\ref{eq:fsefpi}) and (\ref{eq:fsega}) we obtain
to leading one-loop order (i.e.\ $\mathcal{O}(\epsilon^3)$
in the SU(2) SSE~\cite{Hemmert:1997ye})
\begin{equation}
\label{eq:fseratio}
\frac{g_A(L)}{F_{\pi}(L)}=\frac{g_A(\infty)}{F_{\pi}}\frac{1-\frac{g_A^0}{g_A(\infty)}
[4h(L)-D(L,\Delta_0)]}
{1-4h(L)}\,.
\end{equation}
For $F_{\pi}(L)$ also the next-to-leading order and NNLO corrections
are known~\cite{Colangelo:2005gd}, however, to be consistent
in terms of the order of the SSE, we do not add these here.
We set $g_A^0=1.21$ (see below),
$c_A^0= 1.5$~\cite{Gail:2005gz}, $g_1^0=2.2\approx (9/5)g_A^0$ and
$\Delta_0=272\,$MeV~\cite{Svarc:2014sqa}.
In Fig.~\ref{fig_fse2} we show the resulting curves
for the infinite volume pion masses
$m_{\pi}=149.5\,$MeV, $m_{\pi}=289.5\,$MeV and
$m_{\pi}=421.5\,$MeV as functions of $Lm_{\pi}$.
The normalization $g_A(\infty)/F_{\pi}$ will depend on the pion mass
and is adjusted to match the three data sets while
the error band is from varying $g_A^0/g_A(\infty)\in[0.9,1.1]$
within Eq.~(\ref{eq:fseratio}).
Indeed, finite volume effects are much reduced, relative to
those for $g_A$ visible in Fig.~\ref{fig_ga},
and these are also broadly consistent with the predicted behaviour.

\begin{figure}[t]
\centerline{
\includegraphics[width=.48\textwidth,clip=]{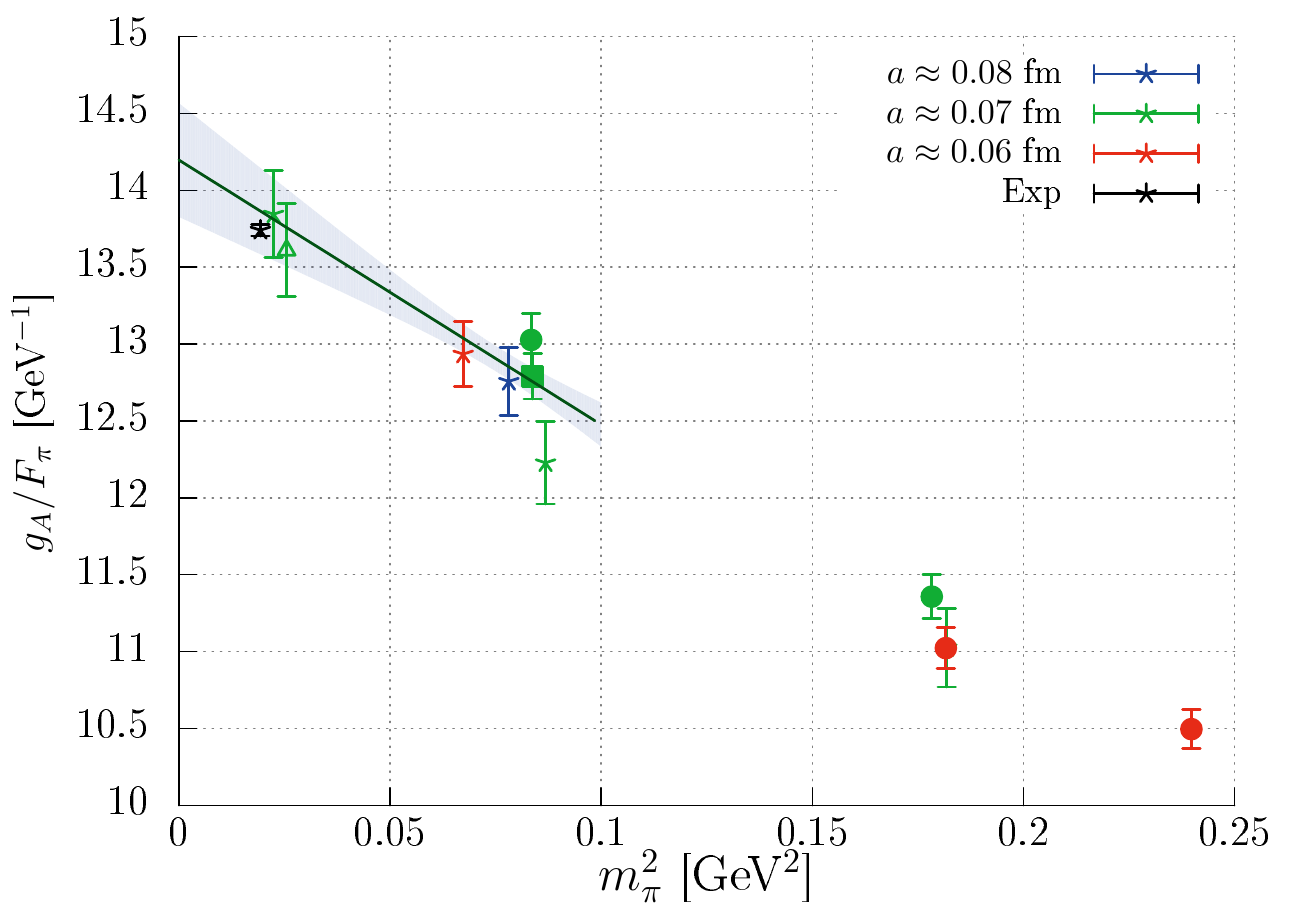}
}
\caption{$g_A/F_{\pi}$ as a function of $m_{\pi}^2$ for all ensembles, together
with a linear fit to the low mass points, omitting the smallest
volume (ensemble VII).
Symbols are as in Fig.~\protect\ref{fig:overview}.}
\label{fig_gafse}
\end{figure}

Finally, in Fig.~\ref{fig_gafse} we show the ratio $g_A(L)/F_{\pi}(L)$ as
a function of the squared pion mass, together with a linear fit
to the $m_{\pi}<300\,$MeV data, omitting the
$Lm_{\pi}<3.4$ data point (ensemble VII). This fit, with a
reduced $\chi^2/N_{\mathrm{DF}}=5.9/4$, gives
$g_A/F_{\pi}=13.88(29)\,\mathrm{GeV}^{-1}$ at $m_{\pi}=135\,$MeV
which compares well with
the experimental result $g_A/F_{\pi}=13.797(34)$.
Using $F_{\pi}=92.21(15)\,$MeV~\cite{Agashe:2014kda} at the physical point
as an input, this gives
$g_A=1.280(27)(35)$, where the second error corresponds to
the overall uncertainty of assigning physical values to our
lattice spacings~\cite{Bali:2012qs} (not shown in
the figure).
We remark that towards the chiral limit $g_A$ decreases with decreasing
pion mass while the observed increase of the ratio $g_A/F_{\pi}$
is entirely due to an also decreasing pion decay constant.
Towards large pion masses $F_{\pi}$ will continue to increase while
$g_A$ eventually starts decreasing again.

From $F_{\pi}/F=1.0744(67)$~\cite{Aoki:2013ldr}
we obtain the ratio $g_A/g_A^0=1.050(14)$, giving
$g_A^0=1.211(16)$ using
$g_A=1.2723(23)$~\cite{Agashe:2014kda}. 
Using the normalization conventions
\begin{align}
\label{eq:bbar}
g_A(m_{\pi})&=g_A^0\left(1+\frac{m_{\pi}^2}{16\pi^2F^2}\overline{b}+\cdots\right)\,,\\
F_{\pi}&=
F\left[1+\frac{m_{\pi}^2}{16\pi^2F^2}\bar{\ell}_4+\cdots\right]
\end{align}
for the leading chiral corrections,
one obtains
\begin{equation}
\frac{g_A(m_{\pi})}{F_{\pi}}=\frac{g_A^0}{F}+\frac{g_A^0}{16\pi^2F^3}(
\overline{b}-\bar{\ell}_4)m_{\pi}^2+\cdots.
\end{equation}
From our fit we find
$\overline{b}-\bar{\ell}_4=-1.41(36)$ and,
using $\bar{\ell}_4=4.62(22)$~\cite{Aoki:2013ldr}, arrive at the value
$\overline{b}=3.21(42)>0$ for this LEC:
$g_A$ increases with the pion mass (as is also obvious from the
ratio $g_A(135\,\mathrm{MeV})/g_A^0>1$ above).
Note however that $g_A$ is expected to start decreasing towards
larger pion masses,
due to the effect of the nearby $\Delta(1232)$
resonance~\cite{Detmold:2002nf,Hemmert:2003cb}. This is
also reflected in the lattice data, see Fig.~\ref{fig_ga}.

We did not detect any lattice spacing effects
within our statistical errors and therefore so far have ignored these.
Not being able to resolve such differences does not mean they are absent
and we will re-address this issue
in the summary Sec.~\ref{conc}.

\section{The scalar, tensor and pseudoscalar charges}
\label{results}
The scalar and tensor couplings can be obtained directly
in the forward limit of Eqs.~(\ref{eq:gs}) and (\ref{eq:gt})
while the 
induced tensor and pseudoscalar charges
are extracted from extrapolating
the respective form factors Eqs.~(\ref{eq:gv}) and (\ref{eq:ga})
to small virtualities. We will also determine the value
of the induced pseudoscalar form factor 
$g_P^*=\tilde{g}_P(Q^2)$ at the virtuality
$Q^2=-q^2=0.88\,m_{\mu}^2\approx 9.82\cdot 10^{-3}\,\mathrm{GeV}^2$,
corresponding to muon capture~\cite{Bernard:1994wn}.
\subsection{The scalar charge $g_S$}
\begin{figure}[t]
\centerline{
\includegraphics[width=.48\textwidth,clip=]{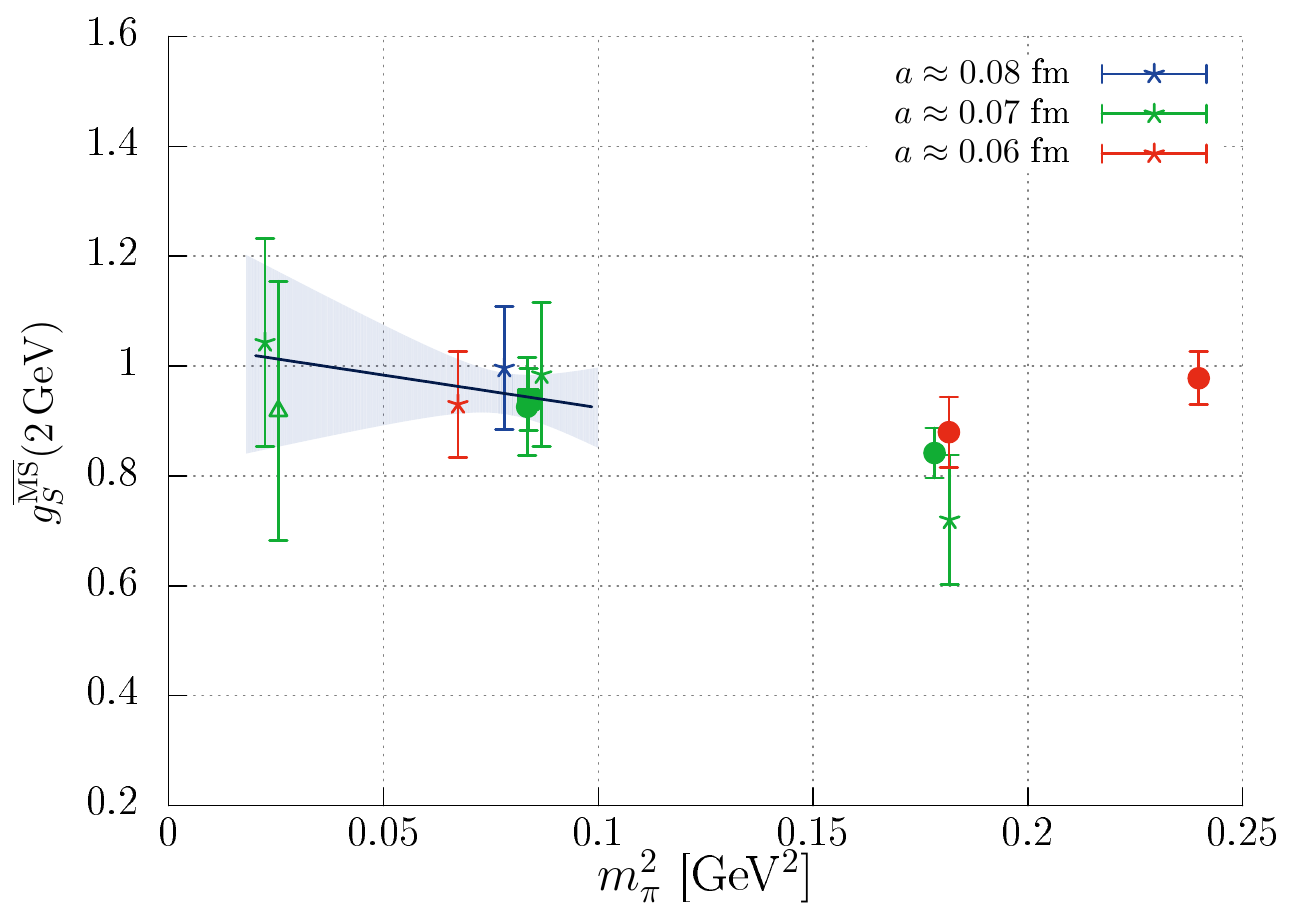}
}
\caption{$g_S^{\MS}(2\,\mathrm{GeV})$ as a function of $m_{\pi}^2$ for all ensembles.
Symbols are as in Fig.~\protect\ref{fig:overview}. Also shown is a linear extrapolation
in $m_{\pi}^2$ to the physical point.}
\label{fig_gs}
\end{figure}
In Fig.~\ref{fig_gs} we show our results for $g_S$ as a function of
$m_{\pi}^2$. Within their large errors the $m_{\pi}<430\,$MeV data are consistent
with a linear extrapolation and we find no lattice spacing or volume
dependence. The result of such an extrapolation to the
physical point, fitting the six $m_{\pi}<300\,$MeV data points with
$Lm_{\pi}>3.4$ is shown in the figure.  We find
$g_S^{\overline{\mathrm{MS}}}(2\,\mathrm{GeV})=1.02(18)$ for a fit with $\chi^2/N_{\mathrm{DF}}=0.48/4$.

The charge $g_S$ can,
via the conserved vector charge relation, also be obtained
as the ratio of the mass splitting of proton and neutron in the
absence of electromagnetic interactions over the difference of
light quark masses. The determination of this requires either
further assumptions or lattice simulations of QCD plus (Q)ED with
electrically charged quarks.
Recently, such lattice input was used in
Ref.~\cite{Gonzalez-Alonso:2013ura} to give $g_S=1.02(11)$.
However, not all systematic uncertainties were accounted for
in the error estimate.
The central value agrees with our direct determination.

\begin{figure}[t]
\centerline{
\includegraphics[width=.48\textwidth,clip=]{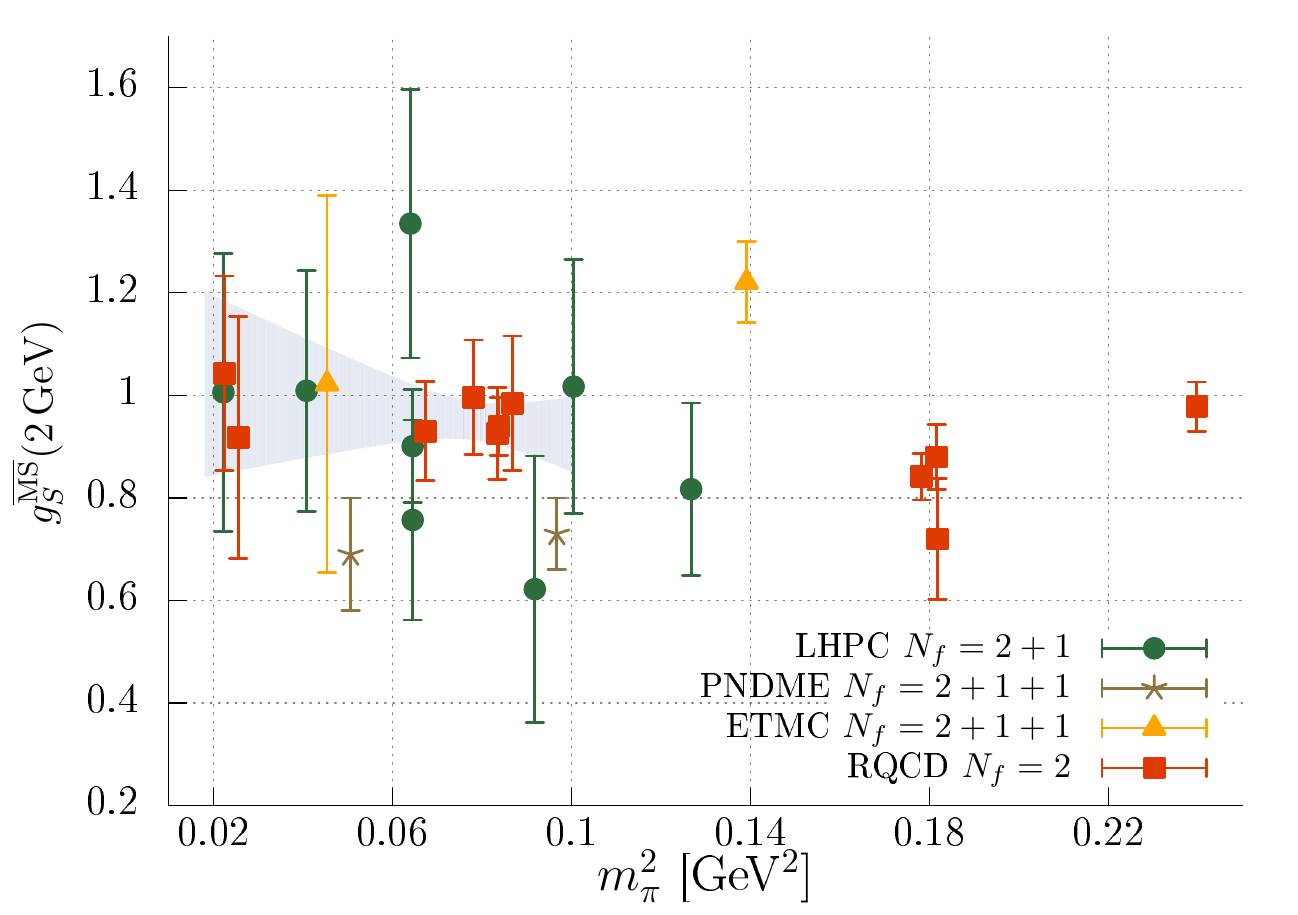}
}
\caption{$g_S^{\MS}(2\,\mathrm{GeV})$ as a function of
$m_{\pi}^2$: our results (RQCD,
NPI Wilson-clover)
in comparison
with other results.
$N_{\mathrm{f}}=2+1$: LHPC~\protect\cite{Green:2012ej} (HEX-smeared Wilson-clover).
$N_{\mathrm{f}}=2+1+1$:
PNDME~\protect\cite{Bhattacharya:2013ehc} (Wilson-clover on a
HISQ staggered sea) and
ETMC\textsuperscript{\protect\ref{fn3}}~\protect\cite{Alexandrou:2014wca}
(twisted mass).
Also included is the linear extrapolation of our data points.}
\label{fig_gs2}
\end{figure}
In Fig.~\ref{fig_gs2} we compare our results on $g_S$ to recent lattice
determinations by other groups, namely LHPC~\cite{Green:2012ej}, employing
$N_{\mathrm{f}}=2+1$ HEX-smeared Wilson-clover fermions,
PNDME~\cite{Bhattacharya:2013ehc},
using clover valence fermions on top of a $N_{\mathrm{f}}=2+1+1$ 
highly improved staggered quark (HISQ) sea and ETMC\footnote{\label{fn3}
At $m_{\pi}\approx 370\,$MeV we show their $t_{\mathrm{f}}=14a\approx 1.14\,$fm
result. In this reference also $N_{\mathrm{f}}=2$ results at
$m_{\pi}\approx 126\,$MeV can be found: $1.01(46)$ at $t=12a\approx 1.13\,$fm and
$1.63(76)$ at $t=14a\approx 1.32\,$fm.}~\cite{Alexandrou:2014wca}, using
$N_{\mathrm{f}}=2+1+1$ twisted mass fermions. The errors
of LHPC are quite large while there appears to be some tension
between our results and those of PNDME.
Notwithstanding this, around any single pion mass value all results are
compatible with each other as well as with our extrapolation
on the level of two standard deviations.

\subsection{The tensor charge $g_T$}
\begin{figure}[t]
\centerline{
\includegraphics[width=.48\textwidth,clip=]{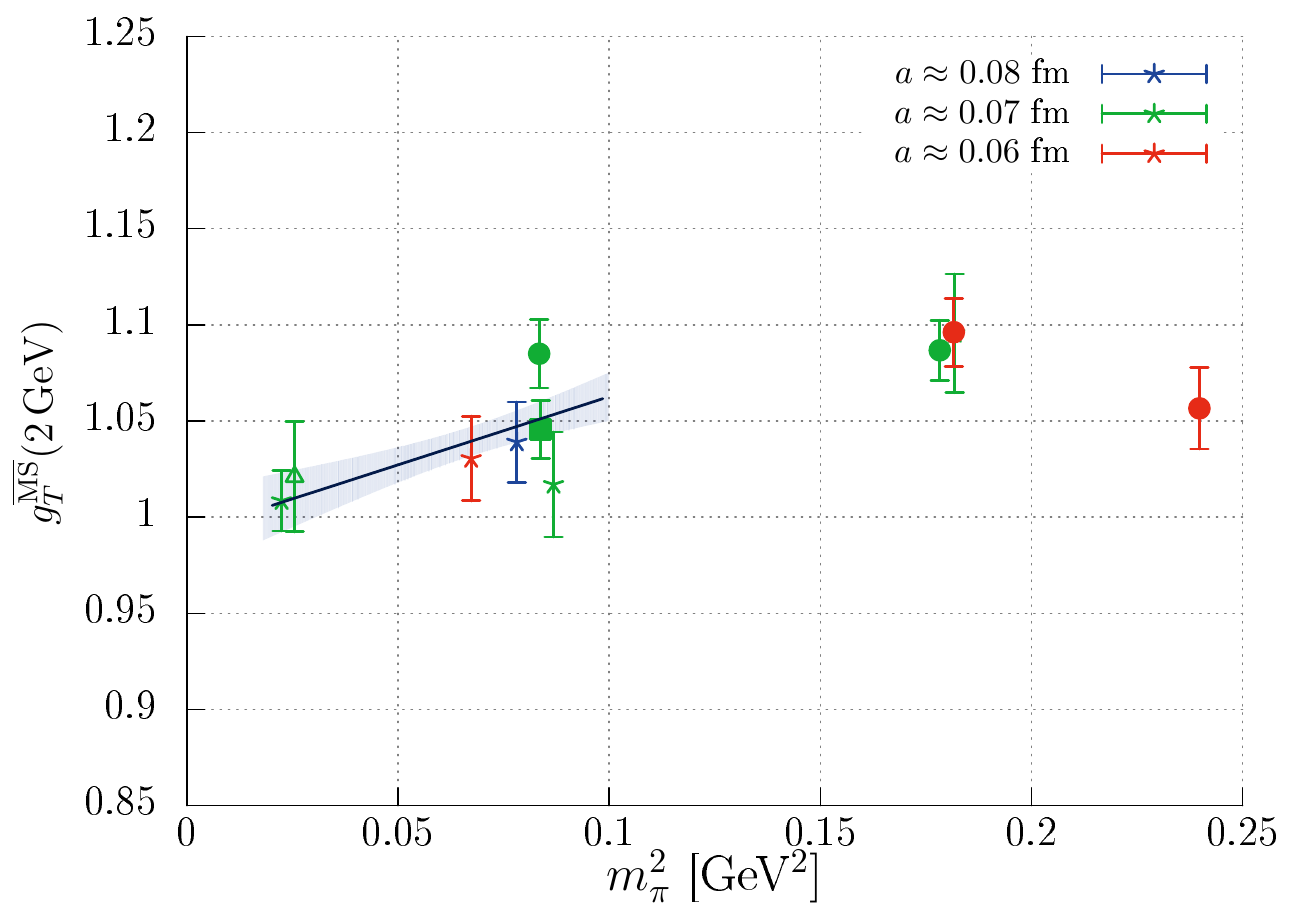}
}
\caption{$g_T^{\MS}(2\,\mathrm{GeV})$ as a
function of $m_{\pi}^2$ for all ensembles.
Symbols are as in Fig.~\protect\ref{fig:overview}.
Also shown is a linear extrapolation
in $m_{\pi}^2$ to the physical point.}
\label{fig_gt}
\end{figure}
\begin{figure}[t]
\centerline{
\includegraphics[width=.48\textwidth,clip=]{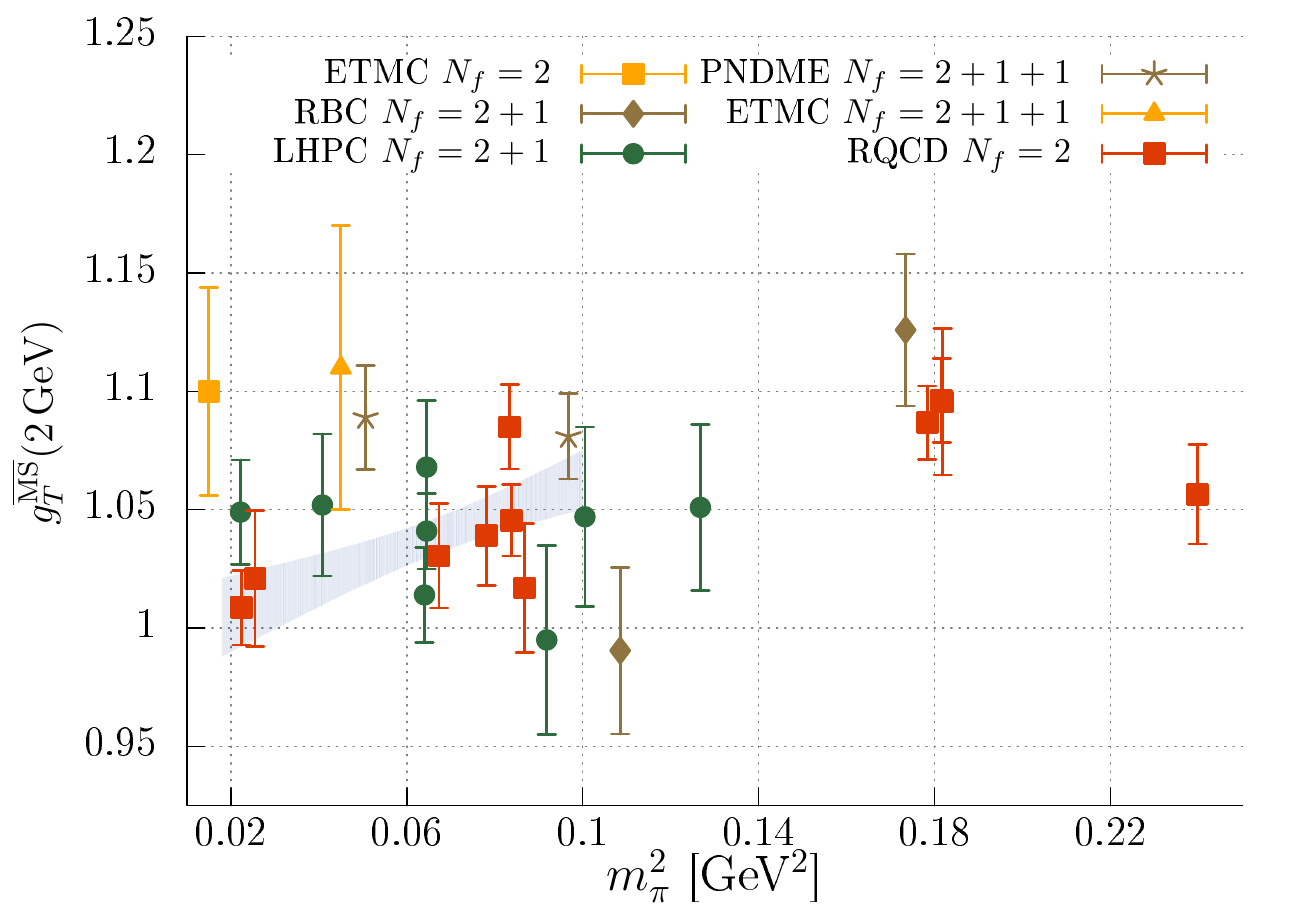}
}
\caption{$g_T^{\MS}(2\,\mathrm{GeV})$ as a function of
$m_{\pi}^2$: our results (RQCD, NPI Wilson-clover)
in comparison
with other results.
$N_{\mathrm{f}}=2$: ETMC~\protect\cite{Alexandrou:2013wka} (twisted mass).
$N_{\mathrm{f}}=2+1$: RBC/UKQCD~\protect\cite{Aoki:2010xg} (domain wall),
LHPC~\protect\cite{Green:2012ej} (HEX-smeared Wilson-clover).
$N_{\mathrm{f}}=2+1+1$: 
PNDME~\protect\cite{Bhattacharya:2013ehc} (Wilson-clover on a
HISQ staggered sea), ETMC~\protect\cite{Alexandrou:2013wka} (twisted mass).
Also included is the linear extrapolation of our data.\\[-.7cm]~}
\label{fig_gt2}
\end{figure}

\begin{figure*}[t]
\centerline{
\includegraphics[width=.9\textwidth,clip=]{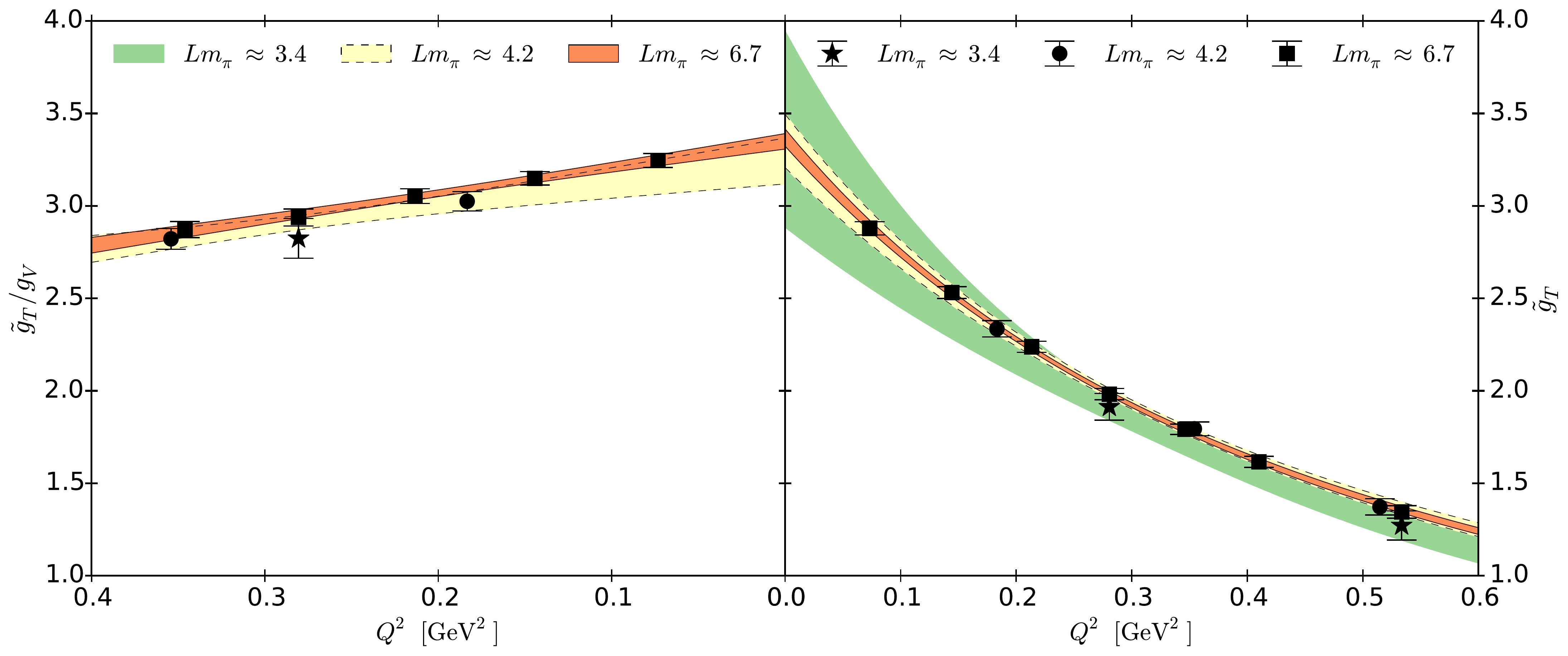}
}
\caption{$\tilde{g}_T(Q^2)/g_V(Q^2)$ (left panel) and $\tilde{g}_T(Q^2)$
as functions of the virtuality
$Q^2$ at $m_{\pi}\approx 290\,$MeV for three volumes
(ensembles IV, V and VI).}
\label{fig_gttextra}
\end{figure*}

In Fig.~\ref{fig_gt}
we show our results
on $g_T$. Again, we cannot detect any
lattice spacing or volume effects. Note that for our three
$a\approx 0.071\,$fm  points at $m_{\pi}\approx 290\,$MeV
($m_{\pi}^2\approx 0.084\,\mathrm{GeV}^2$),
the central value for the largest volume ($Lm_{\pi}\approx 6.7$)
lies inbetween those for the $Lm_{\pi}\approx 3.4$ and $Lm_{\pi}\approx 4.2$ lattices.
Again, we show a linear extrapolation to the physical point which gives
$g_T^{\overline{\mathrm{MS}}}(2\,\mathrm{GeV})=1.005(17)$ with $\chi^2/N_{\mathrm{DF}}=6.0/4$.
Unlike in the case of $g_A$ we regard such an extrapolation of $g_T$
as safe since there are no indications of finite volume effects
and our lowest mass point $m_{\pi}\approx 150\,$MeV is already
very close to the physical pion mass $m_{\pi}=135\,$MeV.
This conclusion is also supported by
Fig.~\ref{fig_gt2} where we compare our results to those of
ETMC~\cite{Alexandrou:2013wka} ($N_{\mathrm{f}}=2$ twisted mass fermions),
RBC/UKQCD~\cite{Aoki:2010xg} ($N_{\mathrm{f}}=2+1$ domain wall fermions),
LHPC~\cite{Green:2012ej} ($N_{\mathrm{f}}=2+1$ HEX-smeared Wilson-clover fermions),
PNDME~\cite{Bhattacharya:2013ehc} (Wilson-clover on a
HISQ staggered $N_{\mathrm{f}}=2+1+1$ sea)
and ETMC~\cite{Alexandrou:2013wka} ($N_{\mathrm{f}}=2+1+1$ twisted mass fermions).
No correlation with the sea quark content, volume,
lattice action or lattice spacing is obvious.
Moreover, all these determinations
are statistically consistent with each other as well as with our extrapolation.

\subsection{The induced tensor charge $\tilde{g}_T$}

The induced tensor coupling
$\tilde{g}_T=\kappa_{u-d}\approx\kappa_p-\kappa_n\approx 3.706$
is well-determined
experimentally.
Computing $\tilde{g}_T$ requires an extrapolation
of lattice data obtained at virtualities $Q^2>0$ to $Q^2=0$. At small
$Q^2$ one can expand
\begin{align}
g_V(Q^2)&=1-\frac{r_1^2}{6} Q^2+\mathcal{O}(Q^4)\,,\\
\tilde{g}_T(Q^2)&=\tilde{g}_T(0)\left[1-\frac{r_2^2}{6} Q^2+\mathcal{O}(Q^4)\right]\,,\label{eq36}
\end{align}
where the proton isovector Dirac and Pauli radii $r_1$ and
$r_2$ diverge as the pion mass approaches
zero.\footnote{
Note that the electric Sachs
form factor reads $G_E(Q^2)=g_V(Q^2)-Q^2/(4m_N^2)\tilde{g}_T(Q^2)$.
Therefore, in the isospin symmetric limit, the
squared charge radius is given as
$r_p^2=r_1^2+3\tilde{g}_T/(2m_N^2)$.}
It is well known that the $Q^2$-dependence
exhibits a substantial curvature, see,
e.g., Refs.~\cite{Gockeler:2003ay,Yamazaki:2009zq,Syritsyn:2009mx,Collins:2011mk,Alexandrou:2013joa,Bhattacharya:2013ehc,Green:2014xba}.
This means
small $Q^2$-values are required for a controlled extrapolation,
in particular at small quark masses where the coefficient
$r_2^2$ of the leading $Q^2$-term becomes large.
We expect this effect to partially cancel from the ratio
\begin{equation}
\label{eq:linear}
\frac{\tilde{g}_T(Q^2)}{g_V(Q^2)}=\frac{\tilde{g}_T^{\lat}(Q^2)}{g_V^{\lat}(Q^2)}
\stackrel{Q^2\rightarrow 0}{\longrightarrow} \tilde{g}_T\,.
\end{equation}
Therefore, one of our strategies is to extrapolate this ratio
as a linear function of $Q^2$ to $Q^2=0$.

Another parametrization that incorporates the curvature
is a dipole fit
\begin{equation}
\label{eq:dipole}
\tilde{g}_T(Q^2)=\frac{\tilde{g}_T(0)}{\left(1+Q^2/m_V^2\right)^2}\,.
\end{equation}
Taylor expanding this expression, the linear approximation
Eq.~(\ref{eq36}) should be valid for
$Q^2\ll m_V^2\equiv 12/r_2^2$.
We show both extrapolations, Eqs.~(\ref{eq:linear}) and (\ref{eq:dipole}),
for our three
$m_{\pi}\approx  290\,$MeV volumes (ensembles IV, V and VI, see
Fig~\ref{fig:overview}) in Fig.~\ref{fig_gttextra}. 
The $\tilde{g}_T/g_V$ data (shown in the left panel) are compatible
with a linear behaviour down to our largest $Q^2\approx 0.6\,\mathrm{GeV}^2
\approx m_{\rho}^2$
value, however, in this case we restrict ourselves to the range
$Q^2<0.4\,\mathrm{GeV}^2$ to keep $Q^2<m_V^2\approx m_{\rho}^2$.
Note that for $Lm_{\pi}=3.4$ only one point lies within this window,
so no extrapolation is possible.
In the right panel we show the corresponding dipole fits to the
$Q^2<0.6\,\mathrm{GeV}^2$ data. We see no
significant volume dependence between the $Lm_{\pi}=3.4, 4.2$ and 6.7 data.
Moreover, all five extrapolated values are consistent with each other.

\begin{figure}[t]
\centerline{
\includegraphics[width=.48\textwidth,clip=]{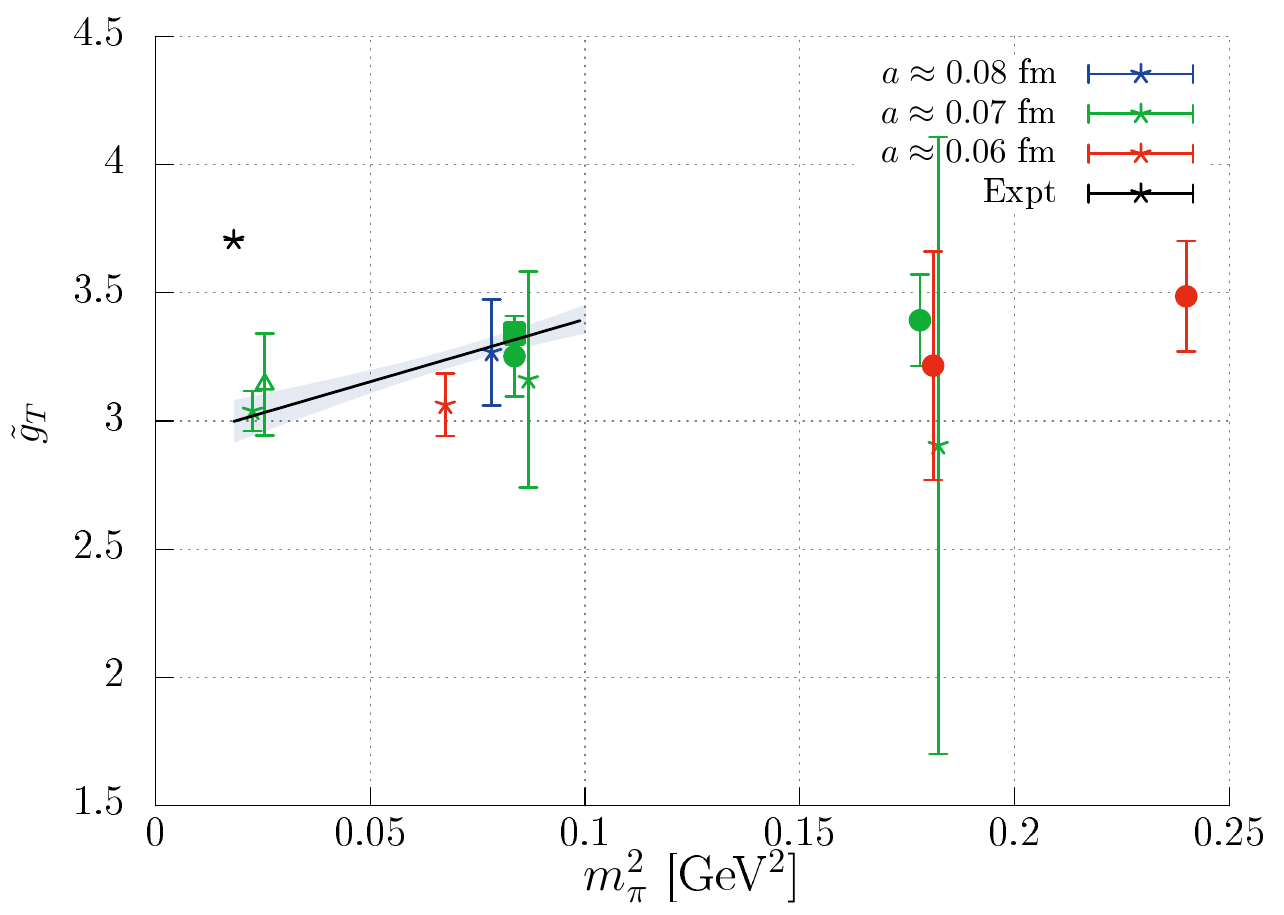}
}
\caption{The isovector induced tensor charge $\tilde{g}_T=\kappa_{u-d}$
as a function of $m_{\pi}^2$.
Symbols are as in Fig.~\protect\ref{fig:overview}.
Also shown is a linear extrapolation
in $m_{\pi}^2$ to the physical point.}
\label{fig_gtt}
\end{figure}

We repeat this procedure for all ensembles and take the central value
from dipole fits, adding in quadrature to the
statistical error an uncertainty from taking the difference between
using the two extrapolation methods and varying the fit range.
The resulting induced tensor charges are shown in
Fig.~\ref{fig_gtt} as a function of $m_{\pi}^2$. Due to the
different volumes the numbers of points within the fit
ranges vary considerably, thus giving rise to
significantly fluctuating error sizes. We
extrapolate the $m_{\pi}<300\,$MeV, $Lm_{\pi}>3.4$ data
linearly to the physical point, obtaining
$\tilde{g}_T=3.00(8)$, which is significantly smaller
than the experimental value $3.706$.
While there could be a deviation between this value
and the one relevant for the isospin symmetric approximation,
one would not expect this to exceed eight of our standard deviations.
It is interesting that results obtained at larger pion masses
are closer to experiment than our lowest mass point, which
dominates the extrapolation. Small volumes result in a larger
low-momentum cut-off and a significant loss of precision
which complicates resolving the volume dependence.
In general, the central values increase with the
lattice size and this deserves further study.

\begin{figure}[t]
\centerline{
\includegraphics[width=.48\textwidth,clip=]{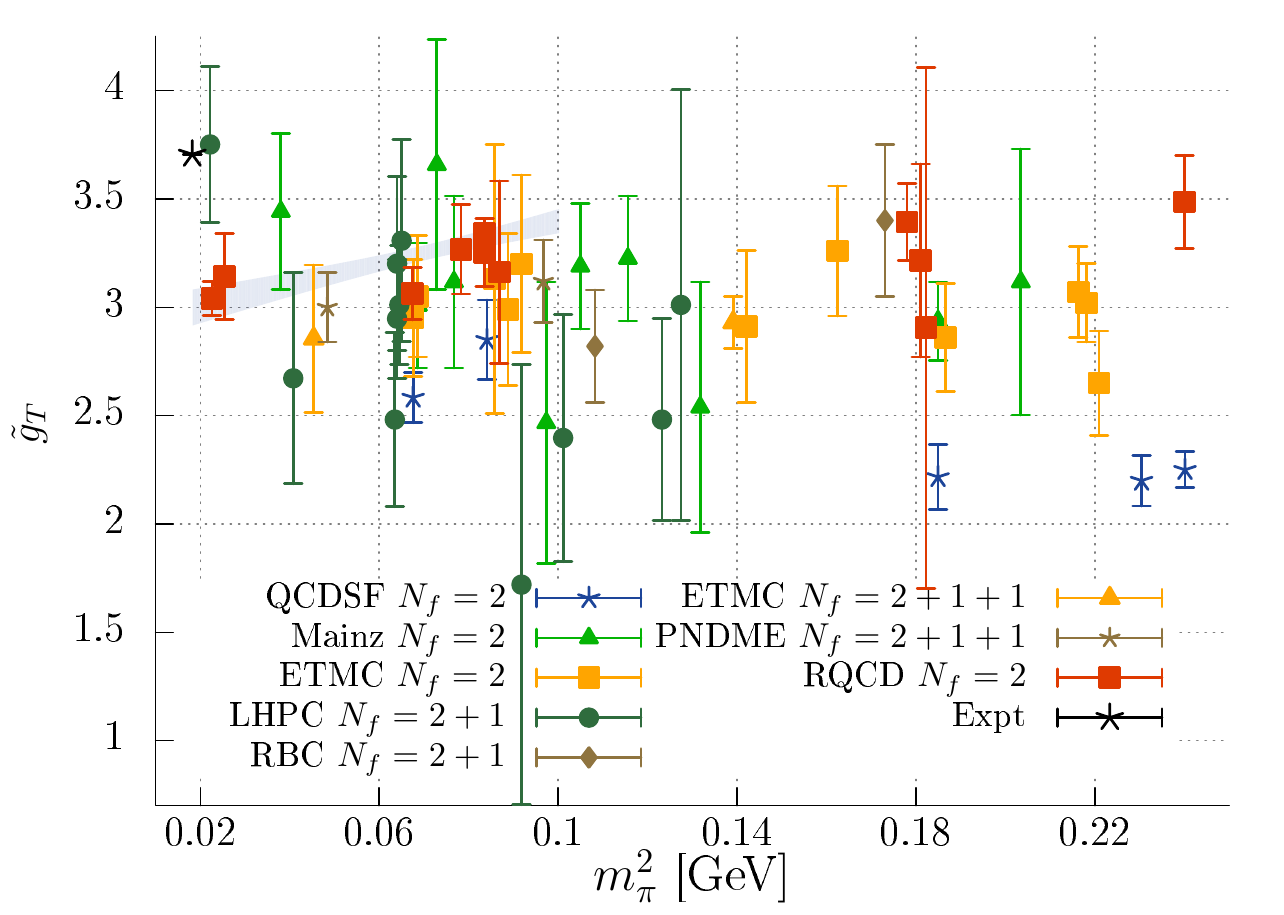}
}
\caption{The isovector anomalous magnetic moment $\tilde{g}_T$ as a function of
$m_{\pi}^2$: our results (RQCD, NPI Wilson-clover)
in comparison
with other results (fermion action used in brackets). $N_{\mathrm{f}}=2$:
QCDSF~\protect\cite{Collins:2011mk} (NPI Wilson-clover), Mainz\textsuperscript{\protect\ref{fn2}}~\protect\cite{Jager:2013kha,vonHippel:2014hla}
(NPI Wilson-clover),
ETMC~\protect\cite{Alexandrou:2011db} (twisted mass).
$N_{\mathrm{f}}=2+1$: LHPC~\protect\cite{Green:2014xba} (HEX-smeared Wilson-clover),
RBC/UKQCD~\protect\cite{Yamazaki:2009zq} (domain wall).
$N_{\mathrm{f}}=2+1+1$: ETMC~\protect\cite{Alexandrou:2013joa} (twisted mass),
PNDME~\protect\cite{Bhattacharya:2013ehc} (Wilson-clover on a
HISQ staggered sea).
Also included is the linear extrapolation of our data.}
\label{fig_gtt2}
\end{figure}

In Fig.~\ref{fig_gtt2}
we compare our results on $\tilde{g}_T$ to recent
lattice determinations by other groups,
namely QCDSF~\cite{Collins:2011mk},
the Mainz group\footnote{\label{fn2}See footnote \protect\ref{fn}.}~\cite{Jager:2013kha,vonHippel:2014hla}
and ETMC~\cite{Alexandrou:2011db} for $N_{\mathrm{f}}=2$,
LHPC~\cite{Green:2014xba} and
RBC/UKQCD~\cite{Yamazaki:2009zq}
for $N_{\mathrm{f}}=2+1$ as well as ETMC~\cite{Alexandrou:2013joa} and
PNDME~\cite{Bhattacharya:2013ehc} 
for $N_{\mathrm{f}}=2+1+1$.
With the exception of one LHPC point, that carries one of the larger error
bars, all the central values are below the experimental result.
The figure does not include 
recent CSSM/QCDSF/UKQCD $N_{\mathrm{f}}=2+1$ stout link
NPI Wilson-clover data that, extrapolated to the physical
point, give $\tilde{g}_T=2.8(3)$~\cite{Shanahan:2014uka}.
Most points with a precision better than 10\% are hard to
reconcile with the experimental value.
At least in part this may be related to finite volume
effects that we are not yet able to resolve sufficiently well.
Discretization effects will be addressed in Sec.~\ref{conc}.

\subsection{The pseudoscalar couplings $g_P^*$ , $g_{\pi NN}$ and $g_P$}
\begin{figure}[t]
\centerline{
\includegraphics[width=.48\textwidth,clip=]{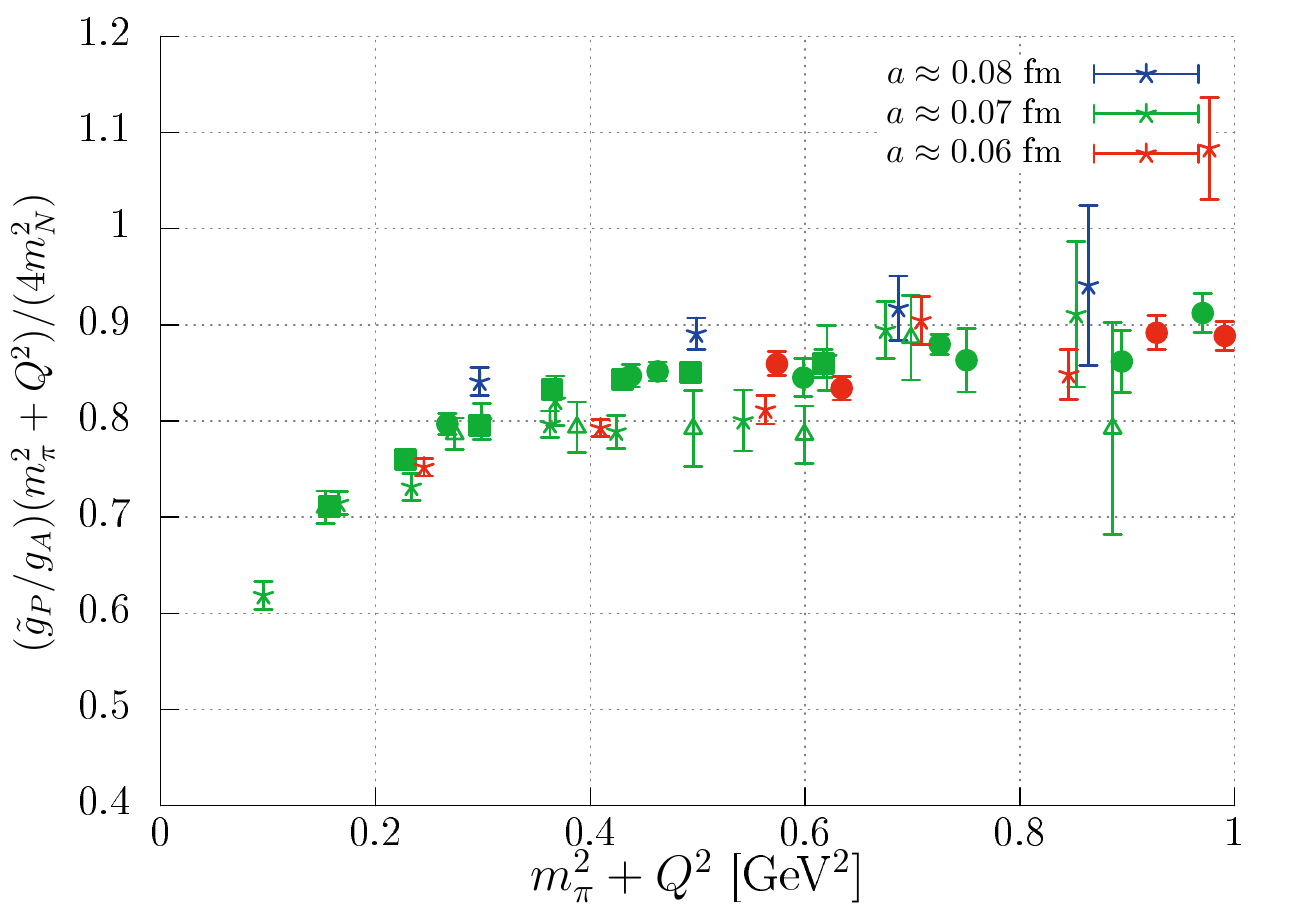}
}
\caption{The ratio of form factors $\tilde{g}_P(Q^2)/g_A(Q^2)$,
normalized with respect to the single pole dominance expectation, as a
function of the virtuality $Q^2$. Data from all 11 ensembles
are plotted on top of each other.
Symbols are as in Fig.~\protect\ref{fig:overview}.
Deviations from unity quantify violations of the pole dominance model.}
\label{fig_gtpga}
\end{figure}

From Eq.~(\ref{eq:gp1}) we expect, up to $\mathcal{O}(aQ)$ discretization
errors,
\begin{equation}
\frac{\tilde{g}_P(Q^2)}{g_A(Q^2)}=
\frac{\tilde{g}_P^{\lat}(Q^2)}{g_A^{\lat}(Q^2)}=
\frac{4c_N^2}{m_{\pi}^2+Q^2}+\cdots\,,
\end{equation}
where $c_N\rightarrow m_N$ as $m_{\pi}\rightarrow 0$ and
the ellipses represent corrections due to singularities at $Q^2<-m_{\pi}^2$,
i.e.\ terms that are regular at $Q^2\geq -m_{\pi}^2$.
Pole dominance implies neglecting these
terms and setting $c_N=m_N$.
In Fig.~\ref{fig_gtpga} we test this model assumption by plotting
the combination
$[\tilde{g}_P(Q^2)/g_A(Q^2)] (m_{\pi}^2+Q^2)/(4m_N^2)$ as
a function of $m_{\pi}^2+Q^2$. The data
obtained at different pion masses, volumes and lattice spacings appear
to follow an almost universal shape, starting out at values around 0.9 at
$m_{\pi}^2+Q^2\approx 1\,\mathrm{GeV}^2$ and decreasing towards 0.6
for $m_{\pi}^2+Q^2\approx 0.1\,\mathrm{GeV}^2$. These deviations
of the ratio from unity illustrate that at small virtualities
terms other than the contribution of the leading pole
cannot be neglected. A similar observation was reported
in Refs.~\cite{Lin:2008uz,Yamazaki:2009zq} where for
$Q^2>0.2\,\mathrm{GeV}^2$ and different quark mass
values $\sim 0.8$ were obtained for this ratio.
Here, we find deviations from single pole dominance
to increase towards low momenta, thereby ruling
out that a dominant part of these violations can be
ascribed to lattice spacing effects.

\begin{figure}[t]
\centerline{
\includegraphics[width=.48\textwidth,clip=]{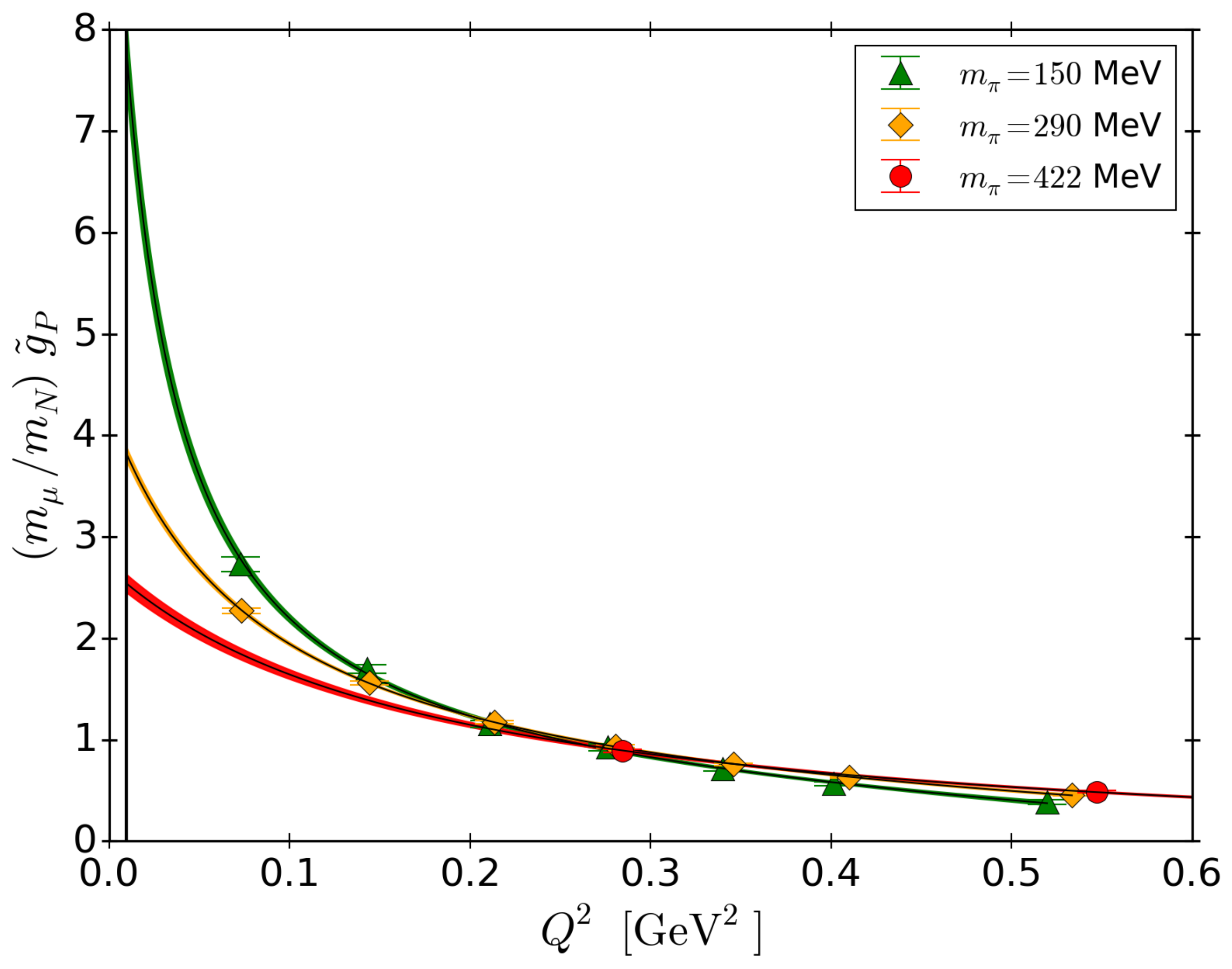}
}
\caption{Extrapolation of the induced pseudoscalar form factor to
the muon capture point $Q^2=0.88 m_{\mu}^2$ (vertical line)
for three values of the
pion mass (ensembles III, VI and VIII). The error bands
correspond to fits according to Eq.~(\protect\ref{eq:exgstar}).}
\label{fig_gstar}
\end{figure}

\begin{figure}[t]
\centerline{
\includegraphics[width=.48\textwidth,clip=]{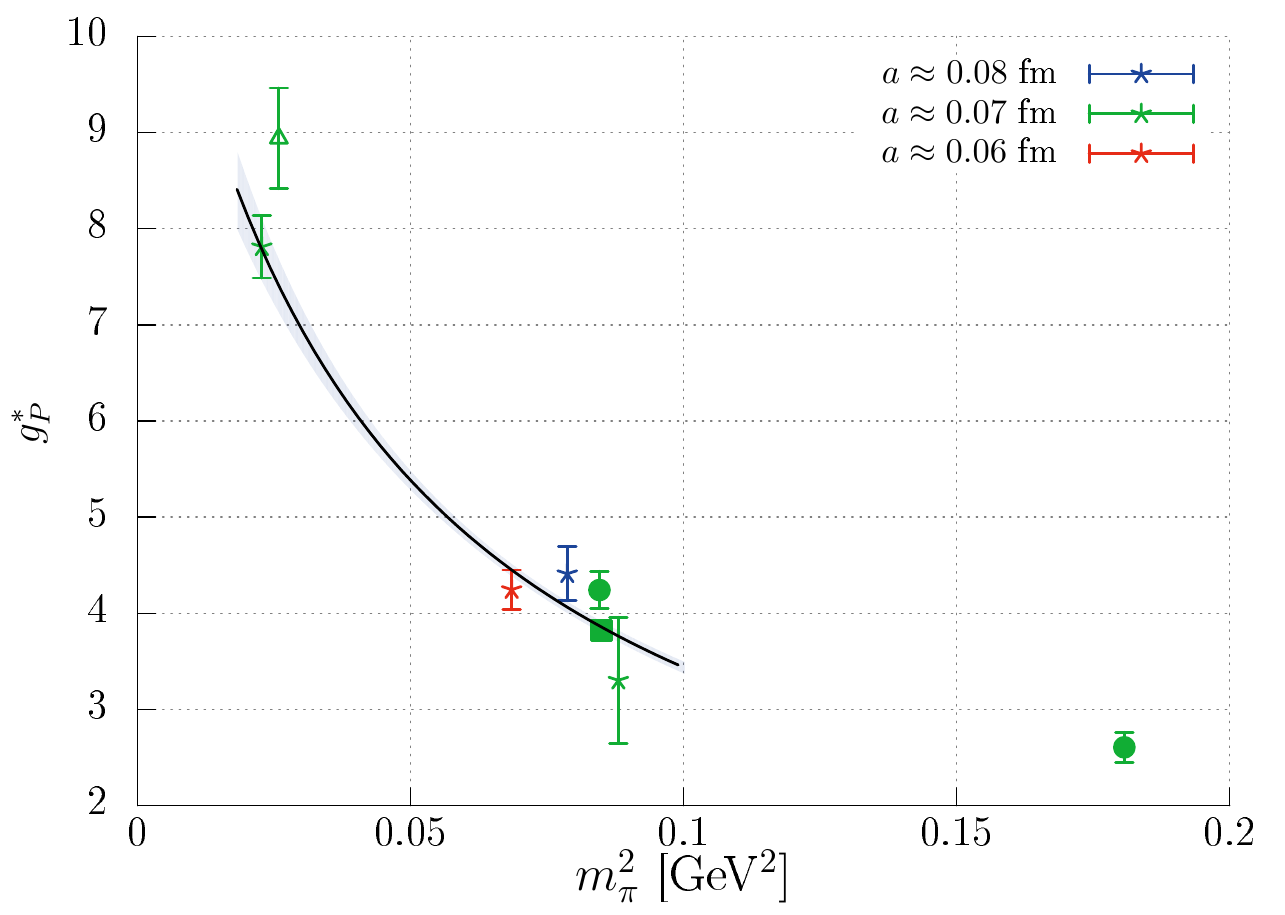}
}
\caption{Chiral extrapolation of the induced pseudoscalar
coupling $g_P^*$. The error band corresponds to the
parametrization Eq.~(\protect\ref{eq:gpstar2}).
Symbols are as in Fig.~\protect\ref{fig:overview}.}
\label{fig_gpstar}
\end{figure}

The induced pseudoscalar coupling for muon capture
$g_P^*$ is defined in Eq.~(\ref{eq:muon}). It can be obtained,
extrapolating the induced pseudoscalar form factor
$(m_{\mu}/m_N)\tilde{g}_P(Q^2)$ to 
$Q^2=9.82\cdot 10^{-3}\,\mathrm{GeV}^2$.
We employ a phenomenological
parametrization that incorporates the leading pole:
\begin{equation}
\label{eq:exgstar}
\frac{m_{\mu}}{m_N}\tilde{g}_P(Q^2)=\frac{c_1}{m_{\pi}^2+Q^2}+c_2+c_3Q^2\,,
\end{equation}
where the parameters $c_1<4m_N^2g_A^0$,
$c_2$ and $c_3$ are fitted separately
for each ensemble. The terms involving $c_2$ and $c_3$
turn out to be necessary to approximate corrections to the pole ansatz,
which are regular at positive virtualities.

We display the resulting extrapolations for three
pion masses (ensembles III, VI and VIII) in Fig.~\ref{fig_gstar}.
We are not able to reliably determine
the above form factor for $Q^2>1\,\mathrm{GeV}^2$ which means
results cannot be obtained for the small volume ensembles
II, IX and X, where less than four data points are within this range.
We show the remaining eight results in Fig.~\ref{fig_gpstar}
as a function of the squared pion mass. A phenomenological
fit of the $m_{\pi}<300\,$MeV, $Lm_{\pi}>3.4$ data
to the functional form
\begin{equation}
\label{eq:gpstar2}
g_P^*(m_{\pi}^2)=\frac{a_1}{m_{\pi}^2+a_2}\,,
\end{equation}
with parameters $a_1$ and $a_2$, gives
$g_P^*=8.40(40)$ at the physical point
with a $\chi^2/N_{\mathrm{DF}}=6.4/4$. Since our
nearly physical $m_{\pi}\approx 150\,$MeV point dominates the
extrapolated value, this is robust against changes of
the parametrization. The number obtained compares well
with the recent experimental determination of the MuCap
Collaboration~\cite{Andreev:2012fj} $g_P^*=8.06(55)$
and also
with the determinations $g_P^*=8.44(23)$~\cite{Bernard:1994wn} or
$g_P^*=8.21(9)$~\cite{Fearing:1997dp}
from heavy baryon chiral perturbation theory
or $g_P^*=8.29^{+24}_{-13}(52)$~\cite{Schindler:2006it}
from covariant baryon chiral perturbation theory.
Previously, the RBC and UKQCD collaborations~\cite{Yamazaki:2009zq}
obtained $g_P^*= 6.6(1.0)$, extrapolating $N_{\mathrm{f}}=2+1$
domain wall fermion results to the physical point.

The flavour changing coupling constant $g_{\pi NN}$ between the nucleon
and the charged pion is defined as the residue of the pole
of the induced pseudoscalar form factor at $Q^2=-m_{\pi}^2$:
\begin{equation}
\label{eq:gpnndef}
g_{\pi NN}\equiv\lim_{Q^2\rightarrow -m_{\pi}^2}\frac{m_{\pi}^2+Q^2}{4m_NF_{\pi}}\tilde{g}_P(Q^2)\,.
\end{equation}
Implementing the
above definition requires an extrapolation
of lattice data, which is limited to positive virtualities.
Figure~\ref{fig_gtpga} demonstrates that corrections
to the pole dominance model become significant towards small
virtualities. Assuming the parametrization
Eq.~(\ref{eq:exgstar}), we obtain $g_{\pi NN}=c_1/(4m_{\mu}F_{\pi})$,
which then needs to be extrapolated to the physical pion mass.
However, it is already obvious from Fig.~\ref{fig_gstar}
that a controlled extrapolation of
$Q^2\gtrsim 0.1\,\mathrm{GeV}^2$ data to negative virtualities
is hardly possible.
Indeed, playing around with different
parametrizations of $\tilde{g}_P(Q^2)$
that assume a pole at $Q^2=-m_{\pi}^2$,
values ranging from
$g_{\pi NN}\sim 8$ up to
$g_{\pi NN}\sim 14$ can easily be produced from our lattice data.

The Goldberger-Treiman relation
$g_{\pi NN}\approx m_Ng_A/F_{\pi}$ does not require such an
extrapolation, however, it is subject to
$\mathcal{O}(m_{\pi}^2)$ corrections.
The relative difference between $g_{\pi NN}$ defined in
Eq.~(\ref{eq:gpnndef}) and this approximation is known as
the Goldberger-Treiman discrepancy
\begin{equation}
\label{eq:defgpnn}
\Delta_{\pi N}=\frac{1}{g_{\pi NN}}\left[g_{\pi NN}-m_N\left.\frac{g_A}{F_{\pi}}\right|_{m_{\pi}=135\,\mathrm{MeV}}\right]\,.
\end{equation}
Using the experimental values of $m_N$, $g_A$ and $F_{\pi}$, the
Goldberger-Treiman relation amounts to
$g_{\pi NN}\approx 12.96(3)$ while determinations of $g_{\pi NN}$
from $N\pi$ scattering data result in values
$g_{\pi NN}=14.11(20)$~\cite{Ericson:2000md},
$g_{\pi NN}=13.76(8)$~\cite{Arndt:2006bf} or
$g_{\pi NN}=13.69(19)$~\cite{Baru:2011bw}.
We remark that obtaining these values also involves
extrapolating in $Q^2$.
Combining the last number quoted above with
the Goldberger-Treiman relation translates into
$\Delta_{\pi N}= 0.053(13)$. Experimental data, both
from nucleon-nucleon scattering and pionic atoms,
have been analysed systematically in the framework of
covariant baryon chiral perturbation theory
in Ref.~\cite{Alarcon:2012kn} (see also
references therein), with the central values obtained
for $g_{\pi NN}$ ranging from 13.0 to 14.1, depending on the
experimental input and the method used
(with or without including the $\Delta$ resonance).

\begin{figure}[t]
\centerline{
\includegraphics[width=.48\textwidth,clip=]{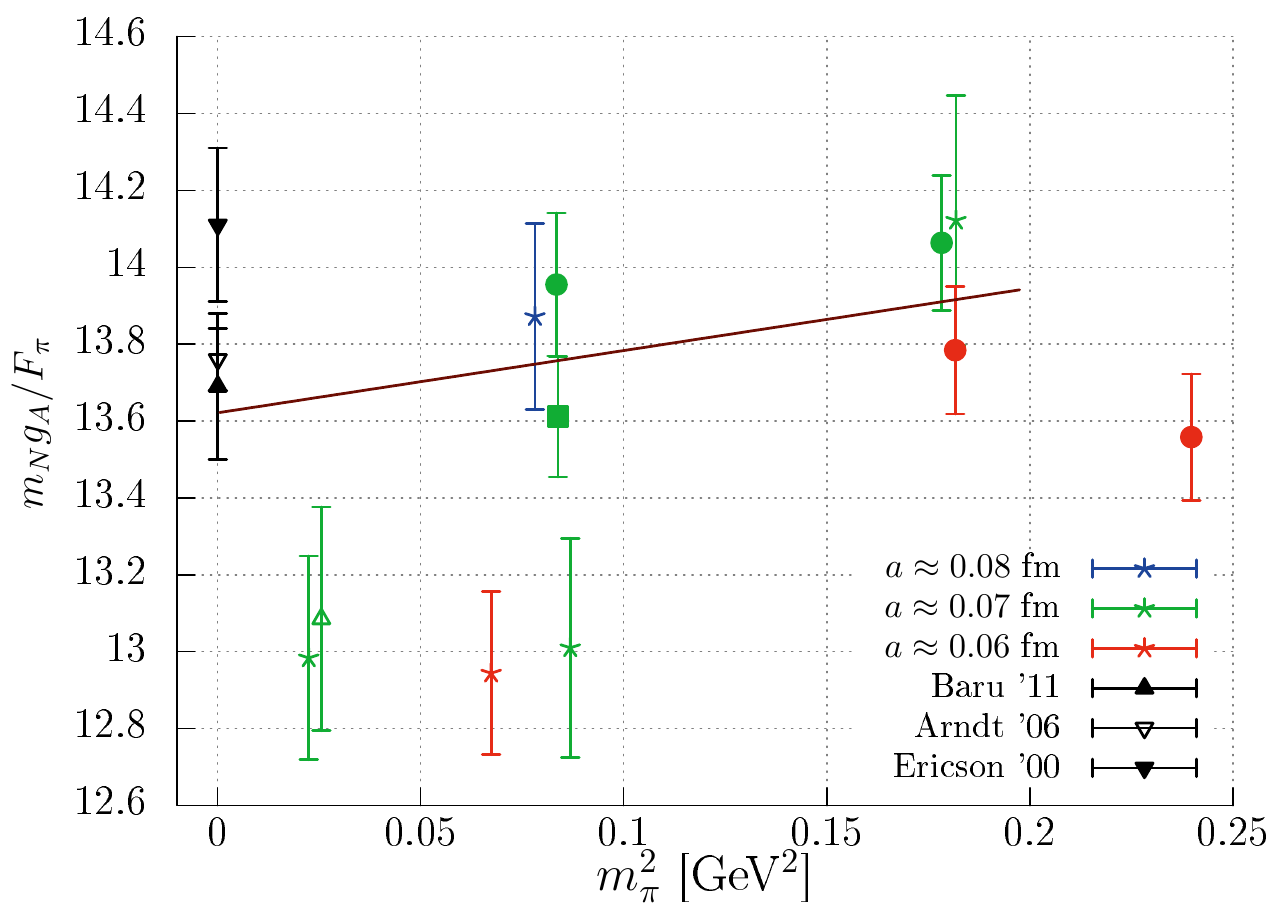}
}
\caption{The Goldberger-Treiman ratio
$m_Ng_A/F_{\pi}$ as a function of the squared pion mass.
Symbols are as in Fig.~\protect\ref{fig:overview}.
The line indicates
a linear extrapolation of $Lm_{\pi}>4.1$ data. The experimental
values for $g_{\pi NN}$ (black triangles) are from
Refs.~\protect\cite{Ericson:2000md,Arndt:2006bf,Baru:2011bw}.}
\label{fig_goldberg}
\end{figure}

In Fig.~\ref{fig_goldberg} we plot the combination
\begin{equation}
\label{eq:gpextra}
m_N\frac{g_A}{F_{\pi}}=
m_N\frac{g_A^{\lat}}{F_{\pi}^{\lat}}=g_{\pi NN}\left[
1+\mathcal{O}\left(m_{\pi}^2\right)\right]
\end{equation}
versus $m_{\pi}^2$, see Eq.~(\ref{eq:gtr}).
As demonstrated in Sec.~\ref{fse},
finite volume effects between $g_A$ and $F_{\pi}$ partially cancel, however,
the nucleon mass adds a new source of volume dependence. Extrapolating
the combination Eq.~(\ref{eq:gpextra}) to the physical
pion mass corresponds to the Goldberger-Treiman approximation
while extrapolating it to $m_{\pi}=0$
gives the pion-nucleon-nucleon coupling in the chiral limit.
A linear fit to the $Lm_{\pi}>4.1$ data (indicated as a line) results in
$g_{\pi NN}(m_{\pi}=0)=13.62(32)$. This is
broadly consistent with the phenomenological
values~\cite{Ericson:2000md,Arndt:2006bf,Baru:2011bw}
that can differ by $\mathcal{O}(m_{\pi}^2)$ terms.
Note, however, that this fit
overestimates the known value 
$m_Ng_A/F_{\pi}=\approx 12.96$ at the physical
pion mass by two standard deviations. We conclude
that while our results are consistent with expectations,
predicting $g_{\pi NN}$ at $m_{\pi}>0$ or determining the
Goldberger-Treiman discrepancy $\Delta_{\pi N}$ requires
different methods, not least due to the significant violations
of single pole dominance illustrated in Fig.~\ref{fig_gtpga}.

\begin{figure}[t]
\centerline{
\includegraphics[width=.48\textwidth,clip=]{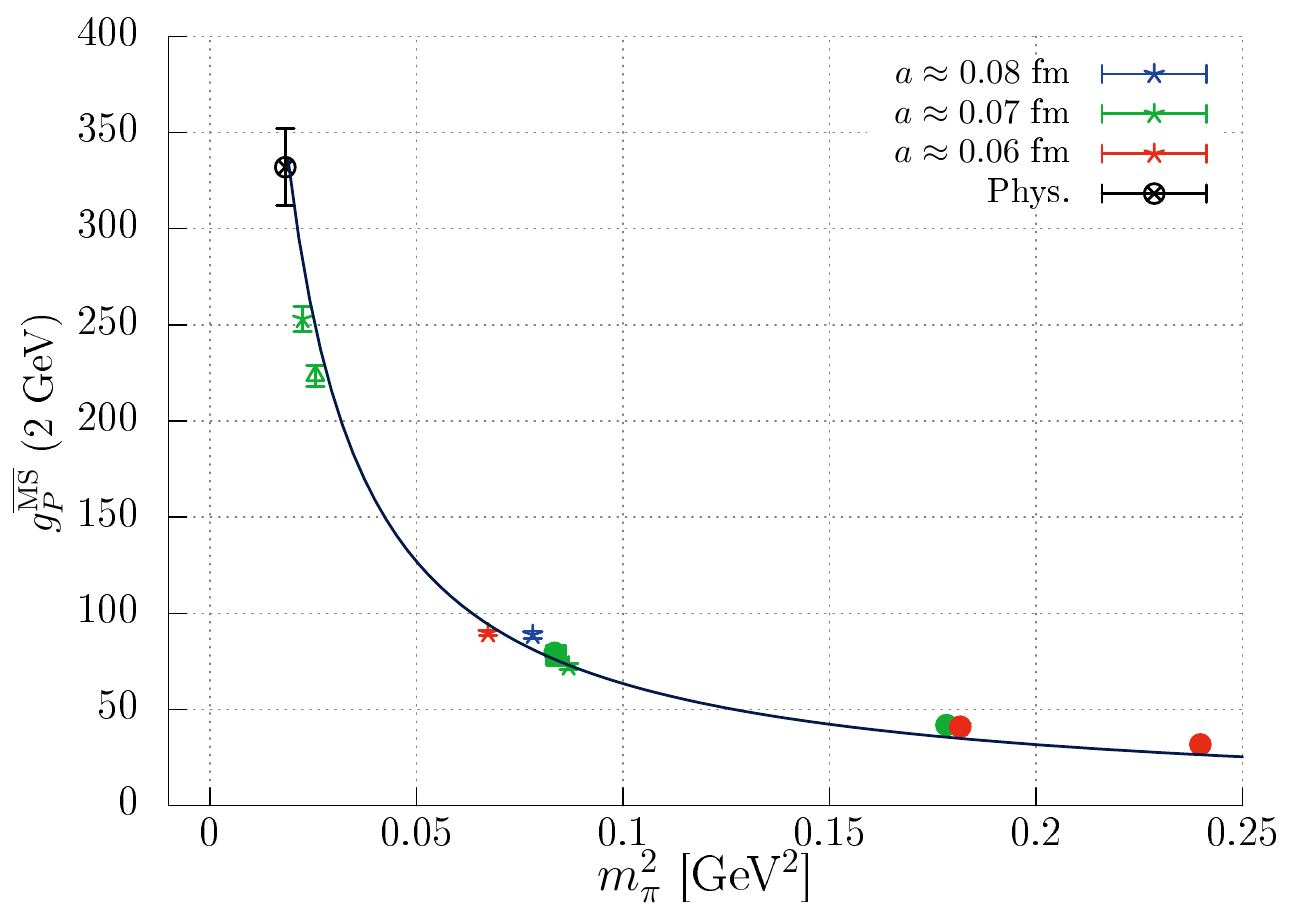}
}
\caption{The pseudoscalar charge $g_P^{\MS}(2\,\mathrm{GeV})$, defined in
Eq.~(\protect\ref{eq:gpplot}), as a function of the squared pion mass.
Symbols are as in Fig.~\protect\ref{fig:overview}.
The physical point (Phys.) is obtained
dividing the experimental value of $m_Ng_A$ by the $\MS$-scheme quark mass
of Ref.~\protect\cite{Aoki:2013ldr}. The $1/m_{\pi}^2$ curve
is drawn to guide the eye.}
\label{fig_gpplot}
\end{figure}

Finally, in Fig.~\ref{fig_gpplot} we show the pseudoscalar charge, obtained
from the first equality in Eq.~(\ref{eq:gtr}):
\begin{equation}
\label{eq:gpplot}
g_P^{\MS}(2\,\mathrm{GeV})=Z_P\frac{m_N}{\tilde{m}}g_A^{\lat}(1+amb_A)\,.
\end{equation}
Note that order-$a$ improvement is already
incorporated into our definition Eq.~(\ref{eq:mpcac})
of the lattice PCAC mass $\tilde{m}$, which is why the coefficient $b_A$
rather than $b_P$ appears above. $Z_P$, $\kappa_{\mathrm{crit}}$
and $P$ [needed to compute $amb_A$, see Eqs.~(\ref{eq:quarkmass}) and
Eq.~(\ref{eq:oneloop})] can be found
in Table~\ref{tab_3} and $g_A^{\lat}$, the nucleon and
lattice PCAC masses in Table~\ref{tab_2}.
We expect $g_P$ to diverge like $1/m_{ud}$ and thus,
using the Gell-Mann-Oakes-Renner relation, to be proportional
to $1/m_{\pi}^2$. Such a curve is drawn to guide the eye.
Using the $N_{\mathrm{f}}=2$
value $m_{ud}^{\MS}(2\,\mathrm{GeV})=3.6(2)\,$MeV
of the FLAG Working group~\cite{Aoki:2013ldr}, from Eq.~(\ref{eq:gtr})
we expect $g_P^{\MS}(2\,\mathrm{GeV})=332(19)$ at the physical point.
Our data are broadly consistent with this value:
obviously our quark mass, extrapolated to $m_{\pi}= 135\,\mathrm{MeV}$,
is consistent with the FLAG average.

\section{Summary}
\label{conc}
We have computed all nucleon charges that may be relevant for
non-standard model (and standard model)
transitions~\cite{Bhattacharya:2011qm,Ivanov:2012qe,Cirigliano:2013xha}
between the neutron and the proton in lattice simulations with
$N_{\mathrm{f}}=2$ mass-degenerate
flavours of sea quarks. These isovector couplings are by 
definition valence quark quantities. Therefore, we
do not expect significant effects from including
strange (or charm) sea quarks. This claim is substantiated by
comparison with lattice results of other groups, some of which
have included more sea quark flavours, see
Figs.~\ref{fig_ga2}, \ref{fig_gs2}, \ref{fig_gt2} and \ref{fig_gtt2}.
In contrast to this, the chiral extrapolation may be an issue.
Therefore, we have
included a point at $m_{\pi}\approx 150\,$MeV, close to
the physical pion mass. Differences between the numbers
obtained at this mass point and our final results, extrapolated
to $m_{\pi}= 135\,$MeV, were all much smaller than the
errors encountered at $m_{\pi}\approx 150\,$MeV.
This means these extrapolations are under control.
Finite volume effects were investigated too and
found to be significant in the case of the axial coupling $g_A$
and, by implication, the pseudoscalar and induced pseudoscalar
form factors. These could be much reduced, considering ratios
over the pion decay constant $F_{\pi}$, which shares a similar
finite volume behaviour. Consistency checks were made,
regarding the renormalization. The known results for
$g_V$ and $g_A$ were reproduced.

\begin{table}
\caption{Summary of results, extrapolated to the physical point.
The first errors contain statistics and systematics. 
The second errors are estimates of lattice spacing effects.
$g_A$ was obtained, dividing by $F_{\pi}$ and therefore
a scale setting error is included in the first error,
that is not subject to further lattice spacing effects.
To determine $g_A^0$ in the chiral limit, the experimental
$g_A$-value was used as an input. The experimental
$g_A$ and $\tilde{g}_T=\kappa_p-\kappa_n$ numbers
are Particle Data Group averages~\cite{Agashe:2014kda}
and $g_P^*$ is from the MuCap Collaboration~\cite{Andreev:2012fj}.}
\label{tab:summary}
\begin{center}
\begin{ruledtabular}
\begin{tabular}{ccc}
& Our result & Experiment\\\hline
$g_A$                      &1.280(44)(46)& 1.2723(23)\\
$g_A^0$                    &1.211(16)(27)& --- \\
$g_S^{\MS}(2\,\mathrm{GeV})$&1.02(18)(30) & --- \\
$g_T^{\MS}(2\,\mathrm{GeV})$&1.005(17)(29)& --- \\
$\tilde{g}_T$              &3.00(08)(31) &3.7058901(5)\\
$g_P^*$                    &8.40(40)(159)&8.06(55)\\
\end{tabular}
\end{ruledtabular}
\end{center}
\end{table}

The charges, extrapolated to the physical point, as well as
$g_A$ in the chiral limit
are summarized in Table~\ref{tab:summary}.
The first errors displayed contain our statistical and
systematic uncertainties related to fit ranges and parametrizations
used. The second errors are estimates of the maximally possible
discretization effects. These were obtained as follows.
To leading order in $a$, assuming $\mathcal{O}(a^n)$
discretization effects, we can write
$g(a)=g(0)+\delta_g a^n/\mathrm{fm}^n=g(0)+\Delta_a g$, where $g(0)$
denotes the continuum limit, $g(a)$ the result for this coupling
determined at a fixed lattice spacing and 
the dimensionless constant $\delta_g$ is unknown.
We varied the lattice constant from $a\approx 0.081\,$fm
down to $a\approx 0.060\,$fm. 
The non-detection of any discretization effect means that our error
on a coupling $g$ is bigger than the associated variation:
$\Delta g> (0.081^n-0.060^n)|\delta_g|$. Our extrapolated results are
dominated by points at $a=0.071\,$fm, meaning that
we cannot exclude lattice corrections
$\Delta_a g =0.071^n|\delta_g|<0.071^n\Delta g/(0.081^n-0.060^n)
\approx 1.7\Delta g$ ($n=2$). Therefore, we multiply our errors by this
factor. For the induced couplings $g^*$ and
$\tilde{g}_T$ the leading discretization
effects are linear in $a$ which is why in these
cases we allow for discretization errors of $3.7\Delta g$.

The errors not related to the lattice spacing vary significantly
between different couplings. Therefore, our estimates of lattice
spacing effects --- if obtained as detailed above ---
become large for some of the channels.
However, there is no obvious
reason why some couplings should carry much larger discretization
effects than others. This means in some cases, in particular for
$g_S$ and $g_P^*$, our discretization error assignment may be
overly conservative. However, in the absence of a real continuum
limit extrapolation, we do not see any way of reliably
estimating this remaining uncertainty.

In addition to the results displayed in Table~\ref{tab:summary},
we find values for the pion-nucleon-nucleon coupling $g_{\pi NN}$, defined
in the chiral limit, consistent
with experimental estimates, which may not be too surprising,
given that $g_A$
comes out correctly. However, violations of the
pole dominance model are found to be large, see Fig.~\ref{fig_gtpga}.
We also quote
$g_P^{\MS}(2\,\mathrm{GeV})=332(19)$, which is no independent
determination as it relies on the FLAG Working Group quark mass
average~\cite{Aoki:2013ldr}. Moreover, we determined
the low energy constant
\begin{equation}
\overline{b}=3.21(42)\,,
\end{equation}
defined
in Eq.~(\ref{eq:bbar}), that encodes the leading order
chiral correction to $g_A^0$.

The disagreement between the anomalous magnetic moment
$\tilde{g}_T=\tilde{g}_T(0)$
and experiment (see Table~\ref{tab:summary})
is puzzling and deserves further attention. The determination
of the induced couplings is less direct than computing
$g_V$, $g_A$, $g_S$ and $g_T$ since it requires extrapolating
form factors to vanishing virtuality, where the momentum
resolution on a finite volume becomes an issue.
The error of this extrapolation to the forward
limit reduces with the minimal momentum available $\pi/L$
while finite volume effects are dominantly functions
of the combination $Lm_{\pi}$. Therefore, $Lm_{\pi}\approx 3.5$
results at $m_{\pi}\approx 290\,$MeV carry much larger errors
than at $m_{\pi}\approx 150\,$MeV, which may hide finite
volume effects. Moreover, we find excited state
contributions to increase with $Q^2$. This behaviour, while
under control at each single value of $Q^2$, may become
amplified in the slope of the form factor and its extrapolation.
We will discuss form factors in detail, including
$\tilde{g}_T(Q^2)$, in a forthcoming publication.

While lattice calculations of baryon structure have not yet reached
the level of precision of computations of quantities related to
meson properties, it is now
possible to obtain predictions, e.g., for the isovector scalar and tensor
charges, with uncertainties that have an impact on beyond-the-standard-model
phenomenology
and in other cases, e.g., for $g_P^*$, to reduce errors to a level
that is competitive with experimental determinations.
The next obvious step is to significantly
vary the lattice spacing, thus enabling a controlled continuum limit
extrapolation, further reducing the remaining uncertainties.

\begin{acknowledgments}
We thank Rajan Gupta and Hartmut Wittig for discussions.
The ensembles were generated primarily  on the QPACE
computer~\cite{Baier:2009yq,Nakamura:2011cd}, which was built as part
of the Deutsche Forschungsgemeinschaft Collaborative Research
Centre/Transregio 55 (SFB/TRR 55) project.
The analyses were performed on the
iDataCool cluster in Regensburg and the SuperMUC system of the Leibniz
Supercomputing Centre in Munich. Additional support was provided by European
Union Initial Training Network Grant No.\  238353 (ITN STRONGnet) and
International Reintegration Grant No.\ 256594.
The BQCD~\cite{Nakamura:2010qh} and
CHROMA~\cite{Edwards:2004sx} software packages were used extensively
along with the locally deflated domain decomposition solver
implementation of openQCD~\cite{luscher3}.
\end{acknowledgments}
\bibliography{charges}

\begin{thebibliography}{101}%
\makeatletter
\providecommand \@ifxundefined [1]{%
 \@ifx{#1\undefined}
}%
\providecommand \@ifnum [1]{%
 \ifnum #1\expandafter \@firstoftwo
 \else \expandafter \@secondoftwo
 \fi
}%
\providecommand \@ifx [1]{%
 \ifx #1\expandafter \@firstoftwo
 \else \expandafter \@secondoftwo
 \fi
}%
\providecommand \natexlab [1]{#1}%
\providecommand \enquote  [1]{``#1''}%
\providecommand \bibnamefont  [1]{#1}%
\providecommand \bibfnamefont [1]{#1}%
\providecommand \citenamefont [1]{#1}%
\providecommand \href@noop [0]{\@secondoftwo}%
\providecommand \href [0]{\begingroup \@sanitize@url \@href}%
\providecommand \@href[1]{\@@startlink{#1}\@@href}%
\providecommand \@@href[1]{\endgroup#1\@@endlink}%
\providecommand \@sanitize@url [0]{\catcode `\\12\catcode `\$12\catcode
  `\&12\catcode `\#12\catcode `\^12\catcode `\_12\catcode `\%12\relax}%
\providecommand \@@startlink[1]{}%
\providecommand \@@endlink[0]{}%
\providecommand \url  [0]{\begingroup\@sanitize@url \@url }%
\providecommand \@url [1]{\endgroup\@href {#1}{\urlprefix }}%
\providecommand \urlprefix  [0]{URL }%
\providecommand \Eprint [0]{\href }%
\providecommand \doibase [0]{http://dx.doi.org/}%
\providecommand \selectlanguage [0]{\@gobble}%
\providecommand \bibinfo  [0]{\@secondoftwo}%
\providecommand \bibfield  [0]{\@secondoftwo}%
\providecommand \translation [1]{[#1]}%
\providecommand \BibitemOpen [0]{}%
\providecommand \bibitemStop [0]{}%
\providecommand \bibitemNoStop [0]{.\EOS\space}%
\providecommand \EOS [0]{\spacefactor3000\relax}%
\providecommand \BibitemShut  [1]{\csname bibitem#1\endcsname}%
\let\auto@bib@innerbib\@empty
\bibitem [{\citenamefont {Brown}(1978)}]{Pauli:1930pc}%
  \BibitemOpen
  \bibfield  {author} {\bibinfo {author} {\bibfnamefont {L.}~\bibnamefont
  {Brown}},\ }\href@noop {} {\bibfield  {journal} {\bibinfo  {journal} {Phys.
  Today}\ }\textbf {\bibinfo {volume} {31}},\ \bibinfo {pages} {27} (\bibinfo
  {year} {1978})}\BibitemShut {NoStop}%
\bibitem [{\citenamefont {Cowan}\ \emph {et~al.}(1956)\citenamefont {Cowan},
  \citenamefont {Reines}, \citenamefont {Harrison}, \citenamefont {Kruse},\
  and\ \citenamefont {McGuire}}]{Cowan:1992xc}%
  \BibitemOpen
  \bibfield  {author} {\bibinfo {author} {\bibfnamefont {C.~L.}\ \bibnamefont
  {Cowan}}, \bibinfo {author} {\bibfnamefont {F.}~\bibnamefont {Reines}},
  \bibinfo {author} {\bibfnamefont {F.~B.}\ \bibnamefont {Harrison}}, \bibinfo
  {author} {\bibfnamefont {H.~W.}\ \bibnamefont {Kruse}}, \ and\ \bibinfo
  {author} {\bibfnamefont {A.~D.}\ \bibnamefont {McGuire}},\ }\href {\doibase
  10.1126/science.124.3212.103} {\bibfield  {journal} {\bibinfo  {journal}
  {Science}\ }\textbf {\bibinfo {volume} {124}},\ \bibinfo {pages} {103}
  (\bibinfo {year} {1956})}\BibitemShut {NoStop}%
\bibitem [{\citenamefont {Olive}\ \emph {et~al.}(2014)\citenamefont {Olive}
  \emph {et~al.}}]{Agashe:2014kda}%
  \BibitemOpen
  \bibfield  {author} {\bibinfo {author} {\bibfnamefont {K.~A.}\ \bibnamefont
  {Olive}} \emph {et~al.} (\bibinfo {collaboration} {Particle Data Group}),\
  }\href {\doibase 10.1088/1674-1137/38/9/090001} {\bibfield  {journal}
  {\bibinfo  {journal} {Chin. Phys.}\ }\textbf {\bibinfo {volume} {C38}},\
  \bibinfo {pages} {090001} (\bibinfo {year} {2014})}\BibitemShut {NoStop}%
\bibitem [{\citenamefont {Kuhn}\ \emph {et~al.}(2009)\citenamefont {Kuhn},
  \citenamefont {Chen},\ and\ \citenamefont {Leader}}]{Kuhn:2008sy}%
  \BibitemOpen
  \bibfield  {author} {\bibinfo {author} {\bibfnamefont {S.~E.}\ \bibnamefont
  {Kuhn}}, \bibinfo {author} {\bibfnamefont {J.-P.}\ \bibnamefont {Chen}}, \
  and\ \bibinfo {author} {\bibfnamefont {E.}~\bibnamefont {Leader}},\ }\href
  {\doibase 10.1016/j.ppnp.2009.02.001} {\bibfield  {journal} {\bibinfo
  {journal} {Prog. Part. Nucl. Phys.}\ }\textbf {\bibinfo {volume} {63}},\
  \bibinfo {pages} {1} (\bibinfo {year} {2009})},\ \Eprint
  {http://arxiv.org/abs/0812.3535} {arXiv:0812.3535 [hep-ph]} \BibitemShut
  {NoStop}%
\bibitem [{\citenamefont {Ademollo}\ and\ \citenamefont
  {Gatto}(1964)}]{Ademollo:1964sr}%
  \BibitemOpen
  \bibfield  {author} {\bibinfo {author} {\bibfnamefont {M.}~\bibnamefont
  {Ademollo}}\ and\ \bibinfo {author} {\bibfnamefont {R.}~\bibnamefont
  {Gatto}},\ }\href {\doibase 10.1103/PhysRevLett.13.264} {\bibfield  {journal}
  {\bibinfo  {journal} {Phys. Rev. Lett.}\ }\textbf {\bibinfo {volume} {13}},\
  \bibinfo {pages} {264} (\bibinfo {year} {1964})}\BibitemShut {NoStop}%
\bibitem [{\citenamefont {Bhattacharya}\ \emph {et~al.}(2012)\citenamefont
  {Bhattacharya}, \citenamefont {Cirigliano}, \citenamefont {Cohen},
  \citenamefont {Filipuzzi}, \citenamefont {Gonzalez-Alonso}, \citenamefont
  {Graesser}, \citenamefont {Gupta},\ and\ \citenamefont
  {Lin}}]{Bhattacharya:2011qm}%
  \BibitemOpen
  \bibfield  {author} {\bibinfo {author} {\bibfnamefont {T.}~\bibnamefont
  {Bhattacharya}}, \bibinfo {author} {\bibfnamefont {V.}~\bibnamefont
  {Cirigliano}}, \bibinfo {author} {\bibfnamefont {S.~D.}\ \bibnamefont
  {Cohen}}, \bibinfo {author} {\bibfnamefont {A.}~\bibnamefont {Filipuzzi}},
  \bibinfo {author} {\bibfnamefont {M.}~\bibnamefont {Gonzalez-Alonso}},
  \bibinfo {author} {\bibfnamefont {M.~L.}\ \bibnamefont {Graesser}}, \bibinfo
  {author} {\bibfnamefont {R.}~\bibnamefont {Gupta}}, \ and\ \bibinfo {author}
  {\bibfnamefont {H.-W.}\ \bibnamefont {Lin}},\ }\href {\doibase
  10.1103/PhysRevD.85.054512} {\bibfield  {journal} {\bibinfo  {journal} {Phys.
  Rev. D}\ }\textbf {\bibinfo {volume} {85}},\ \bibinfo {pages} {054512}
  (\bibinfo {year} {2012})},\ \Eprint {http://arxiv.org/abs/1110.6448}
  {arXiv:1110.6448 [hep-ph]} \BibitemShut {NoStop}%
\bibitem [{\citenamefont {Ivanov}\ \emph {et~al.}(2013)\citenamefont {Ivanov},
  \citenamefont {Pitschmann},\ and\ \citenamefont
  {Troitskaya}}]{Ivanov:2012qe}%
  \BibitemOpen
  \bibfield  {author} {\bibinfo {author} {\bibfnamefont {A.~N.}\ \bibnamefont
  {Ivanov}}, \bibinfo {author} {\bibfnamefont {M.}~\bibnamefont {Pitschmann}},
  \ and\ \bibinfo {author} {\bibfnamefont {N.~I.}\ \bibnamefont {Troitskaya}},\
  }\href {\doibase 10.1103/PhysRevD.88.073002} {\bibfield  {journal} {\bibinfo
  {journal} {Phys. Rev. D}\ }\textbf {\bibinfo {volume} {88}},\ \bibinfo
  {pages} {073002} (\bibinfo {year} {2013})},\ \Eprint
  {http://arxiv.org/abs/1212.0332} {arXiv:1212.0332 [hep-ph]} \BibitemShut
  {NoStop}%
\bibitem [{\citenamefont {Cirigliano}\ \emph {et~al.}(2013)\citenamefont
  {Cirigliano}, \citenamefont {Gardner},\ and\ \citenamefont
  {Holstein}}]{Cirigliano:2013xha}%
  \BibitemOpen
  \bibfield  {author} {\bibinfo {author} {\bibfnamefont {V.}~\bibnamefont
  {Cirigliano}}, \bibinfo {author} {\bibfnamefont {S.}~\bibnamefont {Gardner}},
  \ and\ \bibinfo {author} {\bibfnamefont {B.}~\bibnamefont {Holstein}},\
  }\href {\doibase 10.1016/j.ppnp.2013.03.005} {\bibfield  {journal} {\bibinfo
  {journal} {Prog. Part. Nucl. Phys.}\ }\textbf {\bibinfo {volume} {71}},\
  \bibinfo {pages} {93} (\bibinfo {year} {2013})},\ \Eprint
  {http://arxiv.org/abs/1303.6953} {arXiv:1303.6953 [hep-ph]} \BibitemShut
  {NoStop}%
\bibitem [{\citenamefont {Gasser}\ \emph {et~al.}(1988)\citenamefont {Gasser},
  \citenamefont {Sainio},\ and\ \citenamefont {{\v{S}}varc}}]{Gasser:1987rb}%
  \BibitemOpen
  \bibfield  {author} {\bibinfo {author} {\bibfnamefont {J.}~\bibnamefont
  {Gasser}}, \bibinfo {author} {\bibfnamefont {M.~E.}\ \bibnamefont {Sainio}},
  \ and\ \bibinfo {author} {\bibfnamefont {A.}~\bibnamefont {{\v{S}}varc}},\
  }\href {\doibase 10.1016/0550-3213(88)90108-3} {\bibfield  {journal}
  {\bibinfo  {journal} {Nucl. Phys. B}\ }\textbf {\bibinfo {volume} {307}},\
  \bibinfo {pages} {779} (\bibinfo {year} {1988})}\BibitemShut {NoStop}%
\bibitem [{\citenamefont {Weinberg}(1958)}]{Weinberg:1958ut}%
  \BibitemOpen
  \bibfield  {author} {\bibinfo {author} {\bibfnamefont {S.}~\bibnamefont
  {Weinberg}},\ }\href {\doibase 10.1103/PhysRev.112.1375} {\bibfield
  {journal} {\bibinfo  {journal} {Phys. Rev.}\ }\textbf {\bibinfo {volume}
  {112}},\ \bibinfo {pages} {1375} (\bibinfo {year} {1958})}\BibitemShut
  {NoStop}%
\bibitem [{\citenamefont {Adler}\ and\ \citenamefont
  {Gilman}(1966)}]{Adler:1966gd}%
  \BibitemOpen
  \bibfield  {author} {\bibinfo {author} {\bibfnamefont {S.~L.}\ \bibnamefont
  {Adler}}\ and\ \bibinfo {author} {\bibfnamefont {F.}~\bibnamefont {Gilman}},\
  }\href {\doibase 10.1103/PhysRev.152.1460} {\bibfield  {journal} {\bibinfo
  {journal} {Phys. Rev.}\ }\textbf {\bibinfo {volume} {152}},\ \bibinfo {pages}
  {1460} (\bibinfo {year} {1966})}\BibitemShut {NoStop}%
\bibitem [{\citenamefont {Bernard}\ \emph {et~al.}(2002)\citenamefont
  {Bernard}, \citenamefont {Elouadrhiri},\ and\ \citenamefont
  {Meissner}}]{Bernard:2001rs}%
  \BibitemOpen
  \bibfield  {author} {\bibinfo {author} {\bibfnamefont {V.}~\bibnamefont
  {Bernard}}, \bibinfo {author} {\bibfnamefont {L.}~\bibnamefont
  {Elouadrhiri}}, \ and\ \bibinfo {author} {\bibfnamefont {U.-G.}\ \bibnamefont
  {Meissner}},\ }\href {\doibase 10.1088/0954-3899/28/1/201} {\bibfield
  {journal} {\bibinfo  {journal} {J. Phys. G}\ }\textbf {\bibinfo {volume}
  {28}},\ \bibinfo {pages} {R1} (\bibinfo {year} {2002})},\ \Eprint
  {http://arxiv.org/abs/hep-ph/0107088} {arXiv:hep-ph/0107088 [hep-ph]}
  \BibitemShut {NoStop}%
\bibitem [{\citenamefont {Fuchs}\ and\ \citenamefont
  {Scherer}(2003)}]{Fuchs:2003vw}%
  \BibitemOpen
  \bibfield  {author} {\bibinfo {author} {\bibfnamefont {T.}~\bibnamefont
  {Fuchs}}\ and\ \bibinfo {author} {\bibfnamefont {S.}~\bibnamefont
  {Scherer}},\ }\href {\doibase 10.1103/PhysRevC.68.055501} {\bibfield
  {journal} {\bibinfo  {journal} {Phys. Rev. C}\ }\textbf {\bibinfo {volume}
  {68}},\ \bibinfo {pages} {055501} (\bibinfo {year} {2003})},\ \Eprint
  {http://arxiv.org/abs/nucl-th/0303002} {arXiv:nucl-th/0303002 [nucl-th]}
  \BibitemShut {NoStop}%
\bibitem [{\citenamefont {Goldberger}\ and\ \citenamefont
  {Treiman}(1958)}]{Goldberger:1958vp}%
  \BibitemOpen
  \bibfield  {author} {\bibinfo {author} {\bibfnamefont {M.~L.}\ \bibnamefont
  {Goldberger}}\ and\ \bibinfo {author} {\bibfnamefont {S.~B.}\ \bibnamefont
  {Treiman}},\ }\href {\doibase 10.1103/PhysRev.111.354} {\bibfield  {journal}
  {\bibinfo  {journal} {Phys. Rev.}\ }\textbf {\bibinfo {volume} {111}},\
  \bibinfo {pages} {354} (\bibinfo {year} {1958})}\BibitemShut {NoStop}%
\bibitem [{\citenamefont {Bernard}\ \emph {et~al.}(1995)\citenamefont
  {Bernard}, \citenamefont {Kaiser},\ and\ \citenamefont
  {Meissner}}]{Bernard:1995dp}%
  \BibitemOpen
  \bibfield  {author} {\bibinfo {author} {\bibfnamefont {V.}~\bibnamefont
  {Bernard}}, \bibinfo {author} {\bibfnamefont {N.}~\bibnamefont {Kaiser}}, \
  and\ \bibinfo {author} {\bibfnamefont {U.-G.}\ \bibnamefont {Meissner}},\
  }\href {\doibase 10.1142/S0218301395000092} {\bibfield  {journal} {\bibinfo
  {journal} {Int. J. Mod. Phys. E}\ }\textbf {\bibinfo {volume} {04}},\
  \bibinfo {pages} {193} (\bibinfo {year} {1995})},\ \Eprint
  {http://arxiv.org/abs/hep-ph/9501384} {arXiv:hep-ph/9501384 [hep-ph]}
  \BibitemShut {NoStop}%
\bibitem [{\citenamefont {Fearing}\ \emph {et~al.}(1997)\citenamefont
  {Fearing}, \citenamefont {Lewis}, \citenamefont {Mobed},\ and\ \citenamefont
  {Scherer}}]{Fearing:1997dp}%
  \BibitemOpen
  \bibfield  {author} {\bibinfo {author} {\bibfnamefont {H.~W.}\ \bibnamefont
  {Fearing}}, \bibinfo {author} {\bibfnamefont {R.}~\bibnamefont {Lewis}},
  \bibinfo {author} {\bibfnamefont {N.}~\bibnamefont {Mobed}}, \ and\ \bibinfo
  {author} {\bibfnamefont {S.}~\bibnamefont {Scherer}},\ }\href {\doibase
  10.1103/PhysRevD.56.1783} {\bibfield  {journal} {\bibinfo  {journal} {Phys.
  Rev. D}\ }\textbf {\bibinfo {volume} {56}},\ \bibinfo {pages} {1783}
  (\bibinfo {year} {1997})},\ \Eprint {http://arxiv.org/abs/hep-ph/9702394}
  {arXiv:hep-ph/9702394 [hep-ph]} \BibitemShut {NoStop}%
\bibitem [{\citenamefont {Fettes}\ \emph {et~al.}(1998)\citenamefont {Fettes},
  \citenamefont {Meissner},\ and\ \citenamefont {Steininger}}]{Fettes:1998ud}%
  \BibitemOpen
  \bibfield  {author} {\bibinfo {author} {\bibfnamefont {N.}~\bibnamefont
  {Fettes}}, \bibinfo {author} {\bibfnamefont {U.-G.}\ \bibnamefont
  {Meissner}}, \ and\ \bibinfo {author} {\bibfnamefont {S.}~\bibnamefont
  {Steininger}},\ }\href {\doibase 10.1016/S0375-9474(98)00452-7} {\bibfield
  {journal} {\bibinfo  {journal} {Nucl. Phys. A}\ }\textbf {\bibinfo {volume}
  {640}},\ \bibinfo {pages} {199} (\bibinfo {year} {1998})},\ \Eprint
  {http://arxiv.org/abs/hep-ph/9803266} {arXiv:hep-ph/9803266 [hep-ph]}
  \BibitemShut {NoStop}%
\bibitem [{\citenamefont {Bernard}\ \emph {et~al.}(1994)\citenamefont
  {Bernard}, \citenamefont {Kaiser},\ and\ \citenamefont
  {Meissner}}]{Bernard:1994wn}%
  \BibitemOpen
  \bibfield  {author} {\bibinfo {author} {\bibfnamefont {V.}~\bibnamefont
  {Bernard}}, \bibinfo {author} {\bibfnamefont {N.}~\bibnamefont {Kaiser}}, \
  and\ \bibinfo {author} {\bibfnamefont {U.-G.}\ \bibnamefont {Meissner}},\
  }\href {\doibase 10.1103/PhysRevD.50.6899} {\bibfield  {journal} {\bibinfo
  {journal} {Phys. Rev. D}\ }\textbf {\bibinfo {volume} {50}},\ \bibinfo
  {pages} {6899} (\bibinfo {year} {1994})},\ \Eprint
  {http://arxiv.org/abs/hep-ph/9403351} {arXiv:hep-ph/9403351 [hep-ph]}
  \BibitemShut {NoStop}%
\bibitem [{\citenamefont {Bernard}\ \emph {et~al.}(1998)\citenamefont
  {Bernard}, \citenamefont {Fearing}, \citenamefont {Hemmert},\ and\
  \citenamefont {Meissner}}]{Bernard:1998gv}%
  \BibitemOpen
  \bibfield  {author} {\bibinfo {author} {\bibfnamefont {V.}~\bibnamefont
  {Bernard}}, \bibinfo {author} {\bibfnamefont {H.~W.}\ \bibnamefont
  {Fearing}}, \bibinfo {author} {\bibfnamefont {T.~R.}\ \bibnamefont
  {Hemmert}}, \ and\ \bibinfo {author} {\bibfnamefont {U.-G.}\ \bibnamefont
  {Meissner}},\ }\href {\doibase 10.1016/S0375-9474(98)00175-4} {\bibfield
  {journal} {\bibinfo  {journal} {Nucl. Phys. A}\ }\textbf {\bibinfo {volume}
  {635}},\ \bibinfo {pages} {121} (\bibinfo {year} {1998})},\ \Eprint
  {http://arxiv.org/abs/hep-ph/9801297} {arXiv:hep-ph/9801297 [hep-ph]}
  \BibitemShut {NoStop}%
\bibitem [{\citenamefont {Lin}\ and\ \citenamefont {Cohen}(2011)}]{Lin:2011sa}%
  \BibitemOpen
  \bibfield  {author} {\bibinfo {author} {\bibfnamefont {H.-W.}\ \bibnamefont
  {Lin}}\ and\ \bibinfo {author} {\bibfnamefont {S.~D.}\ \bibnamefont
  {Cohen}},\ }\href@noop {} {\enquote {\bibinfo {title} {{Nucleon and Pion Form
  Factors from $N_f=2+1$ Anisotropic Lattices}},}\ } (\bibinfo {year} {2011}),\
  \Eprint {http://arxiv.org/abs/1104.4319} {arXiv:1104.4319 [hep-lat]}
  \BibitemShut {NoStop}%
\bibitem [{\citenamefont {Dinter}\ \emph {et~al.}(2011)\citenamefont {Dinter},
  \citenamefont {Alexandrou}, \citenamefont {Constantinou}, \citenamefont
  {Drach}, \citenamefont {Jansen},\ and\ \citenamefont
  {Renner}}]{Dinter:2011sg}%
  \BibitemOpen
  \bibfield  {author} {\bibinfo {author} {\bibfnamefont {S.}~\bibnamefont
  {Dinter}}, \bibinfo {author} {\bibfnamefont {C.}~\bibnamefont {Alexandrou}},
  \bibinfo {author} {\bibfnamefont {M.}~\bibnamefont {Constantinou}}, \bibinfo
  {author} {\bibfnamefont {V.}~\bibnamefont {Drach}}, \bibinfo {author}
  {\bibfnamefont {K.}~\bibnamefont {Jansen}}, \ and\ \bibinfo {author}
  {\bibfnamefont {D.~B.}\ \bibnamefont {Renner}} (\bibinfo {collaboration} {ETM
  Collaboration}),\ }\href {\doibase 10.1016/j.physletb.2011.09.002} {\bibfield
   {journal} {\bibinfo  {journal} {Phys. Lett. B}\ }\textbf {\bibinfo {volume}
  {704}},\ \bibinfo {pages} {89} (\bibinfo {year} {2011})},\ \Eprint
  {http://arxiv.org/abs/1108.1076} {arXiv:1108.1076 [hep-lat]} \BibitemShut
  {NoStop}%
\bibitem [{\citenamefont {Capitani}\ \emph {et~al.}(2012)\citenamefont
  {Capitani}, \citenamefont {Della~Morte}, \citenamefont {von Hippel},
  \citenamefont {J{\"a}ger}, \citenamefont {J{\"u}ttner}, \citenamefont
  {Knippschild}, \citenamefont {Meyer},\ and\ \citenamefont
  {Wittig}}]{Capitani:2012gj}%
  \BibitemOpen
  \bibfield  {author} {\bibinfo {author} {\bibfnamefont {S.}~\bibnamefont
  {Capitani}}, \bibinfo {author} {\bibfnamefont {M.}~\bibnamefont
  {Della~Morte}}, \bibinfo {author} {\bibfnamefont {G.}~\bibnamefont {von
  Hippel}}, \bibinfo {author} {\bibfnamefont {B.}~\bibnamefont {J{\"a}ger}},
  \bibinfo {author} {\bibfnamefont {A.}~\bibnamefont {J{\"u}ttner}}, \bibinfo
  {author} {\bibfnamefont {B.}~\bibnamefont {Knippschild}}, \bibinfo {author}
  {\bibfnamefont {H.~B.}\ \bibnamefont {Meyer}}, \ and\ \bibinfo {author}
  {\bibfnamefont {H.}~\bibnamefont {Wittig}},\ }\href {\doibase
  10.1103/PhysRevD.86.074502} {\bibfield  {journal} {\bibinfo  {journal} {Phys.
  Rev. D}\ }\textbf {\bibinfo {volume} {86}},\ \bibinfo {pages} {074502}
  (\bibinfo {year} {2012})},\ \Eprint {http://arxiv.org/abs/1205.0180}
  {arXiv:1205.0180 [hep-lat]} \BibitemShut {NoStop}%
\bibitem [{\citenamefont {Green}\ \emph
  {et~al.}(2014{\natexlab{a}})\citenamefont {Green}, \citenamefont
  {Engelhardt}, \citenamefont {Krieg}, \citenamefont {Negele}, \citenamefont
  {Pochinsky},\ and\ \citenamefont {Syritsyn}}]{Green:2012ud}%
  \BibitemOpen
  \bibfield  {author} {\bibinfo {author} {\bibfnamefont {J.~R.}\ \bibnamefont
  {Green}}, \bibinfo {author} {\bibfnamefont {M.}~\bibnamefont {Engelhardt}},
  \bibinfo {author} {\bibfnamefont {S.}~\bibnamefont {Krieg}}, \bibinfo
  {author} {\bibfnamefont {J.~W.}\ \bibnamefont {Negele}}, \bibinfo {author}
  {\bibfnamefont {A.~V.}\ \bibnamefont {Pochinsky}}, \ and\ \bibinfo {author}
  {\bibfnamefont {S.~N.}\ \bibnamefont {Syritsyn}},\ }\href {\doibase
  10.1016/j.physletb.2014.05.075} {\bibfield  {journal} {\bibinfo  {journal}
  {Phys. Lett. B}\ }\textbf {\bibinfo {volume} {734}},\ \bibinfo {pages} {290}
  (\bibinfo {year} {2014}{\natexlab{a}})},\ \Eprint
  {http://arxiv.org/abs/1209.1687} {arXiv:1209.1687 [hep-lat]} \BibitemShut
  {NoStop}%
\bibitem [{\citenamefont {Lin}\ and\ \citenamefont {Ohta}(2012)}]{Lin:2012nv}%
  \BibitemOpen
  \bibfield  {author} {\bibinfo {author} {\bibfnamefont {M.}~\bibnamefont
  {Lin}}\ and\ \bibinfo {author} {\bibfnamefont {S.}~\bibnamefont {Ohta}}
  (\bibinfo {collaboration} {RBC and UKQCD Collaborations}),\ }\href@noop {}
  {\bibfield  {journal} {\bibinfo  {journal} {Proc. Science}\ }\textbf
  {\bibinfo {volume} {Lattice 2012}},\ \bibinfo {pages} {171} (\bibinfo {year}
  {2012})},\ \Eprint {http://arxiv.org/abs/1212.3235} {arXiv:1212.3235
  [hep-lat]} \BibitemShut {NoStop}%
\bibitem [{\citenamefont {Owen}\ \emph {et~al.}(2013)\citenamefont {Owen},
  \citenamefont {Dragos}, \citenamefont {Kamleh}, \citenamefont {Leinweber},
  \citenamefont {Mahbub}, \citenamefont {Menadue},\ and\ \citenamefont
  {Zanotti}}]{Owen:2012ts}%
  \BibitemOpen
  \bibfield  {author} {\bibinfo {author} {\bibfnamefont {B.~J.}\ \bibnamefont
  {Owen}}, \bibinfo {author} {\bibfnamefont {J.}~\bibnamefont {Dragos}},
  \bibinfo {author} {\bibfnamefont {W.}~\bibnamefont {Kamleh}}, \bibinfo
  {author} {\bibfnamefont {D.~B.}\ \bibnamefont {Leinweber}}, \bibinfo {author}
  {\bibfnamefont {M.~S.}\ \bibnamefont {Mahbub}}, \bibinfo {author}
  {\bibfnamefont {B.~J.}\ \bibnamefont {Menadue}}, \ and\ \bibinfo {author}
  {\bibfnamefont {J.~M.}\ \bibnamefont {Zanotti}},\ }\href {\doibase
  10.1016/j.physletb.2013.04.063} {\bibfield  {journal} {\bibinfo  {journal}
  {Phys. Lett. B}\ }\textbf {\bibinfo {volume} {723}},\ \bibinfo {pages} {217}
  (\bibinfo {year} {2013})},\ \Eprint {http://arxiv.org/abs/1212.4668}
  {arXiv:1212.4668 [hep-lat]} \BibitemShut {NoStop}%
\bibitem [{\citenamefont {Horsley}\ \emph {et~al.}(2014)\citenamefont
  {Horsley}, \citenamefont {Nakamura}, \citenamefont {Nobile}, \citenamefont
  {Rakow}, \citenamefont {Schierholz},\ and\ \citenamefont
  {Zanotti}}]{Horsley:2013ayv}%
  \BibitemOpen
  \bibfield  {author} {\bibinfo {author} {\bibfnamefont {R.}~\bibnamefont
  {Horsley}}, \bibinfo {author} {\bibfnamefont {Y.}~\bibnamefont {Nakamura}},
  \bibinfo {author} {\bibfnamefont {A.}~\bibnamefont {Nobile}}, \bibinfo
  {author} {\bibfnamefont {P.~E.~L.}\ \bibnamefont {Rakow}}, \bibinfo {author}
  {\bibfnamefont {G.}~\bibnamefont {Schierholz}}, \ and\ \bibinfo {author}
  {\bibfnamefont {J.~M.}\ \bibnamefont {Zanotti}},\ }\href {\doibase
  10.1016/j.physletb.2014.03.002} {\bibfield  {journal} {\bibinfo  {journal}
  {Phys. Lett. B}\ }\textbf {\bibinfo {volume} {732}},\ \bibinfo {pages} {41}
  (\bibinfo {year} {2014})},\ \Eprint {http://arxiv.org/abs/1302.2233}
  {arXiv:1302.2233 [hep-lat]} \BibitemShut {NoStop}%
\bibitem [{\citenamefont {Ohta}(2014)}]{Ohta:2013qda}%
  \BibitemOpen
  \bibfield  {author} {\bibinfo {author} {\bibfnamefont {S.}~\bibnamefont
  {Ohta}} (\bibinfo {collaboration} {RBC and UKQCD Collaborations}),\
  }\href@noop {} {\bibfield  {journal} {\bibinfo  {journal} {Proc. Science}\
  }\textbf {\bibinfo {volume} {LATTICE 2013}},\ \bibinfo {pages} {274}
  (\bibinfo {year} {2014})},\ \Eprint {http://arxiv.org/abs/1309.7942}
  {arXiv:1309.7942 [hep-lat]} \BibitemShut {NoStop}%
\bibitem [{\citenamefont {J{\"a}ger}\ \emph {et~al.}(2014)\citenamefont
  {J{\"a}ger}, \citenamefont {Rae}, \citenamefont {Capitani}, \citenamefont
  {Della~Morte}, \citenamefont {Djukanovic}, \citenamefont {von Hippel},
  \citenamefont {Knippschild}, \citenamefont {Meyer},\ and\ \citenamefont
  {Wittig}}]{Jager:2013kha}%
  \BibitemOpen
  \bibfield  {author} {\bibinfo {author} {\bibfnamefont {B.}~\bibnamefont
  {J{\"a}ger}}, \bibinfo {author} {\bibfnamefont {T.~D.}\ \bibnamefont {Rae}},
  \bibinfo {author} {\bibfnamefont {S.}~\bibnamefont {Capitani}}, \bibinfo
  {author} {\bibfnamefont {M.}~\bibnamefont {Della~Morte}}, \bibinfo {author}
  {\bibfnamefont {D.}~\bibnamefont {Djukanovic}}, \bibinfo {author}
  {\bibfnamefont {G.}~\bibnamefont {von Hippel}}, \bibinfo {author}
  {\bibfnamefont {B.}~\bibnamefont {Knippschild}}, \bibinfo {author}
  {\bibfnamefont {H.~B.}\ \bibnamefont {Meyer}}, \ and\ \bibinfo {author}
  {\bibfnamefont {H.}~\bibnamefont {Wittig}},\ }\href@noop {} {\bibfield
  {journal} {\bibinfo  {journal} {Proc. Science}\ }\textbf {\bibinfo {volume}
  {LATTICE 2013}},\ \bibinfo {pages} {272} (\bibinfo {year} {2014})},\ \Eprint
  {http://arxiv.org/abs/1311.5804} {arXiv:1311.5804 [hep-lat]} \BibitemShut
  {NoStop}%
\bibitem [{\citenamefont {Alexandrou}\ \emph
  {et~al.}(2014{\natexlab{a}})\citenamefont {Alexandrou}, \citenamefont
  {Constantinou}, \citenamefont {Drach}, \citenamefont {Jansen}, \citenamefont
  {Kallidonis},\ and\ \citenamefont {Koutsou}}]{Alexandrou:2013jsa}%
  \BibitemOpen
  \bibfield  {author} {\bibinfo {author} {\bibfnamefont {C.}~\bibnamefont
  {Alexandrou}}, \bibinfo {author} {\bibfnamefont {M.}~\bibnamefont
  {Constantinou}}, \bibinfo {author} {\bibfnamefont {V.}~\bibnamefont {Drach}},
  \bibinfo {author} {\bibfnamefont {K.}~\bibnamefont {Jansen}}, \bibinfo
  {author} {\bibfnamefont {C.}~\bibnamefont {Kallidonis}}, \ and\ \bibinfo
  {author} {\bibfnamefont {G.}~\bibnamefont {Koutsou}} (\bibinfo
  {collaboration} {ETM Collaboration}),\ }\href@noop {} {\bibfield  {journal}
  {\bibinfo  {journal} {Proc. Science}\ }\textbf {\bibinfo {volume} {LATTICE
  2013}},\ \bibinfo {pages} {292} (\bibinfo {year} {2014}{\natexlab{a}})},\
  \Eprint {http://arxiv.org/abs/1312.2874} {arXiv:1312.2874 [hep-lat]}
  \BibitemShut {NoStop}%
\bibitem [{\citenamefont {Aoki}\ \emph {et~al.}(2010)\citenamefont {Aoki},
  \citenamefont {Blum}, \citenamefont {Lin}, \citenamefont {Ohta},
  \citenamefont {Sasaki}, \citenamefont {Tweedie}, \citenamefont {Zanotti},\
  and\ \citenamefont {Yamazaki}}]{Aoki:2010xg}%
  \BibitemOpen
  \bibfield  {author} {\bibinfo {author} {\bibfnamefont {Y.}~\bibnamefont
  {Aoki}}, \bibinfo {author} {\bibfnamefont {T.}~\bibnamefont {Blum}}, \bibinfo
  {author} {\bibfnamefont {H.-W.}\ \bibnamefont {Lin}}, \bibinfo {author}
  {\bibfnamefont {S.}~\bibnamefont {Ohta}}, \bibinfo {author} {\bibfnamefont
  {S.}~\bibnamefont {Sasaki}}, \bibinfo {author} {\bibfnamefont
  {R.}~\bibnamefont {Tweedie}}, \bibinfo {author} {\bibfnamefont
  {J.}~\bibnamefont {Zanotti}}, \ and\ \bibinfo {author} {\bibfnamefont
  {T.}~\bibnamefont {Yamazaki}} (\bibinfo {collaboration} {RBC and UKQCD
  Collaborations}),\ }\href {\doibase 10.1103/PhysRevD.82.014501} {\bibfield
  {journal} {\bibinfo  {journal} {Phys. Rev. D}\ }\textbf {\bibinfo {volume}
  {82}},\ \bibinfo {pages} {014501} (\bibinfo {year} {2010})},\ \Eprint
  {http://arxiv.org/abs/1003.3387} {arXiv:1003.3387 [hep-lat]} \BibitemShut
  {NoStop}%
\bibitem [{\citenamefont {Alexandrou}\ \emph
  {et~al.}(2014{\natexlab{b}})\citenamefont {Alexandrou}, \citenamefont
  {Constantinou}, \citenamefont {Jansen}, \citenamefont {Koutsou},\ and\
  \citenamefont {Panagopoulos}}]{Alexandrou:2013wka}%
  \BibitemOpen
  \bibfield  {author} {\bibinfo {author} {\bibfnamefont {C.}~\bibnamefont
  {Alexandrou}}, \bibinfo {author} {\bibfnamefont {M.}~\bibnamefont
  {Constantinou}}, \bibinfo {author} {\bibfnamefont {K.}~\bibnamefont
  {Jansen}}, \bibinfo {author} {\bibfnamefont {G.}~\bibnamefont {Koutsou}}, \
  and\ \bibinfo {author} {\bibfnamefont {H.}~\bibnamefont {Panagopoulos}},\
  }\href@noop {} {\bibfield  {journal} {\bibinfo  {journal} {Proc. Science}\
  }\textbf {\bibinfo {volume} {LATTICE 2013}},\ \bibinfo {pages} {294}
  (\bibinfo {year} {2014}{\natexlab{b}})},\ \Eprint
  {http://arxiv.org/abs/1311.4670} {arXiv:1311.4670 [hep-lat]} \BibitemShut
  {NoStop}%
\bibitem [{\citenamefont {Green}\ \emph {et~al.}(2012)\citenamefont {Green},
  \citenamefont {Negele}, \citenamefont {Pochinsky}, \citenamefont {Syritsyn},
  \citenamefont {Engelhardt},\ and\ \citenamefont {Krieg}}]{Green:2012ej}%
  \BibitemOpen
  \bibfield  {author} {\bibinfo {author} {\bibfnamefont {J.~R.}\ \bibnamefont
  {Green}}, \bibinfo {author} {\bibfnamefont {J.~W.}\ \bibnamefont {Negele}},
  \bibinfo {author} {\bibfnamefont {A.~V.}\ \bibnamefont {Pochinsky}}, \bibinfo
  {author} {\bibfnamefont {S.~N.}\ \bibnamefont {Syritsyn}}, \bibinfo {author}
  {\bibfnamefont {M.}~\bibnamefont {Engelhardt}}, \ and\ \bibinfo {author}
  {\bibfnamefont {S.}~\bibnamefont {Krieg}},\ }\href {\doibase
  10.1103/PhysRevD.86.114509} {\bibfield  {journal} {\bibinfo  {journal} {Phys.
  Rev. D}\ }\textbf {\bibinfo {volume} {86}},\ \bibinfo {pages} {114509}
  (\bibinfo {year} {2012})},\ \Eprint {http://arxiv.org/abs/1206.4527}
  {arXiv:1206.4527 [hep-lat]} \BibitemShut {NoStop}%
\bibitem [{\citenamefont {Bratt}\ \emph {et~al.}(2010)\citenamefont {Bratt}
  \emph {et~al.}}]{Bratt:2010jn}%
  \BibitemOpen
  \bibfield  {author} {\bibinfo {author} {\bibfnamefont {J.~D.}\ \bibnamefont
  {Bratt}} \emph {et~al.} (\bibinfo {collaboration} {LHPC Collaboration}),\
  }\href {\doibase 10.1103/PhysRevD.82.094502} {\bibfield  {journal} {\bibinfo
  {journal} {Phys. Rev. D}\ }\textbf {\bibinfo {volume} {82}},\ \bibinfo
  {pages} {094502} (\bibinfo {year} {2010})},\ \Eprint
  {http://arxiv.org/abs/1001.3620} {arXiv:1001.3620 [hep-lat]} \BibitemShut
  {NoStop}%
\bibitem [{\citenamefont {Alexandrou}\ \emph
  {et~al.}(2011{\natexlab{a}})\citenamefont {Alexandrou}, \citenamefont
  {Brinet}, \citenamefont {Carbonell}, \citenamefont {Constantinou},
  \citenamefont {Harraud}, \citenamefont {Guichon}, \citenamefont {Jansen},
  \citenamefont {Korzec},\ and\ \citenamefont {Papinutto}}]{Alexandrou:2010hf}%
  \BibitemOpen
  \bibfield  {author} {\bibinfo {author} {\bibfnamefont {C.}~\bibnamefont
  {Alexandrou}}, \bibinfo {author} {\bibfnamefont {M.}~\bibnamefont {Brinet}},
  \bibinfo {author} {\bibfnamefont {J.}~\bibnamefont {Carbonell}}, \bibinfo
  {author} {\bibfnamefont {M.}~\bibnamefont {Constantinou}}, \bibinfo {author}
  {\bibfnamefont {P.~A.}\ \bibnamefont {Harraud}}, \bibinfo {author}
  {\bibfnamefont {P.}~\bibnamefont {Guichon}}, \bibinfo {author} {\bibfnamefont
  {K.}~\bibnamefont {Jansen}}, \bibinfo {author} {\bibfnamefont
  {T.}~\bibnamefont {Korzec}}, \ and\ \bibinfo {author} {\bibfnamefont
  {M.}~\bibnamefont {Papinutto}} (\bibinfo {collaboration} {ETM
  Collaboration}),\ }\href {\doibase 10.1103/PhysRevD.83.045010} {\bibfield
  {journal} {\bibinfo  {journal} {Phys. Rev. D}\ }\textbf {\bibinfo {volume}
  {83}},\ \bibinfo {pages} {045010} (\bibinfo {year} {2011}{\natexlab{a}})},\
  \Eprint {http://arxiv.org/abs/1012.0857} {arXiv:1012.0857 [hep-lat]}
  \BibitemShut {NoStop}%
\bibitem [{\citenamefont {Alexandrou}\ \emph {et~al.}(2013)\citenamefont
  {Alexandrou}, \citenamefont {Constantinou}, \citenamefont {Dinter},
  \citenamefont {Drach}, \citenamefont {Jansen}, \citenamefont {Kallidonis},\
  and\ \citenamefont {Koutsou}}]{Alexandrou:2013joa}%
  \BibitemOpen
  \bibfield  {author} {\bibinfo {author} {\bibfnamefont {C.}~\bibnamefont
  {Alexandrou}}, \bibinfo {author} {\bibfnamefont {M.}~\bibnamefont
  {Constantinou}}, \bibinfo {author} {\bibfnamefont {S.}~\bibnamefont
  {Dinter}}, \bibinfo {author} {\bibfnamefont {V.}~\bibnamefont {Drach}},
  \bibinfo {author} {\bibfnamefont {K.}~\bibnamefont {Jansen}}, \bibinfo
  {author} {\bibfnamefont {C.}~\bibnamefont {Kallidonis}}, \ and\ \bibinfo
  {author} {\bibfnamefont {G.}~\bibnamefont {Koutsou}} (\bibinfo
  {collaboration} {ETM Collaboration}),\ }\href {\doibase
  10.1103/PhysRevD.88.014509} {\bibfield  {journal} {\bibinfo  {journal} {Phys.
  Rev. D}\ }\textbf {\bibinfo {volume} {88}},\ \bibinfo {pages} {014509}
  (\bibinfo {year} {2013})},\ \Eprint {http://arxiv.org/abs/1303.5979}
  {arXiv:1303.5979 [hep-lat]} \BibitemShut {NoStop}%
\bibitem [{\citenamefont {Junnarkar}\ \emph {et~al.}(2015)\citenamefont
  {Junnarkar}, \citenamefont {Capitani}, \citenamefont {Djukanovic},
  \citenamefont {von Hippel}, \citenamefont {Hua}, \citenamefont {J{\"a}ger},
  \citenamefont {Meyer}, \citenamefont {Rae},\ and\ \citenamefont
  {Wittig}}]{Junnarkar:2014jxa}%
  \BibitemOpen
  \bibfield  {author} {\bibinfo {author} {\bibfnamefont {P.~M.}\ \bibnamefont
  {Junnarkar}}, \bibinfo {author} {\bibfnamefont {S.}~\bibnamefont {Capitani}},
  \bibinfo {author} {\bibfnamefont {D.}~\bibnamefont {Djukanovic}}, \bibinfo
  {author} {\bibfnamefont {G.}~\bibnamefont {von Hippel}}, \bibinfo {author}
  {\bibfnamefont {J.}~\bibnamefont {Hua}}, \bibinfo {author} {\bibfnamefont
  {B.}~\bibnamefont {J{\"a}ger}}, \bibinfo {author} {\bibfnamefont {H.~B.}\
  \bibnamefont {Meyer}}, \bibinfo {author} {\bibfnamefont {T.~D.}\ \bibnamefont
  {Rae}}, \ and\ \bibinfo {author} {\bibfnamefont {H.}~\bibnamefont {Wittig}},\
  }\href@noop {} {\bibfield  {journal} {\bibinfo  {journal} {Proc. Science}\
  }\textbf {\bibinfo {volume} {LATTICE2014}},\ \bibinfo {pages} {150} (\bibinfo
  {year} {2015})},\ \Eprint {http://arxiv.org/abs/1411.5828} {arXiv:1411.5828
  [hep-lat]} \BibitemShut {NoStop}%
\bibitem [{\citenamefont {Lin}\ \emph {et~al.}(2008)\citenamefont {Lin},
  \citenamefont {Blum}, \citenamefont {Ohta}, \citenamefont {Sasaki},\ and\
  \citenamefont {Yamazaki}}]{Lin:2008uz}%
  \BibitemOpen
  \bibfield  {author} {\bibinfo {author} {\bibfnamefont {H.-W.}\ \bibnamefont
  {Lin}}, \bibinfo {author} {\bibfnamefont {T.}~\bibnamefont {Blum}}, \bibinfo
  {author} {\bibfnamefont {S.}~\bibnamefont {Ohta}}, \bibinfo {author}
  {\bibfnamefont {S.}~\bibnamefont {Sasaki}}, \ and\ \bibinfo {author}
  {\bibfnamefont {T.}~\bibnamefont {Yamazaki}},\ }\href {\doibase
  10.1103/PhysRevD.78.014505} {\bibfield  {journal} {\bibinfo  {journal} {Phys.
  Rev. D}\ }\textbf {\bibinfo {volume} {78}},\ \bibinfo {pages} {014505}
  (\bibinfo {year} {2008})},\ \Eprint {http://arxiv.org/abs/0802.0863}
  {arXiv:0802.0863 [hep-lat]} \BibitemShut {NoStop}%
\bibitem [{\citenamefont {Yamazaki}\ \emph {et~al.}(2009)\citenamefont
  {Yamazaki}, \citenamefont {Aoki}, \citenamefont {Blum}, \citenamefont {Lin},
  \citenamefont {Ohta}, \citenamefont {Sasaki}, \citenamefont {Tweedie},\ and\
  \citenamefont {Zanotti}}]{Yamazaki:2009zq}%
  \BibitemOpen
  \bibfield  {author} {\bibinfo {author} {\bibfnamefont {T.}~\bibnamefont
  {Yamazaki}}, \bibinfo {author} {\bibfnamefont {Y.}~\bibnamefont {Aoki}},
  \bibinfo {author} {\bibfnamefont {T.}~\bibnamefont {Blum}}, \bibinfo {author}
  {\bibfnamefont {H.-W.}\ \bibnamefont {Lin}}, \bibinfo {author} {\bibfnamefont
  {S.}~\bibnamefont {Ohta}}, \bibinfo {author} {\bibfnamefont {S.}~\bibnamefont
  {Sasaki}}, \bibinfo {author} {\bibfnamefont {R.}~\bibnamefont {Tweedie}}, \
  and\ \bibinfo {author} {\bibfnamefont {J.}~\bibnamefont {Zanotti}} (\bibinfo
  {collaboration} {RBC and UKQCD Collaborations}),\ }\href {\doibase
  10.1103/PhysRevD.79.114505} {\bibfield  {journal} {\bibinfo  {journal} {Phys.
  Rev. D}\ }\textbf {\bibinfo {volume} {79}},\ \bibinfo {pages} {114505}
  (\bibinfo {year} {2009})},\ \Eprint {http://arxiv.org/abs/0904.2039}
  {arXiv:0904.2039 [hep-lat]} \BibitemShut {NoStop}%
\bibitem [{\citenamefont {Bhattacharya}\ \emph {et~al.}(2014)\citenamefont
  {Bhattacharya}, \citenamefont {Cohen}, \citenamefont {Gupta}, \citenamefont
  {Joseph}, \citenamefont {Lin},\ and\ \citenamefont
  {Yoon}}]{Bhattacharya:2013ehc}%
  \BibitemOpen
  \bibfield  {author} {\bibinfo {author} {\bibfnamefont {T.}~\bibnamefont
  {Bhattacharya}}, \bibinfo {author} {\bibfnamefont {S.~D.}\ \bibnamefont
  {Cohen}}, \bibinfo {author} {\bibfnamefont {R.}~\bibnamefont {Gupta}},
  \bibinfo {author} {\bibfnamefont {A.}~\bibnamefont {Joseph}}, \bibinfo
  {author} {\bibfnamefont {H.-W.}\ \bibnamefont {Lin}}, \ and\ \bibinfo
  {author} {\bibfnamefont {B.}~\bibnamefont {Yoon}} (\bibinfo {collaboration}
  {PNDME Collaboration}),\ }\href {\doibase 10.1103/PhysRevD.89.094502}
  {\bibfield  {journal} {\bibinfo  {journal} {Phys. Rev. D}\ }\textbf {\bibinfo
  {volume} {89}},\ \bibinfo {pages} {094502} (\bibinfo {year} {2014})},\
  \Eprint {http://arxiv.org/abs/1306.5435} {arXiv:1306.5435 [hep-lat]}
  \BibitemShut {NoStop}%
\bibitem [{\citenamefont {Alexandrou}\ \emph {et~al.}(2015)\citenamefont
  {Alexandrou}, \citenamefont {Constantinou}, \citenamefont {Hadjiyiannakou},
  \citenamefont {Jansen}, \citenamefont {Kallidonis},\ and\ \citenamefont
  {Koutsou}}]{Alexandrou:2014wca}%
  \BibitemOpen
  \bibfield  {author} {\bibinfo {author} {\bibfnamefont {C.}~\bibnamefont
  {Alexandrou}}, \bibinfo {author} {\bibfnamefont {M.}~\bibnamefont
  {Constantinou}}, \bibinfo {author} {\bibfnamefont {K.}~\bibnamefont
  {Hadjiyiannakou}}, \bibinfo {author} {\bibfnamefont {K.}~\bibnamefont
  {Jansen}}, \bibinfo {author} {\bibfnamefont {C.}~\bibnamefont {Kallidonis}},
  \ and\ \bibinfo {author} {\bibfnamefont {G.}~\bibnamefont {Koutsou}},\
  }\href@noop {} {\bibfield  {journal} {\bibinfo  {journal} {Proc. Science}\
  }\textbf {\bibinfo {volume} {LATTICE2014}},\ \bibinfo {pages} {151} (\bibinfo
  {year} {2015})},\ \Eprint {http://arxiv.org/abs/1411.3494} {arXiv:1411.3494
  [hep-lat]} \BibitemShut {NoStop}%
\bibitem [{\citenamefont {Syritsyn}\ \emph {et~al.}(2010)\citenamefont
  {Syritsyn} \emph {et~al.}}]{Syritsyn:2009mx}%
  \BibitemOpen
  \bibfield  {author} {\bibinfo {author} {\bibfnamefont {S.~N.}\ \bibnamefont
  {Syritsyn}} \emph {et~al.} (\bibinfo {collaboration} {LHPC Collaboration}),\
  }\href {\doibase 10.1103/PhysRevD.81.034507} {\bibfield  {journal} {\bibinfo
  {journal} {Phys. Rev. D}\ }\textbf {\bibinfo {volume} {81}},\ \bibinfo
  {pages} {034507} (\bibinfo {year} {2010})},\ \Eprint
  {http://arxiv.org/abs/0907.4194} {arXiv:0907.4194 [hep-lat]} \BibitemShut
  {NoStop}%
\bibitem [{\citenamefont {Collins}\ \emph {et~al.}(2011)\citenamefont {Collins}
  \emph {et~al.}}]{Collins:2011mk}%
  \BibitemOpen
  \bibfield  {author} {\bibinfo {author} {\bibfnamefont {S.}~\bibnamefont
  {Collins}} \emph {et~al.} (\bibinfo {collaboration} {QCDSF Collaboration}),\
  }\href {\doibase 10.1103/PhysRevD.84.074507} {\bibfield  {journal} {\bibinfo
  {journal} {Phys. Rev. D}\ }\textbf {\bibinfo {volume} {84}},\ \bibinfo
  {pages} {074507} (\bibinfo {year} {2011})},\ \Eprint
  {http://arxiv.org/abs/1106.3580} {arXiv:1106.3580 [hep-lat]} \BibitemShut
  {NoStop}%
\bibitem [{\citenamefont {Alexandrou}\ \emph
  {et~al.}(2011{\natexlab{b}})\citenamefont {Alexandrou}, \citenamefont
  {Brinet}, \citenamefont {Carbonell}, \citenamefont {Constantinou},
  \citenamefont {Harraud}, \citenamefont {Guichon}, \citenamefont {Jansen},
  \citenamefont {Korzec},\ and\ \citenamefont {Papinutto}}]{Alexandrou:2011db}%
  \BibitemOpen
  \bibfield  {author} {\bibinfo {author} {\bibfnamefont {C.}~\bibnamefont
  {Alexandrou}}, \bibinfo {author} {\bibfnamefont {M.}~\bibnamefont {Brinet}},
  \bibinfo {author} {\bibfnamefont {J.}~\bibnamefont {Carbonell}}, \bibinfo
  {author} {\bibfnamefont {M.}~\bibnamefont {Constantinou}}, \bibinfo {author}
  {\bibfnamefont {P.~A.}\ \bibnamefont {Harraud}}, \bibinfo {author}
  {\bibfnamefont {P.}~\bibnamefont {Guichon}}, \bibinfo {author} {\bibfnamefont
  {K.}~\bibnamefont {Jansen}}, \bibinfo {author} {\bibfnamefont
  {T.}~\bibnamefont {Korzec}}, \ and\ \bibinfo {author} {\bibfnamefont
  {M.}~\bibnamefont {Papinutto}} (\bibinfo {collaboration} {ETM
  Collaboration}),\ }\href {\doibase 10.1103/PhysRevD.83.094502} {\bibfield
  {journal} {\bibinfo  {journal} {Phys. Rev. D}\ }\textbf {\bibinfo {volume}
  {83}},\ \bibinfo {pages} {094502} (\bibinfo {year} {2011}{\natexlab{b}})},\
  \Eprint {http://arxiv.org/abs/1102.2208} {arXiv:1102.2208 [hep-lat]}
  \BibitemShut {NoStop}%
\bibitem [{\citenamefont {Shanahan}\ \emph {et~al.}(2014)\citenamefont
  {Shanahan}, \citenamefont {Thomas}, \citenamefont {Young}, \citenamefont
  {Zanotti}, \citenamefont {Horsley}, \citenamefont {Nakamura}, \citenamefont
  {Pleiter}, \citenamefont {Rakow}, \citenamefont {Schierholz},\ and\
  \citenamefont {St{\"u}ben}}]{Shanahan:2014uka}%
  \BibitemOpen
  \bibfield  {author} {\bibinfo {author} {\bibfnamefont {P.~E.}\ \bibnamefont
  {Shanahan}}, \bibinfo {author} {\bibfnamefont {A.~W.}\ \bibnamefont
  {Thomas}}, \bibinfo {author} {\bibfnamefont {R.~D.}\ \bibnamefont {Young}},
  \bibinfo {author} {\bibfnamefont {J.~M.}\ \bibnamefont {Zanotti}}, \bibinfo
  {author} {\bibfnamefont {R.}~\bibnamefont {Horsley}}, \bibinfo {author}
  {\bibfnamefont {Y.}~\bibnamefont {Nakamura}}, \bibinfo {author}
  {\bibfnamefont {D.}~\bibnamefont {Pleiter}}, \bibinfo {author} {\bibfnamefont
  {P.~E.~L.}\ \bibnamefont {Rakow}}, \bibinfo {author} {\bibfnamefont
  {G.}~\bibnamefont {Schierholz}}, \ and\ \bibinfo {author} {\bibfnamefont
  {H.}~\bibnamefont {St{\"u}ben}} (\bibinfo {collaboration} {CSSM and
  QCDSF/UKQCD Collaborations}),\ }\href {\doibase 10.1103/PhysRevD.89.074511}
  {\bibfield  {journal} {\bibinfo  {journal} {Phys. Rev. D}\ }\textbf {\bibinfo
  {volume} {89}},\ \bibinfo {pages} {074511} (\bibinfo {year} {2014})},\
  \Eprint {http://arxiv.org/abs/1401.5862} {arXiv:1401.5862 [hep-lat]}
  \BibitemShut {NoStop}%
\bibitem [{\citenamefont {Green}\ \emph
  {et~al.}(2014{\natexlab{b}})\citenamefont {Green}, \citenamefont {Negele},
  \citenamefont {Pochinsky}, \citenamefont {Syritsyn}, \citenamefont
  {Engelhardt},\ and\ \citenamefont {Krieg}}]{Green:2014xba}%
  \BibitemOpen
  \bibfield  {author} {\bibinfo {author} {\bibfnamefont {J.~R.}\ \bibnamefont
  {Green}}, \bibinfo {author} {\bibfnamefont {J.~W.}\ \bibnamefont {Negele}},
  \bibinfo {author} {\bibfnamefont {A.~V.}\ \bibnamefont {Pochinsky}}, \bibinfo
  {author} {\bibfnamefont {S.~N.}\ \bibnamefont {Syritsyn}}, \bibinfo {author}
  {\bibfnamefont {M.}~\bibnamefont {Engelhardt}}, \ and\ \bibinfo {author}
  {\bibfnamefont {S.}~\bibnamefont {Krieg}},\ }\href {\doibase
  10.1103/PhysRevD.90.074507} {\bibfield  {journal} {\bibinfo  {journal} {Phys.
  Rev. D}\ }\textbf {\bibinfo {volume} {90}},\ \bibinfo {pages} {074507}
  (\bibinfo {year} {2014}{\natexlab{b}})},\ \Eprint
  {http://arxiv.org/abs/1404.4029} {arXiv:1404.4029 [hep-lat]} \BibitemShut
  {NoStop}%
\bibitem [{\citenamefont {von Hippel}\ \emph {et~al.}(2015)\citenamefont {von
  Hippel}, \citenamefont {Capitani}, \citenamefont {Djukanovic}, \citenamefont
  {Hua}, \citenamefont {J{\"a}ger}, \citenamefont {Junnakar}, \citenamefont
  {Meyer}, \citenamefont {Rae},\ and\ \citenamefont
  {Wittig}}]{vonHippel:2014hla}%
  \BibitemOpen
  \bibfield  {author} {\bibinfo {author} {\bibfnamefont {G.~M.}\ \bibnamefont
  {von Hippel}}, \bibinfo {author} {\bibfnamefont {S.}~\bibnamefont
  {Capitani}}, \bibinfo {author} {\bibfnamefont {D.}~\bibnamefont
  {Djukanovic}}, \bibinfo {author} {\bibfnamefont {J.}~\bibnamefont {Hua}},
  \bibinfo {author} {\bibfnamefont {B.}~\bibnamefont {J{\"a}ger}}, \bibinfo
  {author} {\bibfnamefont {P.}~\bibnamefont {Junnakar}}, \bibinfo {author}
  {\bibfnamefont {H.~B.}\ \bibnamefont {Meyer}}, \bibinfo {author}
  {\bibfnamefont {T.~D.}\ \bibnamefont {Rae}}, \ and\ \bibinfo {author}
  {\bibfnamefont {H.}~\bibnamefont {Wittig}},\ }\href@noop {} {\bibfield
  {journal} {\bibinfo  {journal} {Proc. Science}\ }\textbf {\bibinfo {volume}
  {LATTICE2014}},\ \bibinfo {pages} {147} (\bibinfo {year} {2015})},\ \Eprint
  {http://arxiv.org/abs/1411.4804} {arXiv:1411.4804 [hep-lat]} \BibitemShut
  {NoStop}%
\bibitem [{\citenamefont {Syritsyn}\ \emph {et~al.}(2015)\citenamefont
  {Syritsyn} \emph {et~al.}}]{Syritsyn:2014xwa}%
  \BibitemOpen
  \bibfield  {author} {\bibinfo {author} {\bibfnamefont {S.}~\bibnamefont
  {Syritsyn}} \emph {et~al.},\ }\href@noop {} {\bibfield  {journal} {\bibinfo
  {journal} {Proc. Science}\ }\textbf {\bibinfo {volume} {LATTICE2014}},\
  \bibinfo {pages} {134} (\bibinfo {year} {2015})},\ \Eprint
  {http://arxiv.org/abs/1412.3175} {arXiv:1412.3175 [hep-lat]} \BibitemShut
  {NoStop}%
\bibitem [{\citenamefont {Syritsyn}(2014)}]{Syritsyn:2014saa}%
  \BibitemOpen
  \bibfield  {author} {\bibinfo {author} {\bibfnamefont {S.}~\bibnamefont
  {Syritsyn}},\ }\href@noop {} {\bibfield  {journal} {\bibinfo  {journal}
  {Proc. Science}\ }\textbf {\bibinfo {volume} {LATTICE 2013}},\ \bibinfo
  {pages} {009} (\bibinfo {year} {2014})},\ \Eprint
  {http://arxiv.org/abs/1403.4686} {arXiv:1403.4686 [hep-lat]} \BibitemShut
  {NoStop}%
\bibitem [{\citenamefont {Alexandrou}(2013)}]{Alexandrou:2013asa}%
  \BibitemOpen
  \bibfield  {author} {\bibinfo {author} {\bibfnamefont {C.}~\bibnamefont
  {Alexandrou}},\ }\href {\doibase 10.1063/1.4826707} {\bibfield  {journal}
  {\bibinfo  {journal} {AIP Conf. Proc.}\ }\textbf {\bibinfo {volume} {1560}},\
  \bibinfo {pages} {3} (\bibinfo {year} {2013})}\BibitemShut {NoStop}%
\bibitem [{\citenamefont {Brambilla}\ \emph {et~al.}(2014)\citenamefont
  {Brambilla} \emph {et~al.}}]{Brambilla:2014aaa}%
  \BibitemOpen
  \bibfield  {author} {\bibinfo {author} {\bibfnamefont {N.}~\bibnamefont
  {Brambilla}} \emph {et~al.},\ }\href {\doibase
  10.1140/epjc/s10052-014-2981-5} {\bibfield  {journal} {\bibinfo  {journal}
  {Eur. Phys. J. C}\ }\textbf {\bibinfo {volume} {74}},\ \bibinfo {pages}
  {2981} (\bibinfo {year} {2014})},\ \Eprint {http://arxiv.org/abs/1404.3723}
  {arXiv:1404.3723 [hep-ph]} \BibitemShut {NoStop}%
\bibitem [{\citenamefont {Constantinou}(2015)}]{Constantinou:2014tga}%
  \BibitemOpen
  \bibfield  {author} {\bibinfo {author} {\bibfnamefont {M.}~\bibnamefont
  {Constantinou}},\ }\href@noop {} {\bibfield  {journal} {\bibinfo  {journal}
  {Proc. Science}\ }\textbf {\bibinfo {volume} {LATTICE2014}},\ \bibinfo
  {pages} {001} (\bibinfo {year} {2015})},\ \Eprint
  {http://arxiv.org/abs/1411.0078} {arXiv:1411.0078 [hep-lat]} \BibitemShut
  {NoStop}%
\bibitem [{\citenamefont {Green}(2014)}]{Green:2014vxa}%
  \BibitemOpen
  \bibfield  {author} {\bibinfo {author} {\bibfnamefont {J.}~\bibnamefont
  {Green}},\ }\href@noop {} {\enquote {\bibinfo {title} {{Hadron Structure from
  Lattice QCD}},}\ } (\bibinfo {year} {2014}),\ \Eprint
  {http://arxiv.org/abs/1412.4637} {arXiv:1412.4637 [hep-lat]} \BibitemShut
  {NoStop}%
\bibitem [{\citenamefont {Bali}\ \emph
  {et~al.}(2014{\natexlab{a}})\citenamefont {Bali}, \citenamefont {Collins},
  \citenamefont {Gl{\"a}{\ss}le}, \citenamefont {G{\"o}ckeler}, \citenamefont
  {Najjar}, \citenamefont {R{\"o}dl}, \citenamefont {Sch{\"afer}},
  \citenamefont {Schiel},\ and\ \citenamefont {S{\"o}ldner}}]{Bali:2013nla}%
  \BibitemOpen
  \bibfield  {author} {\bibinfo {author} {\bibfnamefont {G.~S.}\ \bibnamefont
  {Bali}}, \bibinfo {author} {\bibfnamefont {S.}~\bibnamefont {Collins}},
  \bibinfo {author} {\bibfnamefont {B.}~\bibnamefont {Gl{\"a}{\ss}le}},
  \bibinfo {author} {\bibfnamefont {M.}~\bibnamefont {G{\"o}ckeler}}, \bibinfo
  {author} {\bibfnamefont {J.}~\bibnamefont {Najjar}}, \bibinfo {author}
  {\bibfnamefont {R.~H.}\ \bibnamefont {R{\"o}dl}}, \bibinfo {author}
  {\bibfnamefont {A.}~\bibnamefont {Sch{\"afer}}}, \bibinfo {author}
  {\bibfnamefont {R.~W.}\ \bibnamefont {Schiel}}, \ and\ \bibinfo {author}
  {\bibfnamefont {W.}~\bibnamefont {S{\"o}ldner}},\ }\href@noop {} {\bibfield
  {journal} {\bibinfo  {journal} {Proc. Science}\ }\textbf {\bibinfo {volume}
  {LATTICE 2013}},\ \bibinfo {pages} {290} (\bibinfo {year}
  {2014}{\natexlab{a}})},\ \Eprint {http://arxiv.org/abs/1311.7041}
  {arXiv:1311.7041 [hep-lat]} \BibitemShut {NoStop}%
\bibitem [{\citenamefont {Bali}\ \emph {et~al.}(2013)\citenamefont {Bali} \emph
  {et~al.}}]{Bali:2012qs}%
  \BibitemOpen
  \bibfield  {author} {\bibinfo {author} {\bibfnamefont {G.~S.}\ \bibnamefont
  {Bali}} \emph {et~al.} (\bibinfo {collaboration} {QCDSF Collaboration}),\
  }\href {\doibase 10.1016/j.nuclphysb.2012.08.009} {\bibfield  {journal}
  {\bibinfo  {journal} {Nucl. Phys. B}\ }\textbf {\bibinfo {volume} {866}},\
  \bibinfo {pages} {1} (\bibinfo {year} {2013})},\ \Eprint
  {http://arxiv.org/abs/1206.7034} {arXiv:1206.7034 [hep-lat]} \BibitemShut
  {NoStop}%
\bibitem [{\citenamefont {Capitani}\ \emph {et~al.}(2011)\citenamefont
  {Capitani}, \citenamefont {Della~Morte}, \citenamefont {von Hippel},
  \citenamefont {Knippschild},\ and\ \citenamefont {Wittig}}]{Capitani:2011fg}%
  \BibitemOpen
  \bibfield  {author} {\bibinfo {author} {\bibfnamefont {S.}~\bibnamefont
  {Capitani}}, \bibinfo {author} {\bibfnamefont {M.}~\bibnamefont
  {Della~Morte}}, \bibinfo {author} {\bibfnamefont {G.}~\bibnamefont {von
  Hippel}}, \bibinfo {author} {\bibfnamefont {B.}~\bibnamefont {Knippschild}},
  \ and\ \bibinfo {author} {\bibfnamefont {H.}~\bibnamefont {Wittig}},\
  }\href@noop {} {\bibfield  {journal} {\bibinfo  {journal} {Proc. Science}\
  }\textbf {\bibinfo {volume} {Lattice 2011}},\ \bibinfo {pages} {145}
  (\bibinfo {year} {2011})},\ \Eprint {http://arxiv.org/abs/1110.6365}
  {arXiv:1110.6365 [hep-lat]} \BibitemShut {NoStop}%
\bibitem [{\citenamefont {Fritzsch}\ \emph {et~al.}(2012)\citenamefont
  {Fritzsch}, \citenamefont {Knechtli}, \citenamefont {Leder}, \citenamefont
  {Marinkovic}, \citenamefont {Schaefer}, \citenamefont {Sommer},\ and\
  \citenamefont {Virotta}}]{Fritzsch:2012wq}%
  \BibitemOpen
  \bibfield  {author} {\bibinfo {author} {\bibfnamefont {P.}~\bibnamefont
  {Fritzsch}}, \bibinfo {author} {\bibfnamefont {F.}~\bibnamefont {Knechtli}},
  \bibinfo {author} {\bibfnamefont {B.}~\bibnamefont {Leder}}, \bibinfo
  {author} {\bibfnamefont {M.}~\bibnamefont {Marinkovic}}, \bibinfo {author}
  {\bibfnamefont {S.}~\bibnamefont {Schaefer}}, \bibinfo {author}
  {\bibfnamefont {R.}~\bibnamefont {Sommer}}, \ and\ \bibinfo {author}
  {\bibfnamefont {F.}~\bibnamefont {Virotta}} (\bibinfo {collaboration} {ALPHA
  Collaboration}),\ }\href {\doibase 10.1016/j.nuclphysb.2012.07.026}
  {\bibfield  {journal} {\bibinfo  {journal} {Nucl. Phys. B}\ }\textbf
  {\bibinfo {volume} {865}},\ \bibinfo {pages} {397} (\bibinfo {year}
  {2012})},\ \Eprint {http://arxiv.org/abs/1205.5380} {arXiv:1205.5380
  [hep-lat]} \BibitemShut {NoStop}%
\bibitem [{\citenamefont {Bali}\ \emph
  {et~al.}(2014{\natexlab{b}})\citenamefont {Bali}, \citenamefont {Collins},
  \citenamefont {Gl{\"a\ss{}}le}, \citenamefont {G{\"o}ckeler}, \citenamefont
  {Najjar}, \citenamefont {R{\"o}dl}, \citenamefont {Sch{\"a}fer},
  \citenamefont {Schiel}, \citenamefont {Sternbeck},\ and\ \citenamefont
  {S{\"o}ldner}}]{Bali:2014gha}%
  \BibitemOpen
  \bibfield  {author} {\bibinfo {author} {\bibfnamefont {G.~S.}\ \bibnamefont
  {Bali}}, \bibinfo {author} {\bibfnamefont {S.}~\bibnamefont {Collins}},
  \bibinfo {author} {\bibfnamefont {B.}~\bibnamefont {Gl{\"a\ss{}}le}},
  \bibinfo {author} {\bibfnamefont {M.}~\bibnamefont {G{\"o}ckeler}}, \bibinfo
  {author} {\bibfnamefont {J.}~\bibnamefont {Najjar}}, \bibinfo {author}
  {\bibfnamefont {R.~H.}\ \bibnamefont {R{\"o}dl}}, \bibinfo {author}
  {\bibfnamefont {A.}~\bibnamefont {Sch{\"a}fer}}, \bibinfo {author}
  {\bibfnamefont {R.~W.}\ \bibnamefont {Schiel}}, \bibinfo {author}
  {\bibfnamefont {A.}~\bibnamefont {Sternbeck}}, \ and\ \bibinfo {author}
  {\bibfnamefont {W.}~\bibnamefont {S{\"o}ldner}},\ }\href {\doibase
  10.1103/PhysRevD.90.074510} {\bibfield  {journal} {\bibinfo  {journal} {Phys.
  Rev. D}\ }\textbf {\bibinfo {volume} {90}},\ \bibinfo {pages} {074510}
  (\bibinfo {year} {2014}{\natexlab{b}})},\ \Eprint
  {http://arxiv.org/abs/1408.6850} {arXiv:1408.6850 [hep-lat]} \BibitemShut
  {NoStop}%
\bibitem [{\citenamefont {Bali}\ \emph {et~al.}(2005)\citenamefont {Bali},
  \citenamefont {Neff}, \citenamefont {{D\"ussel}}, \citenamefont {Lippert},\
  and\ \citenamefont {Schilling}}]{Bali:2005fu}%
  \BibitemOpen
  \bibfield  {author} {\bibinfo {author} {\bibfnamefont {G.~S.}\ \bibnamefont
  {Bali}}, \bibinfo {author} {\bibfnamefont {H.}~\bibnamefont {Neff}}, \bibinfo
  {author} {\bibfnamefont {T.}~\bibnamefont {{D\"ussel}}}, \bibinfo {author}
  {\bibfnamefont {T.}~\bibnamefont {Lippert}}, \ and\ \bibinfo {author}
  {\bibfnamefont {K.}~\bibnamefont {Schilling}} (\bibinfo {collaboration}
  {SESAM Collaboration}),\ }\href {\doibase 10.1103/PhysRevD.71.114513}
  {\bibfield  {journal} {\bibinfo  {journal} {Phys. Rev. D}\ }\textbf {\bibinfo
  {volume} {71}},\ \bibinfo {pages} {114513} (\bibinfo {year} {2005})},\
  \Eprint {http://arxiv.org/abs/hep-lat/0505012} {arXiv:hep-lat/0505012
  [hep-lat]} \BibitemShut {NoStop}%
\bibitem [{\citenamefont {{G\"usken}}\ \emph {et~al.}(1989)\citenamefont
  {{G\"usken}}, \citenamefont {{L\"ow}}, \citenamefont {{M\"utter}},
  \citenamefont {Sommer}, \citenamefont {Patel},\ and\ \citenamefont
  {Schilling}}]{Gusken:1989ad}%
  \BibitemOpen
  \bibfield  {author} {\bibinfo {author} {\bibfnamefont {S.}~\bibnamefont
  {{G\"usken}}}, \bibinfo {author} {\bibfnamefont {U.}~\bibnamefont {{L\"ow}}},
  \bibinfo {author} {\bibfnamefont {K.~H.}\ \bibnamefont {{M\"utter}}},
  \bibinfo {author} {\bibfnamefont {R.}~\bibnamefont {Sommer}}, \bibinfo
  {author} {\bibfnamefont {A.}~\bibnamefont {Patel}}, \ and\ \bibinfo {author}
  {\bibfnamefont {K.}~\bibnamefont {Schilling}},\ }\href {\doibase
  10.1016/S0370-2693(89)80034-6} {\bibfield  {journal} {\bibinfo  {journal}
  {Phys. Lett. B}\ }\textbf {\bibinfo {volume} {227}},\ \bibinfo {pages} {266}
  (\bibinfo {year} {1989})}\BibitemShut {NoStop}%
\bibitem [{\citenamefont {Falcioni}\ \emph {et~al.}(1985)\citenamefont
  {Falcioni}, \citenamefont {Paciello}, \citenamefont {Parisi},\ and\
  \citenamefont {Taglienti}}]{Falcioni:1984ei}%
  \BibitemOpen
  \bibfield  {author} {\bibinfo {author} {\bibfnamefont {M.}~\bibnamefont
  {Falcioni}}, \bibinfo {author} {\bibfnamefont {M.~L.}\ \bibnamefont
  {Paciello}}, \bibinfo {author} {\bibfnamefont {G.}~\bibnamefont {Parisi}}, \
  and\ \bibinfo {author} {\bibfnamefont {B.}~\bibnamefont {Taglienti}},\ }\href
  {\doibase 10.1016/0550-3213(85)90280-9} {\bibfield  {journal} {\bibinfo
  {journal} {Nucl. Phys. B}\ }\textbf {\bibinfo {volume} {251}},\ \bibinfo
  {pages} {624} (\bibinfo {year} {1985})}\BibitemShut {NoStop}%
\bibitem [{\citenamefont {Bali}\ \emph
  {et~al.}(2014{\natexlab{c}})\citenamefont {Bali}, \citenamefont {Collins},
  \citenamefont {{Gl\"a\ss{}le}}, \citenamefont {{G\"ockeler}}, \citenamefont
  {Najjar}, \citenamefont {R{\"o}dl}, \citenamefont {Sch{\"a}fer},
  \citenamefont {Sternbeck},\ and\ \citenamefont {S{\"o}ldner}}]{Bali:2013gxx}%
  \BibitemOpen
  \bibfield  {author} {\bibinfo {author} {\bibfnamefont {G.~S.}\ \bibnamefont
  {Bali}}, \bibinfo {author} {\bibfnamefont {S.}~\bibnamefont {Collins}},
  \bibinfo {author} {\bibfnamefont {B.}~\bibnamefont {{Gl\"a\ss{}le}}},
  \bibinfo {author} {\bibfnamefont {M.}~\bibnamefont {{G\"ockeler}}}, \bibinfo
  {author} {\bibfnamefont {J.}~\bibnamefont {Najjar}}, \bibinfo {author}
  {\bibfnamefont {R.~H.}\ \bibnamefont {R{\"o}dl}}, \bibinfo {author}
  {\bibfnamefont {A.}~\bibnamefont {Sch{\"a}fer}}, \bibinfo {author}
  {\bibfnamefont {A.}~\bibnamefont {Sternbeck}}, \ and\ \bibinfo {author}
  {\bibfnamefont {W.}~\bibnamefont {S{\"o}ldner}},\ }\href@noop {} {\bibfield
  {journal} {\bibinfo  {journal} {Proc. Science}\ }\textbf {\bibinfo {volume}
  {LATTICE 2013}},\ \bibinfo {pages} {271} (\bibinfo {year}
  {2014}{\natexlab{c}})},\ \Eprint {http://arxiv.org/abs/1311.1718}
  {arXiv:1311.1718 [hep-lat]} \BibitemShut {NoStop}%
\bibitem [{\citenamefont {Evans}\ \emph {et~al.}(2010)\citenamefont {Evans},
  \citenamefont {Bali},\ and\ \citenamefont {Collins}}]{Evans:2010tg}%
  \BibitemOpen
  \bibfield  {author} {\bibinfo {author} {\bibfnamefont {R.}~\bibnamefont
  {Evans}}, \bibinfo {author} {\bibfnamefont {G.}~\bibnamefont {Bali}}, \ and\
  \bibinfo {author} {\bibfnamefont {S.}~\bibnamefont {Collins}},\ }\href
  {\doibase 10.1103/PhysRevD.82.094501} {\bibfield  {journal} {\bibinfo
  {journal} {Phys. Rev. D}\ }\textbf {\bibinfo {volume} {82}},\ \bibinfo
  {pages} {094501} (\bibinfo {year} {2010})},\ \Eprint
  {http://arxiv.org/abs/1008.3293} {arXiv:1008.3293 [hep-lat]} \BibitemShut
  {NoStop}%
\bibitem [{\citenamefont {Alexandrou}\ \emph
  {et~al.}(2014{\natexlab{c}})\citenamefont {Alexandrou} \emph
  {et~al.}}]{Alexandrou:2013xon}%
  \BibitemOpen
  \bibfield  {author} {\bibinfo {author} {\bibfnamefont {C.}~\bibnamefont
  {Alexandrou}} \emph {et~al.} (\bibinfo {collaboration} {ETM Collaboration}),\
  }\href {\doibase 10.1140/epjc/s10052-013-2692-3} {\bibfield  {journal}
  {\bibinfo  {journal} {Eur. Phys. J. C}\ }\textbf {\bibinfo {volume} {74}},\
  \bibinfo {pages} {2692} (\bibinfo {year} {2014}{\natexlab{c}})},\ \Eprint
  {http://arxiv.org/abs/1302.2608} {arXiv:1302.2608 [hep-lat]} \BibitemShut
  {NoStop}%
\bibitem [{\citenamefont {Maiani}\ \emph {et~al.}(1987)\citenamefont {Maiani},
  \citenamefont {Martinelli}, \citenamefont {Paciello},\ and\ \citenamefont
  {Taglienti}}]{Maiani:1987by}%
  \BibitemOpen
  \bibfield  {author} {\bibinfo {author} {\bibfnamefont {L.}~\bibnamefont
  {Maiani}}, \bibinfo {author} {\bibfnamefont {G.}~\bibnamefont {Martinelli}},
  \bibinfo {author} {\bibfnamefont {M.~L.}\ \bibnamefont {Paciello}}, \ and\
  \bibinfo {author} {\bibfnamefont {B.}~\bibnamefont {Taglienti}},\ }\href
  {\doibase 10.1016/0550-3213(87)90078-2} {\bibfield  {journal} {\bibinfo
  {journal} {Nucl. Phys. B}\ }\textbf {\bibinfo {volume} {293}},\ \bibinfo
  {pages} {420} (\bibinfo {year} {1987})}\BibitemShut {NoStop}%
\bibitem [{\citenamefont {Capitani}\ \emph {et~al.}(2010)\citenamefont
  {Capitani}, \citenamefont {Knippschild}, \citenamefont {Della~Morte},\ and\
  \citenamefont {Wittig}}]{Capitani:2010sg}%
  \BibitemOpen
  \bibfield  {author} {\bibinfo {author} {\bibfnamefont {S.}~\bibnamefont
  {Capitani}}, \bibinfo {author} {\bibfnamefont {B.}~\bibnamefont
  {Knippschild}}, \bibinfo {author} {\bibfnamefont {M.}~\bibnamefont
  {Della~Morte}}, \ and\ \bibinfo {author} {\bibfnamefont {H.}~\bibnamefont
  {Wittig}},\ }\href@noop {} {\bibfield  {journal} {\bibinfo  {journal} {Proc.
  Science}\ }\textbf {\bibinfo {volume} {Lattice 2010}},\ \bibinfo {pages}
  {147} (\bibinfo {year} {2010})},\ \Eprint {http://arxiv.org/abs/1011.1358}
  {arXiv:1011.1358 [hep-lat]} \BibitemShut {NoStop}%
\bibitem [{\citenamefont {Allton}\ \emph {et~al.}(1993)\citenamefont {Allton}
  \emph {et~al.}}]{Allton:1993wc}%
  \BibitemOpen
  \bibfield  {author} {\bibinfo {author} {\bibfnamefont {C.~R.}\ \bibnamefont
  {Allton}} \emph {et~al.} (\bibinfo {collaboration} {UKQCD Collaboration}),\
  }\href {\doibase 10.1103/PhysRevD.47.5128} {\bibfield  {journal} {\bibinfo
  {journal} {Phys. Rev. D}\ }\textbf {\bibinfo {volume} {47}},\ \bibinfo
  {pages} {5128} (\bibinfo {year} {1993})},\ \Eprint
  {http://arxiv.org/abs/hep-lat/9303009} {arXiv:hep-lat/9303009 [hep-lat]}
  \BibitemShut {NoStop}%
\bibitem [{\citenamefont {Della~Morte}\ \emph {et~al.}(2005)\citenamefont
  {Della~Morte}, \citenamefont {Hoffmann},\ and\ \citenamefont
  {Sommer}}]{DellaMorte:2005se}%
  \BibitemOpen
  \bibfield  {author} {\bibinfo {author} {\bibfnamefont {M.}~\bibnamefont
  {Della~Morte}}, \bibinfo {author} {\bibfnamefont {R.}~\bibnamefont
  {Hoffmann}}, \ and\ \bibinfo {author} {\bibfnamefont {R.}~\bibnamefont
  {Sommer}} (\bibinfo {collaboration} {ALPHA Collaboration}),\ }\href {\doibase
  10.1088/1126-6708/2005/03/029} {\bibfield  {journal} {\bibinfo  {journal} {J.
  High Energy Phys.}\ }\textbf {\bibinfo {volume} {03}},\ \bibinfo {pages}
  {029} (\bibinfo {year} {2005})},\ \Eprint
  {http://arxiv.org/abs/hep-lat/0503003} {arXiv:hep-lat/0503003 [hep-lat]}
  \BibitemShut {NoStop}%
\bibitem [{\citenamefont {G{\"o}ckeler}\ \emph {et~al.}(2010)\citenamefont
  {G{\"o}ckeler} \emph {et~al.}}]{Gockeler:2010yr}%
  \BibitemOpen
  \bibfield  {author} {\bibinfo {author} {\bibfnamefont {M.}~\bibnamefont
  {G{\"o}ckeler}} \emph {et~al.} (\bibinfo {collaboration} {QCDSF and UKQCD
  Collaborations}),\ }\href {\doibase 10.1103/PhysRevD.82.114511} {\bibfield
  {journal} {\bibinfo  {journal} {Phys. Rev. D}\ }\textbf {\bibinfo {volume}
  {82}},\ \bibinfo {pages} {114511} (\bibinfo {year} {2010})},\ \Eprint
  {http://arxiv.org/abs/1003.5756} {arXiv:1003.5756 [hep-lat]} \BibitemShut
  {NoStop}%
\bibitem [{\citenamefont {Martinelli}\ \emph {et~al.}(1995)\citenamefont
  {Martinelli}, \citenamefont {Pittori}, \citenamefont {Sachrajda},
  \citenamefont {Testa},\ and\ \citenamefont {Vladikas}}]{Martinelli:1994ty}%
  \BibitemOpen
  \bibfield  {author} {\bibinfo {author} {\bibfnamefont {G.}~\bibnamefont
  {Martinelli}}, \bibinfo {author} {\bibfnamefont {C.}~\bibnamefont {Pittori}},
  \bibinfo {author} {\bibfnamefont {C.~T.}\ \bibnamefont {Sachrajda}}, \bibinfo
  {author} {\bibfnamefont {M.}~\bibnamefont {Testa}}, \ and\ \bibinfo {author}
  {\bibfnamefont {A.}~\bibnamefont {Vladikas}},\ }\href {\doibase
  10.1016/0550-3213(95)00126-D} {\bibfield  {journal} {\bibinfo  {journal}
  {Nucl. Phys. B}\ }\textbf {\bibinfo {volume} {445}},\ \bibinfo {pages} {81}
  (\bibinfo {year} {1995})},\ \Eprint {http://arxiv.org/abs/hep-lat/9411010}
  {arXiv:hep-lat/9411010 [hep-lat]} \BibitemShut {NoStop}%
\bibitem [{\citenamefont {Sint}\ and\ \citenamefont
  {Weisz}(1997)}]{Sint:1997jx}%
  \BibitemOpen
  \bibfield  {author} {\bibinfo {author} {\bibfnamefont {S.}~\bibnamefont
  {Sint}}\ and\ \bibinfo {author} {\bibfnamefont {P.}~\bibnamefont {Weisz}},\
  }\href {\doibase 10.1016/S0550-3213(97)00372-6} {\bibfield  {journal}
  {\bibinfo  {journal} {Nucl. Phys. B}\ }\textbf {\bibinfo {volume} {502}},\
  \bibinfo {pages} {251} (\bibinfo {year} {1997})},\ \Eprint
  {http://arxiv.org/abs/hep-lat/9704001} {arXiv:hep-lat/9704001 [hep-lat]}
  \BibitemShut {NoStop}%
\bibitem [{\citenamefont {Taniguchi}\ and\ \citenamefont
  {Ukawa}(1998)}]{Taniguchi:1998pf}%
  \BibitemOpen
  \bibfield  {author} {\bibinfo {author} {\bibfnamefont {Y.}~\bibnamefont
  {Taniguchi}}\ and\ \bibinfo {author} {\bibfnamefont {A.}~\bibnamefont
  {Ukawa}},\ }\href {\doibase 10.1103/PhysRevD.58.114503} {\bibfield  {journal}
  {\bibinfo  {journal} {Phys. Rev. D}\ }\textbf {\bibinfo {volume} {58}},\
  \bibinfo {pages} {114503} (\bibinfo {year} {1998})},\ \Eprint
  {http://arxiv.org/abs/hep-lat/9806015} {arXiv:hep-lat/9806015 [hep-lat]}
  \BibitemShut {NoStop}%
\bibitem [{\citenamefont {Capitani}\ \emph {et~al.}(2001)\citenamefont
  {Capitani}, \citenamefont {G{\"o}ckeler}, \citenamefont {Horsley},
  \citenamefont {Perlt}, \citenamefont {Rakow}, \citenamefont {Schierholz},\
  and\ \citenamefont {Schiller}}]{Capitani:2000xi}%
  \BibitemOpen
  \bibfield  {author} {\bibinfo {author} {\bibfnamefont {S.}~\bibnamefont
  {Capitani}}, \bibinfo {author} {\bibfnamefont {M.}~\bibnamefont
  {G{\"o}ckeler}}, \bibinfo {author} {\bibfnamefont {R.}~\bibnamefont
  {Horsley}}, \bibinfo {author} {\bibfnamefont {H.}~\bibnamefont {Perlt}},
  \bibinfo {author} {\bibfnamefont {P.~E.~L.}\ \bibnamefont {Rakow}}, \bibinfo
  {author} {\bibfnamefont {G.}~\bibnamefont {Schierholz}}, \ and\ \bibinfo
  {author} {\bibfnamefont {A.}~\bibnamefont {Schiller}},\ }\href {\doibase
  10.1016/S0550-3213(00)00590-3} {\bibfield  {journal} {\bibinfo  {journal}
  {Nucl. Phys. B}\ }\textbf {\bibinfo {volume} {593}},\ \bibinfo {pages} {183}
  (\bibinfo {year} {2001})},\ \Eprint {http://arxiv.org/abs/hep-lat/0007004}
  {arXiv:hep-lat/0007004 [hep-lat]} \BibitemShut {NoStop}%
\bibitem [{\citenamefont {Fritzsch}\ \emph {et~al.}(2010)\citenamefont
  {Fritzsch}, \citenamefont {Heitger},\ and\ \citenamefont
  {Tantalo}}]{Fritzsch:2010aw}%
  \BibitemOpen
  \bibfield  {author} {\bibinfo {author} {\bibfnamefont {P.}~\bibnamefont
  {Fritzsch}}, \bibinfo {author} {\bibfnamefont {J.}~\bibnamefont {Heitger}}, \
  and\ \bibinfo {author} {\bibfnamefont {N.}~\bibnamefont {Tantalo}} (\bibinfo
  {collaboration} {ALPHA Collaboration}),\ }\href {\doibase
  10.1007/JHEP08(2010)074} {\bibfield  {journal} {\bibinfo  {journal} {J. High
  Energy Phys.}\ }\textbf {\bibinfo {volume} {08}},\ \bibinfo {pages} {074}
  (\bibinfo {year} {2010})},\ \Eprint {http://arxiv.org/abs/1004.3978}
  {arXiv:1004.3978 [hep-lat]} \BibitemShut {NoStop}%
\bibitem [{\citenamefont {Della~Morte}\ \emph {et~al.}(2009)\citenamefont
  {Della~Morte}, \citenamefont {Sommer},\ and\ \citenamefont
  {Takeda}}]{DellaMorte:2008xb}%
  \BibitemOpen
  \bibfield  {author} {\bibinfo {author} {\bibfnamefont {M.}~\bibnamefont
  {Della~Morte}}, \bibinfo {author} {\bibfnamefont {R.}~\bibnamefont {Sommer}},
  \ and\ \bibinfo {author} {\bibfnamefont {S.}~\bibnamefont {Takeda}} (\bibinfo
  {collaboration} {ALPHA Collaboration}),\ }\href {\doibase
  10.1016/j.physletb.2009.01.059} {\bibfield  {journal} {\bibinfo  {journal}
  {Phys. Lett. B}\ }\textbf {\bibinfo {volume} {672}},\ \bibinfo {pages} {407}
  (\bibinfo {year} {2009})},\ \Eprint {http://arxiv.org/abs/0807.1120}
  {arXiv:0807.1120 [hep-lat]} \BibitemShut {NoStop}%
\bibitem [{\citenamefont {Hemmert}\ \emph {et~al.}(2003)\citenamefont
  {Hemmert}, \citenamefont {Procura},\ and\ \citenamefont
  {Weise}}]{Hemmert:2003cb}%
  \BibitemOpen
  \bibfield  {author} {\bibinfo {author} {\bibfnamefont {T.~R.}\ \bibnamefont
  {Hemmert}}, \bibinfo {author} {\bibfnamefont {M.}~\bibnamefont {Procura}}, \
  and\ \bibinfo {author} {\bibfnamefont {W.}~\bibnamefont {Weise}},\ }\href
  {\doibase 10.1103/PhysRevD.68.075009} {\bibfield  {journal} {\bibinfo
  {journal} {Phys. Rev. D}\ }\textbf {\bibinfo {volume} {68}},\ \bibinfo
  {pages} {075009} (\bibinfo {year} {2003})},\ \Eprint
  {http://arxiv.org/abs/hep-lat/0303002} {arXiv:hep-lat/0303002 [hep-lat]}
  \BibitemShut {NoStop}%
\bibitem [{\citenamefont {Gasser}\ and\ \citenamefont
  {Leutwyler}(1987)}]{Gasser:1986vb}%
  \BibitemOpen
  \bibfield  {author} {\bibinfo {author} {\bibfnamefont {J.}~\bibnamefont
  {Gasser}}\ and\ \bibinfo {author} {\bibfnamefont {H.}~\bibnamefont
  {Leutwyler}},\ }\href {\doibase 10.1016/0370-2693(87)90492-8} {\bibfield
  {journal} {\bibinfo  {journal} {Phys. Lett. B}\ }\textbf {\bibinfo {volume}
  {184}},\ \bibinfo {pages} {83} (\bibinfo {year} {1987})}\BibitemShut
  {NoStop}%
\bibitem [{\citenamefont {Gasser}\ and\ \citenamefont
  {Leutwyler}(1988)}]{Gasser:1987zq}%
  \BibitemOpen
  \bibfield  {author} {\bibinfo {author} {\bibfnamefont {J.}~\bibnamefont
  {Gasser}}\ and\ \bibinfo {author} {\bibfnamefont {H.}~\bibnamefont
  {Leutwyler}},\ }\href {\doibase 10.1016/0550-3213(88)90107-1} {\bibfield
  {journal} {\bibinfo  {journal} {Nucl. Phys. B}\ }\textbf {\bibinfo {volume}
  {307}},\ \bibinfo {pages} {763} (\bibinfo {year} {1988})}\BibitemShut
  {NoStop}%
\bibitem [{\citenamefont {Aoki}\ \emph {et~al.}(2014)\citenamefont {Aoki} \emph
  {et~al.}}]{Aoki:2013ldr}%
  \BibitemOpen
  \bibfield  {author} {\bibinfo {author} {\bibfnamefont {S.}~\bibnamefont
  {Aoki}} \emph {et~al.} (\bibinfo {collaboration} {FLAG Working Group}),\
  }\href {\doibase 10.1140/epjc/s10052-014-2890-7} {\bibfield  {journal}
  {\bibinfo  {journal} {Eur. Phys. J. C}\ }\textbf {\bibinfo {volume} {74}},\
  \bibinfo {pages} {2890} (\bibinfo {year} {2014})},\ \Eprint
  {http://arxiv.org/abs/1310.8555} {arXiv:1310.8555 [hep-lat]} \BibitemShut
  {NoStop}%
\bibitem [{\citenamefont {Colangelo}\ and\ \citenamefont
  {D{\"u}rr}(2004)}]{Colangelo:2003hf}%
  \BibitemOpen
  \bibfield  {author} {\bibinfo {author} {\bibfnamefont {G.}~\bibnamefont
  {Colangelo}}\ and\ \bibinfo {author} {\bibfnamefont {S.}~\bibnamefont
  {D{\"u}rr}},\ }\href {\doibase 10.1140/epjc/s2004-01593-y} {\bibfield
  {journal} {\bibinfo  {journal} {Eur. Phys. J. C}\ }\textbf {\bibinfo {volume}
  {33}},\ \bibinfo {pages} {543} (\bibinfo {year} {2004})},\ \Eprint
  {http://arxiv.org/abs/hep-lat/0311023} {arXiv:hep-lat/0311023 [hep-lat]}
  \BibitemShut {NoStop}%
\bibitem [{\citenamefont {Colangelo}\ \emph {et~al.}(2005)\citenamefont
  {Colangelo}, \citenamefont {D{\"u}rr},\ and\ \citenamefont
  {Haefeli}}]{Colangelo:2005gd}%
  \BibitemOpen
  \bibfield  {author} {\bibinfo {author} {\bibfnamefont {G.}~\bibnamefont
  {Colangelo}}, \bibinfo {author} {\bibfnamefont {S.}~\bibnamefont {D{\"u}rr}},
  \ and\ \bibinfo {author} {\bibfnamefont {C.}~\bibnamefont {Haefeli}},\ }\href
  {\doibase 10.1016/j.nuclphysb.2005.05.015} {\bibfield  {journal} {\bibinfo
  {journal} {Nucl. Phys. B}\ }\textbf {\bibinfo {volume} {721}},\ \bibinfo
  {pages} {136} (\bibinfo {year} {2005})},\ \Eprint
  {http://arxiv.org/abs/hep-lat/0503014} {arXiv:hep-lat/0503014 [hep-lat]}
  \BibitemShut {NoStop}%
\bibitem [{\citenamefont {Colangelo}\ \emph {et~al.}(2001)\citenamefont
  {Colangelo}, \citenamefont {Gasser},\ and\ \citenamefont
  {Leutwyler}}]{Colangelo:2001df}%
  \BibitemOpen
  \bibfield  {author} {\bibinfo {author} {\bibfnamefont {G.}~\bibnamefont
  {Colangelo}}, \bibinfo {author} {\bibfnamefont {J.}~\bibnamefont {Gasser}}, \
  and\ \bibinfo {author} {\bibfnamefont {H.}~\bibnamefont {Leutwyler}},\ }\href
  {\doibase 10.1016/S0550-3213(01)00147-X} {\bibfield  {journal} {\bibinfo
  {journal} {Nucl. Phys. B}\ }\textbf {\bibinfo {volume} {603}},\ \bibinfo
  {pages} {125} (\bibinfo {year} {2001})},\ \Eprint
  {http://arxiv.org/abs/hep-ph/0103088} {arXiv:hep-ph/0103088 [hep-ph]}
  \BibitemShut {NoStop}%
\bibitem [{\citenamefont {Beane}\ and\ \citenamefont
  {Savage}(2004)}]{Beane:2004rf}%
  \BibitemOpen
  \bibfield  {author} {\bibinfo {author} {\bibfnamefont {S.~R.}\ \bibnamefont
  {Beane}}\ and\ \bibinfo {author} {\bibfnamefont {M.~J.}\ \bibnamefont
  {Savage}} (\bibinfo {collaboration} {NPLQCD Collaboration}),\ }\href
  {\doibase 10.1103/PhysRevD.70.074029} {\bibfield  {journal} {\bibinfo
  {journal} {Phys. Rev. D}\ }\textbf {\bibinfo {volume} {70}},\ \bibinfo
  {pages} {074029} (\bibinfo {year} {2004})},\ \Eprint
  {http://arxiv.org/abs/hep-ph/0404131} {arXiv:hep-ph/0404131 [hep-ph]}
  \BibitemShut {NoStop}%
\bibitem [{\citenamefont {Procura}\ \emph {et~al.}(2006)\citenamefont
  {Procura}, \citenamefont {Musch}, \citenamefont {Wollenweber}, \citenamefont
  {Hemmert},\ and\ \citenamefont {Weise}}]{Procura:2006bj}%
  \BibitemOpen
  \bibfield  {author} {\bibinfo {author} {\bibfnamefont {M.}~\bibnamefont
  {Procura}}, \bibinfo {author} {\bibfnamefont {B.~U.}\ \bibnamefont {Musch}},
  \bibinfo {author} {\bibfnamefont {T.}~\bibnamefont {Wollenweber}}, \bibinfo
  {author} {\bibfnamefont {T.~R.}\ \bibnamefont {Hemmert}}, \ and\ \bibinfo
  {author} {\bibfnamefont {W.}~\bibnamefont {Weise}},\ }\href {\doibase
  10.1103/PhysRevD.73.114510} {\bibfield  {journal} {\bibinfo  {journal} {Phys.
  Rev. D}\ }\textbf {\bibinfo {volume} {73}},\ \bibinfo {pages} {114510}
  (\bibinfo {year} {2006})},\ \Eprint {http://arxiv.org/abs/hep-lat/0603001}
  {arXiv:hep-lat/0603001 [hep-lat]} \BibitemShut {NoStop}%
\bibitem [{\citenamefont {Ali~Khan}\ \emph {et~al.}(2006)\citenamefont
  {Ali~Khan} \emph {et~al.}}]{Khan:2006de}%
  \BibitemOpen
  \bibfield  {author} {\bibinfo {author} {\bibfnamefont {A.}~\bibnamefont
  {Ali~Khan}} \emph {et~al.} (\bibinfo {collaboration} {QCDSF Collaboration}),\
  }\href {\doibase 10.1103/PhysRevD.74.094508} {\bibfield  {journal} {\bibinfo
  {journal} {Phys. Rev. D}\ }\textbf {\bibinfo {volume} {74}},\ \bibinfo
  {pages} {094508} (\bibinfo {year} {2006})},\ \Eprint
  {http://arxiv.org/abs/hep-lat/0603028} {arXiv:hep-lat/0603028 [hep-lat]}
  \BibitemShut {NoStop}%
\bibitem [{\citenamefont {Hemmert}\ \emph {et~al.}(1998)\citenamefont
  {Hemmert}, \citenamefont {Holstein},\ and\ \citenamefont
  {Kambor}}]{Hemmert:1997ye}%
  \BibitemOpen
  \bibfield  {author} {\bibinfo {author} {\bibfnamefont {T.~R.}\ \bibnamefont
  {Hemmert}}, \bibinfo {author} {\bibfnamefont {B.~R.}\ \bibnamefont
  {Holstein}}, \ and\ \bibinfo {author} {\bibfnamefont {J.}~\bibnamefont
  {Kambor}},\ }\href {\doibase 10.1088/0954-3899/24/10/003} {\bibfield
  {journal} {\bibinfo  {journal} {J. Phys.}\ }\textbf {\bibinfo {volume}
  {G24}},\ \bibinfo {pages} {1831} (\bibinfo {year} {1998})},\ \Eprint
  {http://arxiv.org/abs/hep-ph/9712496} {arXiv:hep-ph/9712496 [hep-ph]}
  \BibitemShut {NoStop}%
\bibitem [{\citenamefont {Gail}\ and\ \citenamefont
  {Hemmert}(2006)}]{Gail:2005gz}%
  \BibitemOpen
  \bibfield  {author} {\bibinfo {author} {\bibfnamefont {T.~A.}\ \bibnamefont
  {Gail}}\ and\ \bibinfo {author} {\bibfnamefont {T.~R.}\ \bibnamefont
  {Hemmert}},\ }\href {\doibase 10.1140/epja/i2006-10023-y} {\bibfield
  {journal} {\bibinfo  {journal} {Eur. Phys. J. A}\ }\textbf {\bibinfo {volume}
  {28}},\ \bibinfo {pages} {91} (\bibinfo {year} {2006})},\ \Eprint
  {http://arxiv.org/abs/nucl-th/0512082} {arXiv:nucl-th/0512082 [nucl-th]}
  \BibitemShut {NoStop}%
\bibitem [{\citenamefont {{\v{S}}varc}\ \emph {et~al.}(2014)\citenamefont
  {{\v{S}}varc}, \citenamefont {Had{\v{z}}imehmedovi{\'{c}}}, \citenamefont
  {Osmanovi{\'{c}}}, \citenamefont {Stahov}, \citenamefont {Tiator},\ and\
  \citenamefont {Workman}}]{Svarc:2014sqa}%
  \BibitemOpen
  \bibfield  {author} {\bibinfo {author} {\bibfnamefont {A.}~\bibnamefont
  {{\v{S}}varc}}, \bibinfo {author} {\bibfnamefont {M.}~\bibnamefont
  {Had{\v{z}}imehmedovi{\'{c}}}}, \bibinfo {author} {\bibfnamefont
  {H.}~\bibnamefont {Osmanovi{\'{c}}}}, \bibinfo {author} {\bibfnamefont
  {J.}~\bibnamefont {Stahov}}, \bibinfo {author} {\bibfnamefont
  {L.}~\bibnamefont {Tiator}}, \ and\ \bibinfo {author} {\bibfnamefont {R.~L.}\
  \bibnamefont {Workman}},\ }\href {\doibase 10.1103/PhysRevC.89.065208}
  {\bibfield  {journal} {\bibinfo  {journal} {Phys. Rev. C}\ }\textbf {\bibinfo
  {volume} {89}},\ \bibinfo {pages} {065208} (\bibinfo {year} {2014})},\
  \Eprint {http://arxiv.org/abs/1404.1544} {arXiv:1404.1544 [nucl-th]}
  \BibitemShut {NoStop}%
\bibitem [{\citenamefont {Detmold}\ \emph {et~al.}(2002)\citenamefont
  {Detmold}, \citenamefont {Melnitchouk},\ and\ \citenamefont
  {Thomas}}]{Detmold:2002nf}%
  \BibitemOpen
  \bibfield  {author} {\bibinfo {author} {\bibfnamefont {W.}~\bibnamefont
  {Detmold}}, \bibinfo {author} {\bibfnamefont {W.}~\bibnamefont
  {Melnitchouk}}, \ and\ \bibinfo {author} {\bibfnamefont {A.~W.}\ \bibnamefont
  {Thomas}},\ }\href {\doibase 10.1103/PhysRevD.66.054501} {\bibfield
  {journal} {\bibinfo  {journal} {Phys. Rev. D}\ }\textbf {\bibinfo {volume}
  {66}},\ \bibinfo {pages} {054501} (\bibinfo {year} {2002})},\ \Eprint
  {http://arxiv.org/abs/hep-lat/0206001} {arXiv:hep-lat/0206001 [hep-lat]}
  \BibitemShut {NoStop}%
\bibitem [{\citenamefont {Gonz{\'a}lez-Alonso}\ and\ \citenamefont
  {Camalich}(2014)}]{Gonzalez-Alonso:2013ura}%
  \BibitemOpen
  \bibfield  {author} {\bibinfo {author} {\bibfnamefont {M.}~\bibnamefont
  {Gonz{\'a}lez-Alonso}}\ and\ \bibinfo {author} {\bibfnamefont {J.~M.}\
  \bibnamefont {Camalich}},\ }\href {\doibase 10.1103/PhysRevLett.112.042501}
  {\bibfield  {journal} {\bibinfo  {journal} {Phys. Rev. Lett.}\ }\textbf
  {\bibinfo {volume} {112}},\ \bibinfo {pages} {042501} (\bibinfo {year}
  {2014})},\ \Eprint {http://arxiv.org/abs/1309.4434} {arXiv:1309.4434
  [hep-ph]} \BibitemShut {NoStop}%
\bibitem [{\citenamefont {G{\"o}ckeler}\ \emph {et~al.}(2005)\citenamefont
  {G{\"o}ckeler}, \citenamefont {Hemmert}, \citenamefont {Horsley},
  \citenamefont {Pleiter}, \citenamefont {Rakow}, \citenamefont {Sch{\"afer}},\
  and\ \citenamefont {Schierholz}}]{Gockeler:2003ay}%
  \BibitemOpen
  \bibfield  {author} {\bibinfo {author} {\bibfnamefont {M.}~\bibnamefont
  {G{\"o}ckeler}}, \bibinfo {author} {\bibfnamefont {T.~R.}\ \bibnamefont
  {Hemmert}}, \bibinfo {author} {\bibfnamefont {R.}~\bibnamefont {Horsley}},
  \bibinfo {author} {\bibfnamefont {D.}~\bibnamefont {Pleiter}}, \bibinfo
  {author} {\bibfnamefont {P.~E.~L.}\ \bibnamefont {Rakow}}, \bibinfo {author}
  {\bibfnamefont {A.}~\bibnamefont {Sch{\"afer}}}, \ and\ \bibinfo {author}
  {\bibfnamefont {G.}~\bibnamefont {Schierholz}} (\bibinfo {collaboration}
  {QCDSF Collaboration}),\ }\href {\doibase 10.1103/PhysRevD.71.034508}
  {\bibfield  {journal} {\bibinfo  {journal} {Phys. Rev. D}\ }\textbf {\bibinfo
  {volume} {71}},\ \bibinfo {pages} {034508} (\bibinfo {year} {2005})},\
  \Eprint {http://arxiv.org/abs/hep-lat/0303019} {arXiv:hep-lat/0303019
  [hep-lat]} \BibitemShut {NoStop}%
\bibitem [{\citenamefont {Andreev}\ \emph {et~al.}(2013)\citenamefont {Andreev}
  \emph {et~al.}}]{Andreev:2012fj}%
  \BibitemOpen
  \bibfield  {author} {\bibinfo {author} {\bibfnamefont {V.~A.}\ \bibnamefont
  {Andreev}} \emph {et~al.} (\bibinfo {collaboration} {MuCap Collaboration}),\
  }\href {\doibase 10.1103/PhysRevLett.110.012504} {\bibfield  {journal}
  {\bibinfo  {journal} {Phys. Rev. Lett.}\ }\textbf {\bibinfo {volume} {110}},\
  \bibinfo {pages} {012504} (\bibinfo {year} {2013})},\ \Eprint
  {http://arxiv.org/abs/1210.6545} {arXiv:1210.6545 [nucl-ex]} \BibitemShut
  {NoStop}%
\bibitem [{\citenamefont {Schindler}\ \emph {et~al.}(2007)\citenamefont
  {Schindler}, \citenamefont {Fuchs}, \citenamefont {Gegelia},\ and\
  \citenamefont {Scherer}}]{Schindler:2006it}%
  \BibitemOpen
  \bibfield  {author} {\bibinfo {author} {\bibfnamefont {M.~R.}\ \bibnamefont
  {Schindler}}, \bibinfo {author} {\bibfnamefont {T.}~\bibnamefont {Fuchs}},
  \bibinfo {author} {\bibfnamefont {J.}~\bibnamefont {Gegelia}}, \ and\
  \bibinfo {author} {\bibfnamefont {S.}~\bibnamefont {Scherer}},\ }\href
  {\doibase 10.1103/PhysRevC.75.025202} {\bibfield  {journal} {\bibinfo
  {journal} {Phys. Rev. C}\ }\textbf {\bibinfo {volume} {75}},\ \bibinfo
  {pages} {025202} (\bibinfo {year} {2007})},\ \Eprint
  {http://arxiv.org/abs/nucl-th/0611083} {arXiv:nucl-th/0611083 [nucl-th]}
  \BibitemShut {NoStop}%
\bibitem [{\citenamefont {Ericson}\ \emph {et~al.}(2002)\citenamefont
  {Ericson}, \citenamefont {Loiseau},\ and\ \citenamefont
  {Thomas}}]{Ericson:2000md}%
  \BibitemOpen
  \bibfield  {author} {\bibinfo {author} {\bibfnamefont {T.~E.~O.}\
  \bibnamefont {Ericson}}, \bibinfo {author} {\bibfnamefont {B.}~\bibnamefont
  {Loiseau}}, \ and\ \bibinfo {author} {\bibfnamefont {A.~W.}\ \bibnamefont
  {Thomas}},\ }\href {\doibase 10.1103/PhysRevC.66.014005} {\bibfield
  {journal} {\bibinfo  {journal} {Phys. Rev. C}\ }\textbf {\bibinfo {volume}
  {66}},\ \bibinfo {pages} {014005} (\bibinfo {year} {2002})},\ \Eprint
  {http://arxiv.org/abs/hep-ph/0009312} {arXiv:hep-ph/0009312 [hep-ph]}
  \BibitemShut {NoStop}%
\bibitem [{\citenamefont {Arndt}\ \emph {et~al.}(2006)\citenamefont {Arndt},
  \citenamefont {Briscoe}, \citenamefont {Strakovsky},\ and\ \citenamefont
  {Workman}}]{Arndt:2006bf}%
  \BibitemOpen
  \bibfield  {author} {\bibinfo {author} {\bibfnamefont {R.~A.}\ \bibnamefont
  {Arndt}}, \bibinfo {author} {\bibfnamefont {W.~J.}\ \bibnamefont {Briscoe}},
  \bibinfo {author} {\bibfnamefont {I.~I.}\ \bibnamefont {Strakovsky}}, \ and\
  \bibinfo {author} {\bibfnamefont {R.~L.}\ \bibnamefont {Workman}},\ }\href
  {\doibase 10.1103/PhysRevC.74.045205} {\bibfield  {journal} {\bibinfo
  {journal} {Phys. Rev. C}\ }\textbf {\bibinfo {volume} {74}},\ \bibinfo
  {pages} {045205} (\bibinfo {year} {2006})},\ \Eprint
  {http://arxiv.org/abs/nucl-th/0605082} {arXiv:nucl-th/0605082 [nucl-th]}
  \BibitemShut {NoStop}%
\bibitem [{\citenamefont {Baru}\ \emph {et~al.}(2011)\citenamefont {Baru},
  \citenamefont {Hanhart}, \citenamefont {Hoferichter}, \citenamefont {Kubis},
  \citenamefont {Nogga},\ and\ \citenamefont {Philips}}]{Baru:2011bw}%
  \BibitemOpen
  \bibfield  {author} {\bibinfo {author} {\bibfnamefont {V.}~\bibnamefont
  {Baru}}, \bibinfo {author} {\bibfnamefont {C.}~\bibnamefont {Hanhart}},
  \bibinfo {author} {\bibfnamefont {M.}~\bibnamefont {Hoferichter}}, \bibinfo
  {author} {\bibfnamefont {B.}~\bibnamefont {Kubis}}, \bibinfo {author}
  {\bibfnamefont {A.}~\bibnamefont {Nogga}}, \ and\ \bibinfo {author}
  {\bibfnamefont {D.~R.}\ \bibnamefont {Philips}},\ }\href {\doibase
  10.1016/j.nuclphysa.2011.09.015} {\bibfield  {journal} {\bibinfo  {journal}
  {Nucl. Phys. A}\ }\textbf {\bibinfo {volume} {872}},\ \bibinfo {pages} {69}
  (\bibinfo {year} {2011})},\ \Eprint {http://arxiv.org/abs/1107.5509}
  {arXiv:1107.5509 [nucl-th]} \BibitemShut {NoStop}%
\bibitem [{\citenamefont {Alarc{\'o}n}\ \emph {et~al.}(2013)\citenamefont
  {Alarc{\'o}n}, \citenamefont {Martin~Camalich},\ and\ \citenamefont
  {Oller}}]{Alarcon:2012kn}%
  \BibitemOpen
  \bibfield  {author} {\bibinfo {author} {\bibfnamefont {J.~M.}\ \bibnamefont
  {Alarc{\'o}n}}, \bibinfo {author} {\bibfnamefont {J.~M.}\ \bibnamefont
  {Martin~Camalich}}, \ and\ \bibinfo {author} {\bibfnamefont {J.}~\bibnamefont
  {Oller}},\ }\href {\doibase 10.1016/j.aop.2013.06.001} {\bibfield  {journal}
  {\bibinfo  {journal} {Annals of Physics}\ }\textbf {\bibinfo {volume}
  {336}},\ \bibinfo {pages} {413} (\bibinfo {year} {2013})},\ \Eprint
  {http://arxiv.org/abs/1210.4450} {arXiv:1210.4450 [hep-ph]} \BibitemShut
  {NoStop}%
\bibitem [{\citenamefont {Baier}\ \emph {et~al.}(2009)\citenamefont {Baier}
  \emph {et~al.}}]{Baier:2009yq}%
  \BibitemOpen
  \bibfield  {author} {\bibinfo {author} {\bibfnamefont {H.}~\bibnamefont
  {Baier}} \emph {et~al.},\ }\href@noop {} {\bibfield  {journal} {\bibinfo
  {journal} {Proc. Science}\ }\textbf {\bibinfo {volume} {LAT2009}},\ \bibinfo
  {pages} {001} (\bibinfo {year} {2009})},\ \Eprint
  {http://arxiv.org/abs/0911.2174} {arXiv:0911.2174 [hep-lat]} \BibitemShut
  {NoStop}%
\bibitem [{\citenamefont {Nakamura}\ \emph {et~al.}(2011)\citenamefont
  {Nakamura}, \citenamefont {Nobile}, \citenamefont {Pleiter}, \citenamefont
  {Simma}, \citenamefont {Streuer}, \citenamefont {Wettig},\ and\ \citenamefont
  {Winter}}]{Nakamura:2011cd}%
  \BibitemOpen
  \bibfield  {author} {\bibinfo {author} {\bibfnamefont {Y.}~\bibnamefont
  {Nakamura}}, \bibinfo {author} {\bibfnamefont {A.}~\bibnamefont {Nobile}},
  \bibinfo {author} {\bibfnamefont {D.}~\bibnamefont {Pleiter}}, \bibinfo
  {author} {\bibfnamefont {H.}~\bibnamefont {Simma}}, \bibinfo {author}
  {\bibfnamefont {T.}~\bibnamefont {Streuer}}, \bibinfo {author} {\bibfnamefont
  {T.}~\bibnamefont {Wettig}}, \ and\ \bibinfo {author} {\bibfnamefont
  {F.}~\bibnamefont {Winter}},\ }\href@noop {} {\bibfield  {journal} {\bibinfo
  {journal} {Procedia Comput. Sci.}\ }\textbf {\bibinfo {volume} {4}},\
  \bibinfo {pages} {841} (\bibinfo {year} {2011})},\ \Eprint
  {http://arxiv.org/abs/1103.1363} {arXiv:1103.1363 [hep-lat]} \BibitemShut
  {NoStop}%
\bibitem [{\citenamefont {Nakamura}\ and\ \citenamefont
  {St{\"u}ben}(2010)}]{Nakamura:2010qh}%
  \BibitemOpen
  \bibfield  {author} {\bibinfo {author} {\bibfnamefont {Y.}~\bibnamefont
  {Nakamura}}\ and\ \bibinfo {author} {\bibfnamefont {H.}~\bibnamefont
  {St{\"u}ben}},\ }\href@noop {} {\bibfield  {journal} {\bibinfo  {journal}
  {Proc. Science}\ }\textbf {\bibinfo {volume} {Lattice 2010}},\ \bibinfo
  {pages} {040} (\bibinfo {year} {2010})},\ \Eprint
  {http://arxiv.org/abs/1011.0199} {arXiv:1011.0199 [hep-lat]} \BibitemShut
  {NoStop}%
\bibitem [{\citenamefont {Edwards}\ and\ \citenamefont
  {Jo\'o}(2005)}]{Edwards:2004sx}%
  \BibitemOpen
  \bibfield  {author} {\bibinfo {author} {\bibfnamefont {R.~G.}\ \bibnamefont
  {Edwards}}\ and\ \bibinfo {author} {\bibfnamefont {B.}~\bibnamefont {Jo\'o}}
  (\bibinfo {collaboration} {SciDAC, LHPC and UKQCD Collaborations}),\ }\href
  {\doibase 10.1016/j.nuclphysbps.2004.11.254} {\bibfield  {journal} {\bibinfo
  {journal} {Nucl. Phys. B Proc. Suppl.}\ }\textbf {\bibinfo {volume} {140}},\
  \bibinfo {pages} {832} (\bibinfo {year} {2005})},\ \Eprint
  {http://arxiv.org/abs/hep-lat/0409003} {arXiv:hep-lat/0409003 [hep-lat]}
  \BibitemShut {NoStop}%
\bibitem [{\citenamefont
  {\url{http://luscher.web.cern.ch/luscher/openQCD/}}()}]{luscher3}%
  \BibitemOpen
  \bibfield  {author} {\bibinfo {author} {\bibfnamefont {S.}~\bibnamefont
  {\url{http://luscher.web.cern.ch/luscher/openQCD/}}}\ }\href@noop {}
  {}\BibitemShut {NoStop}%
\end{thebibliography}%
\end{document}